\documentclass[11pt,a4paper]{article}

\usepackage[pdftex]{graphics}
\usepackage{jheppub}

\usepackage{amsmath,amssymb,amsfonts}

\usepackage{enumitem}
\usepackage{multirow}

\usepackage{array,booktabs}

\usepackage{slashed}
\usepackage{mathtools}
\allowdisplaybreaks
\usepackage{tcolorbox}
\usepackage{float}
\usepackage{tikz,subcaption}
\allowdisplaybreaks
\newcommand{\be}{\begin{eqnarray}}
\newcommand{\ee}{\end{eqnarray}}
\newcommand{\eqa}{\begin{eqnarray}}
\newcommand{\eqae}{\end{eqnarray}}
\newcommand{\nn}{\nonumber}
\newcommand{\bn}{\begin{enumerate}}
\newcommand{\en}{\end{enumerate}}
\newcommand{\bl}{\begin{align}}
\newcommand{\el}{\end{align}}
\newcommand{\eq}{\begin{equation}}
\newcommand{\eqe}{\end{equation}}
\newcommand{\Pv}[1]{|\vec{p}|^{#1}}
\newcommand{\IterCoeff}[4]{\mathcal{A}_{#1, #2; #3, #4}}

\def\identity{{\rlap{1} \hskip 1.6pt \hbox{1}}}
\def\iden{\identity}

\def\CI{{\cal I}}

\def\CO{{\cal O}}

\def\a{\alpha}
\def\b{\beta}
\def\g{\gamma}

\def\e{\epsilon}
\def\ve{\varepsilon}

\def\th{\theta}

\def\l{\lambda}
\def\m{\mu}
\def\n{\nu}

\def\s{\sigma}

\def\t{\tau}

\def\w{\omega}
\def\G{\Gamma}
\def\D{\Delta}

\def\half{\frac{1}{2}}

\def\p{\partial}

\def\identity{{\rlap{1} \hskip 1.6pt \hbox{1}}}

\newcommand{\bfig}{\begin{figure}}
\newcommand{\efig}{\end{figure}}

\def\la{{\langle}}
\def\ra{{\rangle}}
\def\abs#1{{\left| #1 \right|}}

\def\bl#1\el{\begin{align} #1 \end{align}}
\def\bg#1\eg{\begin{gather} #1 \end{gather}}
\def\bld#1\eld{\begin{aligned} #1 \end{aligned}}
\def\bgd#1\egd{\begin{gathered} #1 \end{gathered}}

\newcommand{\bra}[1]{\langle{#1}|}
\newcommand{\ket}[1]{|{#1}\rangle}

\newcommand{\sbra}[1]{ [{#1} |}
\newcommand{\sket}[1]{ | {#1} ]}
\newcommand{\AB}[1]{\langle #1\rangle}

\newcommand{\aPs}[3]{\langle #1|#2|#3]}

\newcommand{\cSquare}{\,\rotatebox{90}{\scalebox{0.6}[0.9]{$\bowtie$}}}

\def\jmath{{j}}
\def\bl#1\el{\begin{align} #1 \end{align}}
\def\bg#1\eg{\begin{gather} #1 \end{gather}}
\newcommand{\fig}[1]{figure \ref{#1}}
\def\bld#1\eld{\begin{aligned} #1 \end{aligned}}
\def\bgd#1\egd{\begin{gathered} #1 \end{gathered}}

\newcommand{\eqc}[1]{eq.(\ref{#1})}

\newcommand{\cbubble}{{\scalebox{0.8}[0.5]{$\bigcirc$}}}
\usetikzlibrary{arrows.meta,bending,calc,decorations.pathmorphing,decorations.markings,shapes.misc,shapes.geometric}
\pgfkeys{/tikz/.cd,
  circle color/.initial=black,
  circle color/.get=\circlecolor,
  circle color/.store in=\circlecolor,
}
\tikzset{/pgf/decoration/.cd,
    number of sines/.initial=10,
    angle step/.initial=20,
}
\newdimen\tmpdimen\pgfdeclaredecoration{complete sines}{initial}
{
    \state{initial}[
        width=+0pt,
        next state=move,
        persistent precomputation={
            \pgfmathparse{\pgfkeysvalueof{/pgf/decoration/angle step}}%
            \let\anglestep=\pgfmathresult%
            \let\currentangle=\pgfmathresult%
            \pgfmathsetlengthmacro{\pointsperanglestep}%
                {(\pgfdecoratedremainingdistance/\pgfkeysvalueof{/pgf/decoration/number of sines})/360*(\anglestep)}%
        }] {}
    \state{move}[width=+\pointsperanglestep, next state=draw]{
        \pgfpathmoveto{\pgfpointorigin}
    }
    \state{draw}[width=+\pointsperanglestep, switch if less than=1.25*\pointsperanglestep to final, 
        persistent postcomputation={
        \pgfmathparse{mod(\currentangle+\anglestep, 360)}%
        \let\currentangle=\pgfmathresult%
    }]{%
        \pgfmathsin{+\currentangle}%
        \tmpdimen=\pgfdecorationsegmentamplitude%
        \tmpdimen=\pgfmathresult\tmpdimen%
        \divide\tmpdimen by2\relax%
        \pgfpathlineto{\pgfqpoint{0pt}{\tmpdimen}}%
    }
    \state{final}{
        \ifdim\pgfdecoratedremainingdistance>0pt\relax
            \pgfpathlineto{\pgfpointdecoratedpathlast}
        \fi
   }
}

\title{The 2PM Hamiltonian for binary Kerr to quartic in spin}
\author[1,2]{Wei-Ming Chen} 
\author[3]{Ming-Zhi Chung} 
\author[3,4]{Yu-tin Huang}
\author[5]{Jung-Wook Kim}

\affiliation[1]{Institute of Physics, University of Amsterdam, Amsterdam, 1098 XH, The Netherlands}
\affiliation[2]{Department of Physics, Kobe University, Kobe 657-8501, Japan}
\affiliation[3]{Department of Physics and Astronomy, National Taiwan University, Taipei 10617, Taiwan}
\affiliation[4]{Physics Division, National Center for Theoretical Sciences, National Tsing-Hua University, No.101, Section 2, Kuang-Fu Road, Hsinchu, Taiwan}
\affiliation[5]{Centre for Theoretical Physics, Department of Physics and Astronomy,\\ Queen Mary University of London, Mile End Road, London E1 4NS, United Kingdom}

\emailAdd{tainist@gmail.com}
\emailAdd{dchung0741@gmail.com}
\emailAdd{yutin@phys.ntu.edu.tw}
\emailAdd{jung-wook.kim@qmul.ac.uk}

\abstract{From the S-matrix of spinning particles, we extract the 2 PM conservative potential for binary spinning black holes up to quartic order in spin operators. An important ingredient is the exponentiated gravitational Compton amplitude in the classical spin-limit for all graviton helicity sectors. The validity of the resulting Hamiltonian is verified by matching to known lower spin order results, as well as direct computation of the 2PM impulse and spin kicks from the eikonal phase and that from the test black hole scattering based on Mathisson-Papapetrou-Dixon equations.}

\begin{document}
\begin{flushright}
\vspace{10pt} \hfill{KOBE-COSMO-21-17\,,QMUL-PH-21-51} \vspace{20mm}
\end{flushright}

\maketitle

\newpage
\section{Introduction}
The detection of gravitational waves by LIGO and Virgo collaboration ~\cite{LIGOScientific:2016aoc, LIGOScientific:2017vwq} has brought upon rich opportunity for deeper understanding of gravitational dynamics, galaxy and star evolutions, as well as opening up new alleys in the era of multi-messenger astronomy. The potential for great discoveries also calls for a vigorous pursuit of precision theoretical predictions. The dynamics of the in-spiraling phase are conventionally captured by the post-Newtonian (PN) expansion, which describes the dynamics as Newtonian gravity perturbed by special ($v^2$) and general ($\frac{GM}{r}$) relativistic corrections~\cite{Einstein:1938yz, Ohta:1973je, Jaranowski:1997ky, Damour:1999cr, Blanchet:2000nv, Damour:2001bu, Damour:2014jta, Jaranowski:2015lha}, where the two dimensionless numbers controlling the expansion have similar order of magnitude ($v^2 \sim \frac{GM}{r}$) due to the virial theorem. The approach has recently been augmented by effective field theory methods~\cite{Goldberger:2004jt,Kol:2007rx, Kol:2007bc, Gilmore:2008gq, Foffa:2016rgu, Porto:2017dgs, Foffa:2019hrb, Blumlein:2019zku, Foffa:2019rdf}. 

An alternative organization of the perturbative expansion is to consider special relativistic dynamics as the unperturbed theory, which is corrected by general relativity ($GM/r$). This is known as the post-Minkowskian (PM) expansion~\cite{Bertotti:1956pxu, Kerr:1959zlt, Bertotti:1960wuq, Portilla:1979xx, Westpfahl:1979gu, Portilla:1980uz, Bel:1981be, Westpfahl:1985tsl, Ledvinka:2008tk,  Damour:2017zjx}. The revival of interest in PM dynamics is partly due to their application to effective one-body formalism~\cite{Damour:2016gwp,Bini:2017xzy}, which is one of the major frameworks for constructing gravitational waveform templates used in LIGO~\cite{LIGOScientific:2016vbw}. Another reason the PM expansion has gathered interest is because this is a natural context for applying scattering amplitude methods to classical gravitational dynamics~\cite{Cachazo:2017jef,Guevara:2017csg,Luna:2017dtq,Bjerrum-Bohr:2018xdl,Cheung:2018wkq,Caron-Huot:2018ape,Kosower:2018adc,Guevara:2018wpp,Chung:2018kqs,Bern:2019nnu,Bern:2019crd,Brandhuber:2019qpg,Maybee:2019jus,AccettulliHuber:2019jqo,KoemansCollado:2019ggb,Cristofoli:2019neg,Cristofoli:2019ewu,Guevara:2019fsj,Bjerrum-Bohr:2019kec,Cheung:2020gyp,Parra-Martinez:2020dzs,Cheung:2020sdj,Haddad:2020que,Bern:2020uwk,Aoude:2020ygw,AccettulliHuber:2020dal,Brandhuber:2021eyq,Bern:2021dqo,Kosmopoulos:2021zoq,DiVecchia:2021bdo,Bjerrum-Bohr:2021vuf,Bjerrum-Bohr:2021din,Carrillo-Gonzalez:2021mqj,Damgaard:2021ipf,Aoude:2021oqj,Kol:2021jjc,Bjerrum-Bohr:2021wwt}.

One major hurdle for analytic high-precision calculations in general relativity is the proliferation of high rank tensors; there are simply too many cranks and handles to coordinate in a computation. On the other hand, recent progress in scattering amplitudes has shown that conventional approaches contain redundant calculations which can be avoided by a judicious choice of computation methods. It was soon realised that the insights of modern amplitude techniques can also be applied to classical gravitational dynamics~\cite{Neill:2013wsa}, and the idea was subsequently developed~\cite{Bjerrum-Bohr:2018xdl, Cheung:2018wkq, Kosower:2018adc} and led to the breakthrough computation of 3 PM gravitational scattering~\cite{Bern:2019nnu,Bern:2019crd}. The current state-of-the-art result is classical gravitational scattering at 4 PM order~\cite{Bern:2021dqo} based on various tools of modern amplitude computations. Our aim is to add another item to this arsenal of tools; efficient reorganization of spin degrees of freedom.

In the EFT approach, spins are introduced as extra degrees of freedom on the worldline~\cite{Porto:2005ac}, accompanied by an infinite number of spin-induced multipole operators whose strengths are parametrised by the corresponding Wilson coefficients~\cite{Porto:2008jj, Levi:2015msa}. These coefficients can be fixed to all orders in spin at 1 PM by matching to the 1 PM potential~\cite{Vines:2016qwa}, or solving for the effective stress-energy tensor through the linearized field equations~\cite{Vines:2017hyw}. The on-shell S-matrix viewpoint alluded to the possibility of a simple organisation scheme for these seemingly complex infinitely many operators, through the observation that these infinitely many operators for a Kerr black hole corresponds to the minimally coupled gravitational three-point amplitude~\cite{Guevara:2018wpp, Chung:2018kqs, Arkani-Hamed:2019ymq, Aoude:2020onz} which only consists of a single term. The relation between black holes and minimal coupling was confirmed through reproduction of the full 1 PM potential, impulse and spin-deflections~\cite{Maybee:2019jus,Chung:2019duq,Guevara:2019fsj,Aoude:2021oqj}. 

The kinematics we are interested in has three length scales separated by large orders of magnitude; the de Broglie wavelength $\lambda_{dB} \sim |\vec{p}|^{-1}$ and the Compton wavelength $\l_C \sim m^{-1}$ controlling the quantum properties of the bodies,\footnote{In relativistic kinematics $v \sim \CO(1)$, therefore $\lambda_{dB} \sim \l_{C}$.} the spin length $a \sim |\vec{S}| m^{-1}$ controlling the ``size'' of the bodies,\footnote{We are interested in compact stars whose typical surface velocities lie in the relativistic regime $v \sim \CO(1)$, therefore the spin scales as $|\vec{S}| \sim m R_0$ where $R_0$ is the radius of the body.} and the impact parameter $|\vec{b}|$ controlling the separation between the bodies.
\bl
\lambda_{dB} \sim \l_{C} \ll a \ll |\vec{b}| 
 \,.
\el
Collectively denoting the wavelengths as $\l$, this separation of scales gives us two expansion parameters; ${\l}/a$ (${\l}/{|\vec{b}|}$) and ${a}/{|\vec{b}|}$. It is more convenient to recast these small numbers in the form
\bl
\frac{\l_C}{a} 
\sim \frac{\hbar}{|\vec{S}|} \sim \frac{1}{s} \,,\quad \frac{\l_{dB}}{|\vec{b}|} 
\sim \frac{\hbar}{|\vec{L}|} \sim \frac{1}{l} \,,\quad \frac{a}{|\vec{b}|} \sim \frac{|\vec{S}|}{m |\vec{b}|} \,, \label{eq:expansion_numbers}
\el
where $\vec{L} = \vec{b} \times \vec{p}$ is the orbital angular momentum, $s$ is the spin quantum number, and $l$ is the orbital angular momentum quantum number, with powers of $\hbar$ restored to match the dimensions. The scaling of small numbers \eqc{eq:expansion_numbers} shows that the classical limit corresponds to leading order kinematics for the expansion parameter(s) ${\l}/{a} \sim 1/s$ (${\l}/{|\vec{b}|} \sim 1/l$), while the spin expansion ${a}/{|\vec{b}|}$ is an expansion in the classical regime. An additional source of $\hbar$ is provided by virtual gravitons inside loops, which is tracked by restoring $\hbar$ to the coupling constant $G \rightarrow \frac{G}{\hbar}$. The inverse power of $\hbar$ is cancelled by transferred momentum $q^\m$, which should be counted as wavenumbers $\bar{q}^\m \equiv q^\m / \hbar$ in the classical limit~\cite{Kosower:2018adc}. Thus the classical limit corresponds to the following $\hbar$ dependence for the variables:
\begin{equation}\label{eq:PMexpparam}
q^{\mu} \rightarrow \hbar q^{\mu}, \quad
L^{\mu\nu}, S^{\mu} \rightarrow \frac{L^{\mu\nu}}{\hbar},\;\frac{S^{\mu}}{\hbar}, \quad
m_{a/b} \rightarrow m_{a/b}, \quad
p^{\mu} \rightarrow p^{\mu}\,, \quad G \rightarrow \frac{G}{\hbar} \,.
\end{equation}
Of the one-loop master integrals relevant to $\CO({G^2})$ order, only the triangle integral ($\propto q^{-1}$) has the right scaling to have direct classical contributions: The bubble integral ($\propto \log q^2$) has \emph{too many} powers of $q$ and corresponds to quantum corrections, while the box integral ($\propto q^{-2} \log q^2$) has \emph{too less} powers of $q$ and the behaviour was termed \emph{superclassical} in ref.\cite{Bern:2019crd}. For classical dynamics the bubble integral is irrelevant for obvious reasons, while the box integrals partially cancel between themselves and the remaining IR divergent part is given an interpretation as the iteration term from the lower perturbation order~\cite{Cheung:2018wkq,Bern:2019crd}. For the spin expansion we perform the subsitution $|\vec{b}| \to |\vec{q}|^{-1}$, since the impact parameter $\vec{b}$ is interpreted as the dual Fourier variable of transferred momentum $\vec{q}$. Therefore the classical spin expansion is an expansion by the small number
\bl
\frac{a}{|\vec{b}|} \sim \frac{|\vec{S}|}{m |\vec{b}|} \sim \frac{|\vec{S}| |\vec{q}|}{m} \,, \label{eq:spinexpparam}
\el
which also could be understood as the ratio of spin to orbital angular momentum $|\vec{S}|/|\vec{L}|$. We perform the expansion in kinematic expansion parameter \eqc{eq:spinexpparam} on top of the classical $G$ expansion provided by the scaling \eqc{eq:PMexpparam}.

In computing the 2 PM potential from scattering amplitudes, one of the important ingredients for the on-shell method is the gravitational compton amplitude, which contributes to the one-loop unitarity cut. In ref.\cite{Chung:2018kqs} the massive spinor helicity form of the amplitude was utilized accompanied by the \textit{holomorphic classical limit} (HCL)~\cite{Guevara:2017csg}, which misses certain spin coupling terms that contribute to the classical potential. This is not a huge limitation as missing terms become irrelevant in the phenomenologically interesting case of aligned-spin configuration~\cite{Guevara:2018wpp}, but it would be preferable to have a complete answer. Recent results at linear and quadratic order in spin~\cite{Bern:2020buy, Kosmopoulos:2021zoq} uses Feynman rules in the effective field theory framework~\cite{Bern:2020buy}. For higher order in spin, comparable (one-loop) order results for effective action are available in the context of PN EFT~\cite{Levi:2019kgk,Levi:2020lfn} (for one-loop order results for lower orders in spin, see refs.\cite{Porto:2005ac,Porto:2006bt, Porto:2008tb, Levi:2008nh, Porto:2010tr, Levi:2010zu}).

To simplify the complexity of the computation beyond quadratic in spin order, we instead utilize the exponentiated form of the Compton amplitude. Indeed it was first observed that the three-point amplitude for gravitational minimally coupled spinning particle exponentiates in the classical spin limit and reproduces the spin multipoles of the Kerr solution~\cite{Guevara:2018wpp, Chung:2018kqs, Arkani-Hamed:2019ymq}. Similarly, using soft factors such exponentiated form was presented for the opposite helicity graviton compton amplitude in~\cite{Guevara:2018wpp,Bautista:2019tdr,Guevara:2019fsj}. Here we instead utilize BCFW recursion~\cite{Britto:2005fq} on the three-point exponentiated form to derive the gravitational Compton amplitude for all helicity sectors, which also matches with previously known results. Note that the non-uniqueness of the Compton amplitude for minimal couplings with spins greater than 2 is reflected in the nonlocality of the exponentiated amplitude when expanded beyond quartic order in spins. This result reaches the limit constrained by our knowledge of the gravitational Compton that matches with Kerr.

Equipped with the gravitational compton amplitude, we proceed to compute the one-loop amplitude of gravitationally coupled spinning particles and extract its classical contributions. The one-loop amplitude is presented in scalar integral basis with its coefficients computed via unitarity methods. Classical contributions only arrises from scalar box and triangles, with the former carrying infrared divergences. We expand the integral coefficients to quartic order in total spin operators and compute the effective Hamiltonian based on the effective field theory (EFT) matching technique~\cite{Cheung:2018wkq} extended to spinning objects~\cite{Bern:2020buy}. In the EFT approach both 1 PM and 2 PM potentials contribute to the $\mathcal{O}(G^2)$ amplitude, where the former appears as iteration terms containing IR divergences. The matching of IR divergence serves as an initial consistency test, and subtracting off the iteration terms gives us the 2 PM potential up to quartic order in spin. For the 1 PM potential, we use the all order in spin 1 PM potential presented in ref.\cite{Chung:2018kqs} in a compact form.

Besides matching to earlier results at lower order in spin, we compute gauge invariant physical observables such as the impulse $\Delta \vec{p}$ and spin-kick $\Delta \vec{S}$ to test the validity of the Hamiltonian. Based on earlier works at 1 PM~\cite{Kosower:2018adc, Maybee:2019jus}, Bern et.~al.~\cite{Bern:2020buy} proposed a general relation between these physical observables and the derivatives of the eikonal phase with respect to impact parameters and spin-vectors. These relations were verified up to quadratic order in spin at $\CO(G^2)$ in ref.\cite{Kosmopoulos:2021zoq} (analogous relations in slightly different contexts appear in refs.\cite{Liu:2021zxr, Jakobsen:2021zvh}), which we extend to the quartic order by matching to our Hamiltonian calculations. Furthermore, we compute these observables in the probe limit using the Mathisson-Papapetrou-Dixon (MPD) equations, which describe motion of a spinning body on a curved background. We compute observables to quartic order in background spin and linear order in test body spin with orthogonal spin orientations to verify our results beyond the well-studied aligned-spin scattering.

\section{The classical spinning Compton}
In this section, we review the exponentiated gravitational Compton amplitude valid to quartic spin order $\CO(S^4)$, which is equivalent to the known spinor-helicity expressions for minimally coupled particles with quantized spin $s\leq2$~\cite{Arkani-Hamed:2017jhn,Johansson:2019dnu}. While similar expressions have already appeared in the literature~\cite{Guevara:2018wpp,Bautista:2019tdr,Guevara:2019fsj,Aoude:2020onz}, our approach differs in that i) we use BCFW recursion in place of soft theorems, ii) we seek an expression independent of spin representations, bypassing the need to introduce generalized expectation value (GEV) used in refs.\cite{Guevara:2018wpp,Guevara:2019fsj}, and iii) we present the same helicity Compton amplitude in exponentiated form, which to the best of authors' knowledge has only appeared in the context of heavy particle effective theory~\cite{Aoude:2020onz}.\footnote{The authors would like to thank Kays Haddad for bringing this paper to attention.} We also present extension of the exponentiated form to non-minimal couplings in appendix \ref{app:genWC}. 

\subsection{The exponentiated representation of matter-graviton coupling} \label{sec:expmin3pt}
The 3-pt amplitude that describes a spin-$s$ particle minimally coupled to helicity $h$ boson is given as\footnote{We will take the coupling constants to be real. Complex couplings correspond to dyons for electromagnetic couplings and Taub-NUT probes for gravitational coupling~\cite{Huang:2019cja,Emond:2020lwi,Kim:2020cvf}. Spinor conventions follow appendix A of ref.\cite{Chung:2018kqs}.}
\bl
M_{3,min}^{+\abs{h}} = g_0 x^{\abs{h}} \la \mathbf{21} \ra^{2s} \,,\, M_{3,min}^{-\abs{h}} = g_0 x^{-\abs{h}} [ \mathbf{21} ]^{2s} \,.
\el
Here minimal coupling is defined kinematically by matching to the helicity conserving amplitude in the high energy limit~\cite{Arkani-Hamed:2017jhn}. Note that plus helicity couplings are simplest in the chiral basis, while negative helicity are simplest in the anti-chiral basis. The two expressions can be combined into a single expression\footnote{$M^s_{3,min} = M^0_{3,min} \left[ \ve_{2} \cdot \exp \left( - 2i \left[ \frac{p_1 \cdot \ve_{3}^\ast}{m^2} \right] k_3^\m \ve_{3}^\n J_{\m\n} \right) \cdot \ve_{1} \right]_{\text{Rep}(s)}$ is another valid expression, where $\ve_{3\pm}^\ast = \ve_{3\mp}$ is the ``complex conjugate'' of the polarisation vector. This expression is singular in the $m \to 0$ limit and 4pt amplitudes constructed from this expression are plagued by non-physical $t$-channel poles, so we do not consider this expression.}~\cite{Bautista:2019tdr, Guevara:2019fsj}
\bl
M^s_{3,min} &= M^0_{3,min} \left[ \ve_{2} \cdot \exp \left( -i \frac{k_3^\m \ve_{3}^\n J_{\m\n}}{p_1 \cdot \ve_{3}} \right) \cdot \ve_{1} \right]_{\text{Rep}(s)} \,, \label{eq:minampexp0}
\el
where $M^0_{3,min}$ is the minimal coupling of spinless matter ($g_0 x^{\abs{h}}$ or $g_0 x^{-\abs{h}}$), which we refer to as the \emph{scalar factor}, and adopt the convention that massive leg 2 is interpreted as the final state particle with all momenta in the incoming convention. The subscript $\text{Rep}(s)$ denotes the spin-$s$ Lorentz representation for the polarisation spinors/tensors $\ve_{1}$ and $\ve_{2}$, and the associated representation for the Lorentz generator $J_{\m\n}$. We emphasise that the representation $\text{Rep}(s)$ could be \emph{any} spin-$s$ Lorentz representation,\footnote{However, the spin representation of the spinning particle will matter when deviations from minimal coupling are considered as in appendix \ref{app:genWC}.} i.e. any of $(s,0)$, $(\frac{2s-1}{2},\half)$, $\cdots$, $(0,s)$ can be used as $\text{Rep}(s)$. The importance of \emph{representation independence} in classical scattering of spinning particles has been emphasised in ref.\cite{Bern:2020buy}, and this feature follows from the relations \eqc{eq:anglebkexp} and \eqc{eq:sqrbkexp} reviewed in appendix \ref{app:spinprod}. 

For finite spin representations, expansion of the exponential as a Taylor series terminates at finite orders~\cite{Guevara:2018wpp,Chung:2018kqs} because the exponent is a nilpotent operator, reflecting the trace relations of generators for finite dimensional representations. In practice we are considering the classical spin limit, where the classical spin $S$ is held fixed:
\eq \label{eq:cslimdef}
s\rightarrow \infty, \quad \hbar \rightarrow 0,\quad S=\hbar s \quad{\rm fixed}\,.
\eqe
In this limit the nilpotency of the exponent is pushed to infinite order and all terms in the expansion becomes relevant. We use the equality $\stackrel{\cdot}{=}$ whenever the representation of polarisation spinors/tensors is irrelevant, and omit the unnecessary details. For example, \eqc{eq:minampexp0} is recast as
\bl
M^s_{3,min} &\stackrel{\cdot}{=} M^0_{3,min} \exp \left( -i \frac{k_3^\m \ve_{3}^\n J_{\m\n}}{p_1 \cdot \ve_{3}} \right) \,. \label{eq:minampexp}
\el

An expression equivalent to \eqc{eq:minampexp0} was used in refs.\cite{Guevara:2018wpp,Guevara:2019fsj} for studying binary black hole scattering, where generalized expectation value (GEV) was introduced to account for the chiral nature of the spin representation used in the computations. The main distinction in our approach is avoiding specific spin representations as in \eqc{eq:minampexp} based on representation independence of \eqc{eq:minampexp0}, which unnecessitates the introduction of GEV\footnote{For non-minimal coupling considered in appendix \ref{app:genWC} the amplitudes will be dependent on the used spin representation, but GEV is avoided by use of non-chiral spin representations.} and allows us to interpret the results using standard quantum field theory. The difference will become transparent for Compton amplitudes presented in the following sections.

The main advantage of an expression of the form \eqc{eq:minampexp} is that the Lorentz generators appear in an \emph{exponentiated} form. The little group of massive spinning particles in 4d is $SU(2)$, where all finite dimensional representations are given as symmetric tensor products of the smallest nontrivial representation. Given a symmetric product spin-$s$ state\footnote{Spin coherent states form a subset of symmetric product states. Since coherent states are considered as ``classical'' states, we may restrict to symmetric product states without any loss of generality. Note that the \textbf{bold} notation of massive spinor helicity variables implicitly uses symmetric product states as wavefunctions for the particles.} $\ket{\psi_{s}} = \ket{\psi_{1/2}}^{\otimes 2s}$, the action of symmetry group element $\exp (- i \vec{\th} \cdot \vec{S})$ is given as;
\bl
e^{- i \vec{\th} \cdot \vec{S}_{s}} \ket{\psi_{s}} &= e^{- i \vec{\th} \cdot \vec{S}_{s}} \left( \ket{\psi_{1/2}} \right)^{\otimes 2s} = \left( e^{- i \vec{\th} \cdot \vec{S}_{1/2}} \ket{\psi_{1/2}} \right)^{\otimes 2s} \,,
\el
i.e. we can work with generators of the smallest representation instead of wrestling with generators of the actual representation our spinning matter has. This feature is especially advantageous for large spin representations; working with $2 \times 2$ matrices is considerably simpler than working with $(2s + 1) \times (2s + 1)$ matrices with large $s$. For example, full evaluation using the Baker-Campbell-Hausdorff (BCH) formula often becomes complicated due to nested commutators, but $2 \times 2$ matrices are simple enough that matrix logarithms can be computed explicitly.

Moreover, exponentiation automatically symmetrises generator products when expanded into a polynomial of generators. This feature simplifies identification of spin products, which we adopt the conventions of refs.\cite{Bern:2020buy,Kosmopoulos:2021zoq}
\bl
\ve_2 (p_2) \cdot \left( \w_{\m\n} J^{\m\n} \right)^n  \cdot \ve_1 (p_1) &= \left( \w_{\m\n} S^{\m\n} (p_1) \right)^n \ve_2 (p_2) \cdot \ve_1 (p_1) \,, \label{eq:spinfactoriden}
\el
where $S^{\m\n} (p_1)$ is the spin tensor for momentum $p_1^\m$ in covariant spin supplementary condition (SSC) $S^{\m\n} (p) p_{\n} = 0$. The spin tensor can be rewritten using the spin vector $S^\m (p)$ for momentum $p^\m$ by
\bl
S^{\m\n} (p) &= - \frac{1}{m} \e^{\m\n\a\b} p_\a S_{\b} (p) \,,
\el
where the spin vector is identified with the mass-rescaled Pauli-Lubanski pseudovector
\bl
S^\m (p) = \frac{W^\m (p)}{m} = - \frac{1}{2m} \e^{\m\a\b\g} p_\a J_{\b\g} \,.
\el
After extracting the spin contributions using \eqc{eq:spinfactoriden}, the remaining polarisation spinor/tensor contraction $\ve_2 (p_2) \cdot \ve_1 (p_1)$ requires further processing due to the difference of little group frames between the initial and final states. Identification of this contribution is essential when constructing the Hamiltonian, and the procedure was named \emph{Hilbert space matching} in ref.\cite{Chung:2019duq}. The frame difference causes the Thomas-Wigner rotation effect, which results in additional spin factors that were computed to full relativistic order in ref.\cite{Chung:2020rrz}; see also ref.\cite{Bern:2020buy} for its connections to different choices of SSCs. The full form of the factor can be found in \eqc{Ufactor}. 

\subsection{BCFW construction of Compton amplitude (opposite helicity)} \label{sec:mixhelComp}
The 3pt amplitude \eqc{eq:minampexp} can be used as the seed amplitude to construct 4pt amplitudes. We aim for the 4pt amplitude $M_4(p_1,k_2^+,k_3^-,p_4)$ using the BCFW shift $\ket{2} \to \ket{2} + w\ket{3}$ and $\sket{3} \to \sket{3}-w\sket{2}$ (the $\la 23 ]$ shift). Definitions for the Mandelstam variables are $s = (p_1 + k_2)^2$, $t = (k_2 + k_3)^2$, and $u = (p_1 + k_3)^2$. 

The new ingredient in this construction is the treatment of spin factors appearing in \eqc{eq:minampexp}. These factors are operators, and in general their order of composition matters because spin operators are non-abelian. We adopt the convention that leg 4 is the final state particle, so the spin factor associated with the leg 4 is always put to the left of the spin factor associated with the leg 1. On the $s$-channel cut ($w_s = - \frac{\bra{2}p_1\sket{2}}{\bra{3}p_1\sket{2}}$), the spin factors combine into the following simple expression\footnote{The state sum over the intermediate $s$-channel leg forms the identity element, so it can be dropped.}
\bg
\bgd
\exp \left( i \frac{\hat{k}_3^\m \hat{\ve}_{3-}^\n J_{\m\n}}{\hat{P} \cdot \hat{\ve}_{3-}} \right) \exp \left( i \frac{\hat{k}_2^\m \hat{\ve}_{2+}^\n J_{\m\n}}{\hat{P} \cdot \hat{\ve}_{2+}} \right) = \exp \left( - i K^\m L^\n J_{\m\n} \right)\,,
\\ K^\m = k_2^\m + k_3^\m \,,~~~~~~~\, L^\m = \frac{\bra{3} \s^\m \sket{2}}{\bra{3} p_1 \sket{2}} \,,
\egd \label{eq:spinfactprod}
\eg
where hatted variables denote BCFW-shifted variables and $\hat{P} = - p_1 - \hat{k}_2 (w_s)$. The above result can be checked by explicitly comparing the matrix elements for chiral spin-$\half$ representation and anti-chiral spin-$\half$ representation, which we provide the details in appendix \ref{app:spinprod}. Since matrix elements of other spin representations follow from their tensor products, checking consistency of the two cases is sufficient for generic spin representations. Interestingly, the $u$-channel cut ($w_u = - \frac{\bra{2}p_4\sket{2}}{\bra{3}p_4\sket{2}}$) results in the same spin factor
\bg
\bgd
\exp \left( - i \frac{\hat{k'}_2^\m \hat{\ve'}_{2+}^\n J_{\m\n}}{\hat{P'} \cdot \hat{\ve'}_{2+}} \right) \exp \left( - i \frac{\hat{k'}_3^\m \hat{\ve'}_{3-}^\n J_{\m\n}}{\hat{P'} \cdot \hat{\ve'}_{3-}} \right) = \exp \left( - i K^\m L^\n J_{\m\n} \right)\,,
\egd
\eg
where $\hat{P'} = - p_4 - \hat{k}_2 (w_u)$, and the relevant shifted variables marked by hats are primed to denote that they correspond to the $u$-channel factorisation conditions. 
We do not present composition of scalar factors as it is not different from the usual BCFW construction~\cite{Johansson:2019dnu,Chung:2019duq,Aoude:2020onz}. Adding up all contributions, the 4pt amplitude $M_4(p_1,k_2^+,k_3^-,p_4)$ is given as~\cite{Bautista:2019tdr}
\bl
M_4^{s} &\stackrel{\cdot}{=} M_4^{0} 
\exp \left( -i K^\m L^\n J_{\m\n} \right) \,, \nn
\\ M_4^{0} &= \left\{ \begin{aligned}
- \frac{\sbra{2}p_1\ket{3}^2}{(s-m^2)(u-m^2)} \,,&& \abs{h} = 1\,, \\
\frac{\sbra{2}p_1\ket{3}^4}{(s-m^2)t(u-m^2)}\,, && \abs{h} = 2\,,
\end{aligned} \right. \label{eq:BCFWmixCompmin}
\el
with definitions for variables in \eqc{eq:spinfactprod}, which we present an extension to non-minimal coupling in appendix \ref{app:mixedComp}. The expression \eqc{eq:BCFWmixCompmin} appears in a slightly different form as eq.(56) in ref.\cite{Bautista:2019tdr} or as eq.(5.14) in ref.\cite{Aoude:2020onz}, and differs from the analogous expresison eq.(2.34) in ref.\cite{Guevara:2018wpp} by the spin factor \eqc{eq:spinfactprod}. For $s \leq 2$, all the expressions are equivalent to the spinor-helicity expression for the same amplitude first presented as eq.(5.36) in ref.\cite{Arkani-Hamed:2017jhn}. We make the following remarks regarding the expression \eqc{eq:BCFWmixCompmin}:
\bn
\item All physical channel residues including the $t$-channel for graviton coupling are correctly captured by this expression to all orders in $J$.

\item Spurious poles start to appear at $(J)^5$ order for graviton coupling ( $J^3$ for photon), due to the denominator of the vector $L^\m$. Since we are aiming for the potential up to order $J^4$, this is not a concern. However the consistent removal of spurious poles and its associated polynomial ambiguity, will be crucial for pushing towards higher spin order.

\item While ref.\cite{Bautista:2019tdr} used the BCH formula to arrive at the results, \eqc{eq:spinfactprod} was obtained by solving matrix equations. The advantage of \emph{not} resorting to the BCH formula will become clear for the same helicity Compton amplitude.
\en
Once again, as we are only interested up to spin-quartic order, the spurious pole appearing at higher orders in $J$ will be irrelevant to us.

\subsection{BCFW construction of Compton amplitude (same helicity)}
Following the previous section, we construct the 4pt amplitude $M_4(p_1,k_2^+,k_3^+,p_4)$ using the $\la 23 ]$ shift, with the same definitions for shifted variables used in the previous section. The spin factors turn out to be quite involved for this set-up, where the $s$-channel spin factor is given as 
\bl
\bgd
\exp \left( i \frac{\hat{k}_3^\m \hat{\ve}_{3+}^\n J_{\m\n}}{\hat{P} \cdot \hat{\ve}_{3+}} \right) \exp \left( i \frac{\hat{k}_2^\m \hat{\ve}_{2+}^\n J_{\m\n}}{\hat{P} \cdot \hat{\ve}_{2+}} \right) = \exp \left( - \frac{i}{2} b_{min}^{\m\n} J_{\m\n} \right)\,,
\\ b_{min}^{\m\n} = 4 f \left( \frac{t}{m^2} \right)
 \left[ a_1 \left( k_2^{[\m} k_3^{\n]} + l^{[\m} \bar{l}^{\n]} \right) + a_2 k_2^{[\m} \bar{l}^{\n]} + a_3k_3^{[\m} l^{\n]} \right] \,,\,
\\ l^\m = \frac{\bra{2} \s^\m \sket{3}}{2} \,,\, \bar{l}^\m = \frac{\bra{3} \s^\m \sket{2}}{2} \,,
\\ a_1 = \frac{s-u}{2 t m^2} \,,\, a_2 = -\frac{\sbra{3}p_1\ket{2}}{t m^2} \,, a_3 = -\frac{\sbra{2}p_1\ket{3}}{t m^2} \,,
\egd \label{eq:BCFWsamehelSFschanmin}
\el
with $f(x)= \sum_{n=0}^{\infty} \frac{(n!)^2 x^n}{(2n+1)!}$. This infinite sum have a radius of convergence of $\abs{x} < 4$, reflecting the threshold condition $t = 4m^2$. When restoring factors of $\hbar$, in the classical limit one simply takes $f(x)=1$.  The spin factor composition \eqc{eq:BCFWsamehelSFschanmin} is obtained as the minimal coupling limit for the generic case presented in appendix \ref{app:sameComp}, and the details of the computation are given in appendix \ref{app:spinprod}. Similar to the opposite helicity case, the $u$-channel spin factor also turns out to be the same;
\bl
\bgd
\exp \left( - i \frac{\hat{k'}_2^\m \hat{\ve'}_{2+}^\n J_{\m\n}}{\hat{P'} \cdot \hat{\ve'}_{2+}} \right) \exp \left( - i \frac{\hat{k'}_3^\m \hat{\ve'}_{3+}^\n J_{\m\n}}{\hat{P'} \cdot \hat{\ve'}_{3+}} \right) = \exp \left( - \frac{i}{2} b_{min}^{\m\n} J_{\m\n} \right)\,,
\egd
\el
which is a property specific to minimal coupling.
Summing up, we get the following expression for $M_4(p_1,k_2^+,k_3^+,p_4)$ using BCFW construction.
\bl
M_4^{s} &\stackrel{\cdot}{=} M_4^{0} 
\exp \left( - \frac{i}{2} b_{min}^{\m\n} J_{\m\n} \right) \,, \nn
\\ M_4^{0} &= \left\{ \begin{aligned}
- \frac{m^2 [23]^2}{(s-m^2)(u-m^2)}\,, && \abs{h} = 1 \,,\\
\frac{m^4[23]^4}{(s-m^2)t(u-m^2)}\,, && \abs{h} = 2\,.
\end{aligned} \right.  \label{eq:samehelBCFWmin}
\el
In contrast to opposite helicity case \eqc{eq:BCFWmixCompmin}, the expression \eqc{eq:samehelBCFWmin} has no spurious poles. This is consistent with spinor-helicity expression for the same amplitude~\cite{Johansson:2019dnu}, which is equivalent due to the relation
\bl
\left[ \ve_{4} \cdot \exp \left( - \frac{i}{2} b_{min}^{\m\n} J_{\m\n} \right)  \cdot \ve_{1} \right]_{\text{Rep}(1/2)} &= \left\{
\begin{aligned}
\la \mathbf{41} \ra ~~~~&& &~~~\text{(chiral)}
\\- \frac{\sbra{\mathbf{4}}  p_4 p_1
\sket{\mathbf{1}}}{m^2}  && &\text{(anti-chiral)}
\end{aligned}
\right.\,\,, \label{eq:mincoupComptonbrackets}
\el
which holds when $m^{-1}$ is treated as a formal expansion parameter. We make following remarks regarding the expression \eqc{eq:samehelBCFWmin}:
\bn
\item The non-local $t$-channel pole in $a_i$ coefficients actually cancel on the $t$-channel, so all physical channel residues are correctly captured by this expression to all orders in $J$.
\item Similar to \eqc{eq:BCFWmixCompmin}, the expression \eqc{eq:samehelBCFWmin} has a factorisation into the scalar factor and the spin factor. This is not true for generic couplings, as shown in appendix \ref{app:sameComp}.
\item The spin factor in \eqc{eq:BCFWsamehelSFschanmin} was obtained by solving the matrix equations explicitly without resorting to the BCH formula, as explained in appendix \ref{app:spinprod}. The all-order resummation would have been impossibly complicated using the BCH formula.
\item A similar exponentiated form exists in the context of heavy particle effective theory~\cite{Aoude:2020onz} 
where representation independence is attained by off-shell continuation of polarisation spinors, in contrast to purely on-shell expression \eqc{eq:samehelBCFWmin}.
\en

The same helicity Compton amplitudes do not contribute to the triple cut integrand at one-loop for minimal coupling~\cite{Guevara:2017csg}, but they do contribute to the quadruple cut integrand. The quadruple cuts are needed for checking IR cancellations.


\begin{figure}[t]
\begin{tikzpicture}[line width=1. pt, scale=2,
sines/.style={
        line width=1pt,
        line join=round, 
        draw=black, 
        decorate, 
        decoration={complete sines, number of sines=4, amplitude=4pt}
    }
]
\draw[black, line width=1.pt] (-0.2,0) -- (0.7,0);
\draw[black, line width=1.pt] (-0.2,0.5) -- (0.7,0.5);
\draw[black, line width=1.pt] (0,0) -- (0,0.5);
\draw[black, line width=1.pt] (0.5,0) -- (0.5,0.5);
\node[scale=1] at (-0.5,0.25) {$c_{\,\square}$};
\node[scale=1] at (-1.3,0.28) {$M^{1-\rm loop}_{\rm classical}~=$};
\node[scale=0.8] at (-0.2,-0.1) {$1$};
\node[scale=0.8] at (-0.2,0.6) {$2$};
\node[scale=0.8] at (0.7,0.6) {$3$};
\node[scale=0.8] at (0.7,-0.1) {$4$};
\node[scale=0.8] at (0.1,0.23) {$\ell$};
\draw [-stealth](0,0.27) -- (0,0.272);
\draw [-stealth](-0.08,0) --(-0.05,0);
\draw [-stealth](-0.08,0.5) --(-0.05,0.5);
\draw [-stealth](0.62,0) --(0.64,0);
\draw [-stealth](0.62,0.5) --(0.64,0.5);
\end{tikzpicture}~\begin{tikzpicture}[line width=1. pt, scale=2,
sines/.style={
        line width=1pt,
        line join=round, 
        draw=black, 
        decorate, 
        decoration={complete sines, number of sines=4, amplitude=4pt}
    }
]
\draw[black, line width=1.pt] (-0.2,0) -- (0.7,0);
\draw[black, line width=1.pt] (-0.2,0.5) -- (0.7,0.5);
\draw[black, line width=1.pt] (0,0) -- (0.5,0.5);
\filldraw[color=white, fill=,  thick](0.25,0.25) circle[radius=0.035] ;
\draw[black, line width=1.pt](0.5,0) -- (0,0.5);
\node[scale=1] at (-0.5,0.25) {$+\,c_{\,\rotatebox{90}{\scalebox{0.6}[0.9]{$\bowtie$}}}$};
\node[scale=0.8] at (-0.2,-0.1) {$1$};
\node[scale=0.8] at (-0.2,0.6) {$2$};
\node[scale=0.8] at (0.7,0.6) {$3$};
\node[scale=0.8] at (0.7,-0.1) {$4$};
\node[scale=0.8] at (0.25,0.4) {$\ell$};
\draw [-stealth](0.1,0.4) -- (0.08,0.42);
\draw [-stealth](-0.08,0) --(-0.05,0);
\draw [-stealth](-0.08,0.5) --(-0.05,0.5);
\draw [-stealth](0.62,0) --(0.64,0);
\draw [-stealth](0.62,0.5) --(0.64,0.5);
\end{tikzpicture}~\begin{tikzpicture}[line width=1. pt, scale=2,
sines/.style={
        line width=1pt,
        line join=round, 
        draw=black, 
        decorate, 
        decoration={complete sines, number of sines=4, amplitude=4pt}
    }
]
\draw[black, line width=1.pt] (-0.2,0) -- (0.7,0);
\draw[black, line width=1.pt] (-0.2,0.5) -- (0.7,0.5);
\draw[black, line width=1.pt] (0,0) -- (0.25,0.5);
\draw[black, line width=1.pt](0.5,0) -- (0.25,0.5);
\node[scale=1] at (-0.5,0.25) {$+\,c_{\bigtriangleup}$};
\node[scale=0.8] at (-0.2,-0.1) {$1$};
\node[scale=0.8] at (-0.2,0.6) {$2$};
\node[scale=0.8] at (0.7,0.6) {$3$};
\node[scale=0.8] at (0.7,-0.1) {$4$};
\node[scale=0.8] at (0.23,0.22) {$\ell$};
\draw [-stealth](0.1235,0.25) -- (0.128,0.26);
\draw [-stealth](-0.08,0) --(-0.05,0);
\draw [-stealth](-0.08,0.5) --(-0.05,0.5);
\draw [-stealth](0.62,0) --(0.64,0);
\draw [-stealth](0.62,0.5) --(0.64,0.5);
\end{tikzpicture}~\begin{tikzpicture}[line width=1. pt, scale=2,
sines/.style={
        line width=1pt,
        line join=round, 
        draw=black, 
        decorate, 
        decoration={complete sines, number of sines=4, amplitude=4pt}
    }
]
\draw[black, line width=1.pt] (-0.2,0) -- (0.7,0);
\draw[black, line width=1.pt] (-0.2,0.5) -- (0.7,0.5);
\draw[black, line width=1.pt] (0,0.5) -- (0.25,0);
\draw[black, line width=1.pt](0.5,0.5) -- (0.25,0);
\node[scale=1] at (-0.5,0.25) {$+\,c_{\bigtriangledown}$};
\node[scale=0.8] at (-0.2,-0.1) {$1$};
\node[scale=0.8] at (-0.2,0.6) {$2$};
\node[scale=0.8] at (0.7,0.6) {$3$};
\node[scale=0.8] at (0.7,-0.1) {$4$};
\node[scale=0.8] at (0.25,0.25) {$\ell$};
\draw [-stealth](0.1148,0.27) -- (0.11,0.28);
\draw [-stealth](-0.08,0) --(-0.05,0);
\draw [-stealth](-0.08,0.5) --(-0.05,0.5);
\draw [-stealth](0.62,0) --(0.64,0);
\draw [-stealth](0.62,0.5) --(0.64,0.5);
\end{tikzpicture}
\caption{The relevant part of 1-loop amplitude contributes to classical effects. The diagrams are scalar integrals and $c_i$ are their associated integral coefficients. 
}
\label{fig:1-loop_basis}
\end{figure}
\section{Integral coefficients and the classical potential} \label{sec:IntCoeff}
We evaluate the 2PM conservative potential in this section.  We start by computing the one-loop amplitude relevant to classical contribution, which is conveniently expanded in terms of scalar integral basis, 
\eqa
\label{1loopb}M^{1-\rm{loop}}_{\rm classical}=c_{\, \square}\mathcal{I}_{\, \square }+c_{\cSquare}\mathcal{I}_{ \cSquare}+c_{\bigtriangleup}\mathcal{I}_{\bigtriangleup}+c_{\bigtriangledown}\mathcal{I}_{\bigtriangledown}\,,
\eqae
where we only include the terms relevant to classical physics.  The explicit forms of the box and triangle integrals are 
\eqa
 \mathcal{I}_{\, \square }&=&\int \frac{d^d \ell}{(2\pi)^d} \frac{1}{\ell^2  [(\ell{-}p_1)^2 - m_a^2](\ell{-}q)^2[(\ell{+}p_2)^2 - m_b^2]}\,,\nonumber\\
\mathcal{I}_{ \cSquare}&=&\int \frac{d^d \ell}{(2\pi)^d} \frac{1}{\ell^2  [(\ell{+}p_4)^2 - m_a^2](\ell{-}q)^2[(\ell{+}p_2)^2 - m_b^2]}\,,\\
\mathcal{I}_{\bigtriangleup} &=& \int \frac{d^d \ell}{(2\pi)^d} \frac{1}{\ell^2 (\ell-q)^2 [(\ell{-}p_1)^2 - m_a^2]}\,,\quad\quad \mathcal{I}_{\bigtriangledown} = \int \frac{d^d \ell}{(2\pi)^d} \frac{1}{\ell^2 (\ell{-}q)^2 [(\ell{+}p_2)^2 - m_b^2]}\,.\nonumber
\eqae

We set the kinematics in the centre of momentum (COM) frame as:
\begin{equation}\label{eq:Kinematics}
p_1 = \left(E_a, \vec{p}\right), \;
p_2 = \left(E_b, -\vec{p}\right), \;
p_3 = \left(E_b, -\vec{p}^{\,\prime}\right), \;
p_4 = \left(E_a, \vec{p}^{\,\prime}\right),
\end{equation}
where $\vec{p}^{\,\prime}=\vec{p}{-}\vec{q}$ and $\vec{q}$ is the momentum transfer between the two spinning particles. Expanding in $|\vec{q}|$ the basis integrals yields~\cite{Donoghue:1996mt, Bjerrum-Bohr:2002gqz}
\eqa\label{IntResult}
\mathcal{I}_{\, \square }\Big|_{\rm classical}&=&- \frac{i}{16\pi^2 |\vec{q}|^2}\left[\frac{1}{\epsilon} - \log|\vec{q}|^2\right]
\left[
\frac{\log\left[\sigma - \sqrt{\sigma^2 - 1} \right]}{m_a m_b\sqrt{\sigma^2 - 1}} + \frac{i\pi}{ |\vec{p}|(E_a+E_b)}
\right]
\,,\nonumber\\
\mathcal{I}_{ \cSquare }\Big|_{\rm classical}&=&+ \frac{i}{16\pi^2 |\vec{q}|^2}\left[\frac{1}{\epsilon} - \log|\vec{q}|^2\right]
\frac{\log\left[\sigma - \sqrt{\sigma^2 - 1} \right]}{m_a m_b\sqrt{\sigma^2 - 1}} 
\,,\\
\mathcal{I}_{\bigtriangleup, \bigtriangledown }\Big|_{\rm classical}&=&{-}\frac{i}{32 m_{a,b}|\vec{q}|}
\,,\nonumber
\eqae
where $\sigma = \frac{p_1\cdot p_2}{m_a m_b}$. The infrared divergent terms with $\epsilon^{-1}$ from the box and the crossed box integrals cancel up to a term proportional to $\left[ |\vec{q}|^2|p|(E_a{+}E_b) \right]^{-1}$, and the remaining IR divergent terms should cancel against iteration terms. This cancellation serves as a consistency check to our final result. 

In summary, we use the formula:
\begin{equation}\label{eq:2PM_formula}
\small
H_{2\text{PM}}
= 
\int \frac{d^3 \vec{q}}{(2\pi)^3} e^{i\vec{q} \cdot \vec{r}}\left[
\frac{-1}{4E_a E_b} \left(M^{1-\rm{loop}}_{\rm classical}
\right)U_a U_b 
-
\int \frac{d^3 \vec{k}}{(2\pi)^3} 
\frac{\hat{V}_{1 \text{PM}}(\vec{p}', \vec{k})
\hat{V}_{1 \text{PM}}(\vec{k}, \vec{p})}{E - \sqrt{|\vec{k}|^2 + m_a^2} - \sqrt{|\vec{k}|^2 + m_b^2}} 
\right]\\[2mm]
\,,
\end{equation}
where $U_a$, $U_b$ are the Thomas-Wigner rotation factors~\cite{Chung:2020rrz,Bern:2020buy} discussed in section \ref{sec:expmin3pt}.
\eqa
 \label{Ufactor}\ve_4 (p_4) \cdot \ve_1 (p_1)= U_a\,,&&~~~
U_a\equiv\exp\left[\frac{{-}i\mathcal{E}_a}{m_a E(m_a+E_a)}\right]\,,\\
 \notag  E\equiv E_a+E_b\,,&&~~~\mathcal{E}_{a} \equiv \epsilon^{\mu\nu\rho\sigma}p_{1\mu}p_{2 \nu}q_{\rho} S_{a, \sigma} 
 \,.
\eqae
$U_b$ is obtained by a simple exchange of labels $a \leftrightarrow b$. This factor ensures that the incoming states and the outgoing states are described in the same Hilbert space, where particles are labelled by their momenta and their little group indices in the COM frame; the iteration of the Hamiltonian makes sense only when the domain and the range are the same. $\hat{V}_{1\text{PM}}$ is the 1PM potential reported in ref.\cite{Chung:2020rrz}. We first derive the relevant scalar integral coefficients using unitarity methods, and then compute the iteration terms. Using eq.(\ref{eq:2PM_formula}) we derive the 2PM Hamiltonian up to quartic order in total spin vectors, where the covariant spin vector is defined as~\cite{Bern:2020buy}:
\begin{equation}\label{eq:Spin_Vec_3D}
S_a^{\mu} = \left( \frac{\vec{p} \cdot \vec{S}_a}{m_a}, \vec{S}_a + \frac{\vec{p}\cdot \vec{S}_a}{(E_a+m_a)m_a} \vec{p}\right),\;
S_b^{\mu} = \left( -\frac{\vec{p} \cdot \vec{S}_b}{m_b}, \vec{S}_b + \frac{\vec{p}\cdot \vec{S}_b}{(E_b+m_b)m_b} \vec{p}\right)\,.
\end{equation}
We test consistency of the results by matching physical observables in the next section.

\subsection{Integral Coefficients} \label{subsec:Int_Coeff}
\begin{figure}[t]
\begin{subfigure}{.5\textwidth}
\centering
\begin{tikzpicture}[line width=1. pt, scale=2,
sines/.style={
        line width=1pt,
        line join=round, 
        draw=black, 
        decorate, 
        decoration={complete sines, number of sines=4, amplitude=4pt}
    }
]
\draw[black, line width=1.pt] (-0.5,0) -- (1.5,0);
\draw[black, line width=1.pt] (-0.5,1) -- (1.5,1);
\draw[white,postaction={sines}] (0,0) -- (0,1);
\draw[white,postaction={sines}] (1,0) -- (1,1);
\filldraw[color=black, fill=gray!70,  thick](0,0) circle[radius=0.1] ;
\filldraw[color=black, fill=gray!70,  thick](0,1) circle[radius=0.1] ;
\filldraw[color=black, fill=gray!70,  thick](1,0) circle[radius=0.1] ;
\filldraw[color=black, fill=gray!70,  thick](1,1) circle[radius=0.1] ;
\node[scale=1] at (-0.7,0) {$1$};
\node[scale=1] at (-0.7,1) {$2$};
\node[scale=1] at (1.7,1) {$3$};
\node[scale=1] at (1.7,0) {$4$};
\draw [-stealth](-0.25,0) -- (-0.24,0);
\draw [-stealth](-0.25,1) -- (-0.24,1);
\draw [-stealth](1.34,0)--(1.35,0);
\draw [-stealth](1.34,1) -- (1.35,1);
\draw [-stealth](0.016,0.23) -- (-0.026,0.28);
\node[scale=1] at (-0.2,0.25) {$\ell$};
\draw[dashed,color=red] (-0.2,0.5)--(0.2,0.5);
\draw[dashed,color=red] (0.8,0.5)--(1.2,0.5);
\draw[densely dotted,color=red] (0.5,0.8)--(0.5,1.2);
\draw[densely dotted,color=red] (0.5,-0.2)--(0.5,0.2);
\end{tikzpicture}
\caption{quadruple cut}
\end{subfigure}
\begin{subfigure}{.5\textwidth}
\centering
\begin{tikzpicture}[line width=1. pt, scale=2,
sines/.style={
        line width=1pt,
        line join=round, 
        draw=black, 
        decorate, 
        decoration={complete sines, number of sines=4, amplitude=4pt}
    }
]
\draw[black, line width=1.pt] (-0.5,0) -- (1.5,0);
\draw[black, line width=1.pt] (-0.5,1) -- (1.5,1);

\draw[white,postaction={sines}] (0,0) -- (0,1);
\draw[white,postaction={sines}] (1,0) -- (1,1);
\filldraw[color=black, fill=gray!70,  thick](0,0) circle[radius=0.1] ;
\filldraw[color=black, fill=gray!70,  thick](1,0) circle[radius=0.1] ;
\filldraw[color=black, fill=gray!70,  thick](0.5,1) circle[radius=0.1] ;
\node[scale=1] at (-0.7,0) {$1$};
\node[scale=1] at (-0.7,1) {$2$};
\node[scale=1] at (1.7,1) {$3$};
\node[scale=1] at (1.7,0) {$4$};
\draw [-stealth](-0.25,0) -- (-0.24,0);
\draw [-stealth](-0.25,1) -- (-0.24,1);
\draw [-stealth](1.34,0)--(1.35,0);
\draw [-stealth](1.34,1) -- (1.35,1);
\draw [-stealth](0.016,0.23) -- (-0.026,0.28);
\node[scale=1] at (-0.2,0.25) {$\ell$};
\node[ellipse,fill=gray!70,  thick,draw = black,
    minimum width = 2.5cm, 
    minimum height = 0.5cm] (e) at (0.5,1) {};
    \draw[dashed,color=red] (-0.2,0.5)--(0.2,0.5);
\draw[dashed,color=red] (0.8,0.5)--(1.2,0.5);
\draw[densely dotted,color=red] (0.5,-0.2)--(0.5,0.2);
\end{tikzpicture}
\caption{triple cut}
\end{subfigure}
\caption{Here are the unitraity cuts used to extract the corresponding integral coefficients. The dashed lines represents cuts from gluing two four-point tree Compton amplitudes (the double cut) and the dotted line is the additional cuts imposed to get the desired cuts. The solid and wavy lines represent spinning bodies and gravitons,  respectively.}
\end{figure}
We evaluate these integral coefficients in eq.\eqref{1loopb} using the unitarity cut methods provided in ref.\cite{Forde:2007mi}. The unitarity cuts of loop amplitudes can be obtained by sewing tree amplitudes.
For example,  triple and quadruple cut are given by sewing of three and four corresponding tree amplitudes, respectively.  
In order to generate the triple and quadruple cuts, we start by sewing the $t$-channel double cut $C_2$ with classical Compton amplitudes eq.(\ref{eq:BCFWmixCompmin}) and \eqc{eq:samehelBCFWmin} as our basic input,\footnote{Alternatively, the triple and quadruple cuts can be constructed from gluing on-shell Compton and 3pt amplitudes. In this construction of the cut integrand the products of spin factors are computed as operator products as in appendix \ref{app:spinprod}.}
\begin{equation}
\label{dbcut}C_{2} = \sum_{h_0, h_1 = \pm 2}M_4( p_1^{s_a}, -p_4^{s_a}, -\ell^{h_0}, \ell_1^{h_1}) M_4( p_2^{s_b}, -p_3^{s_b}, \ell^{-h_0}, -\ell_1^{-h_1})\,.
\end{equation}
The sewing imposes two cut conditions on the loop momentum, leaving two free parameters in four dimensions. Imposing another cut condition yields the triple cut and the number of free parameters reduces to one. For a suitable choice of loop momentum parametrisation the triangle coefficient is determined as the part of the triple cut independent of the free parameter, which can be extracted as the constant term of the Laurent expansion at infinity. Imposing a further cut condition yields the quadruple cut, fixing all parameters of the loop momentum. The (crossed) box coefficient is determined by summing over all possible quadruple cut solutions of the loop momentum.

\subsubsection{The triangle coefficients}\label{subsubsec:Triangle_Coefficient_Results}
The triple cuts $C_{3,\#}$ can be obtained from the double cut $C_2$, eq.\eqref{dbcut},  as 
\begin{equation}
\label{tricut}C_{3, \bigtriangleup} =(-2 \ell \cdot  p_1)C_2\big|_{\ell\cdot p_1 = 0}\,,~~~C_{3, \bigtriangledown} = (2 \ell\cdot p_2)C_2\big|_{\ell\cdot p_2 = 0}\,.
\end{equation}
We describe the evaluation of the triangle coefficient $c_{\bigtriangleup}$ from the cut $C_{3, \bigtriangleup}$. The other coefficient $c_{\bigtriangledown}$ and its associated cut $C_{3, \bigtriangledown}$ can be easily obtained by the substitutions
\begin{equation}
\label{trir}c_{\bigtriangledown} 
= 
c_{\bigtriangleup}\big|_{m_a \leftrightarrow m_b, \; S_a \leftrightarrow S_b, \; p_1 \leftrightarrow p_2, \; q \rightarrow -q}\,.
\end{equation}
The triple cut conditions relevant to $c_\bigtriangleup$\,,
\eqa
\ell^2=(\ell+q)^2=(\ell-p_1)^2-m_a^2=0\,,
\eqae
are solved by the loop parametrisation with one free parameter $z$ as \cite{Bern:2020buy},
\begin{equation}
\label{triloop}
\ell^{\mu}_\pm(z) = \alpha q^{\mu} + \beta p_1^{\mu} + z u_\pm^{\mu} + 
 \frac{ \alpha^2\gamma_{\pm}}{16}\frac{v^{\mu}_\pm}{z}\,,
\end{equation}
where the $\pm$ label the two solutions that satisfies the triple cut conditions and
\eqa
\notag && 
\hspace{-1cm} u^{\mu}_\pm = \aPs{Q^{\flat}_\pm}{\sigma^{\mu}}{P^{\flat}_\pm}\,, ~~
v^{\mu} _\pm= \aPs{P^{\flat}_\pm}{\sigma^{\mu}}{Q^{\flat}_\pm}\,, ~~
P^{\flat\mu}_\pm = p_1^{\mu} + \frac{q^{\mu}}{\gamma_\pm}\,,~~
Q^{\flat\mu}_\pm = q^{\mu} + \frac{q^2}{m_a^2\gamma_\pm} p_1^{\mu}\,,\\[1mm]
&&~~~ \alpha = \frac{2m_a^2}{4m_a^2 - q^2}\,, \quad
\beta = - \frac{q^2}{4m_a^2 - q^2}\,, \quad\,
\gamma_{\pm}= - \frac{q^2 \pm \sqrt{q^2(q^2 - 4m_a^2)} }{2m_a^2}\,.
\eqae
Inserting eq.\eqref{tricut} into eq.\eqref{triloop} yields $C_{3, \bigtriangleup}$ as a function of $z$. The triangle coefficient is determined as~\cite{Forde:2007mi}
\eqa
c_{\bigtriangleup} =  \frac{1}{2}\left[\sum_{\ell = \ell_\pm(z)}{\rm Inf}_z C_{3, \bigtriangleup}(\ell)\right]\,,
\eqae
where 
${\rm Inf}_z C_{3, \bigtriangleup}(l)$ is the constant term of the Laurent expansion of $C_{3, \bigtriangleup}$ at $z=\infty$. 
The other coefficient $c_{\bigtriangledown}$ is obtained via eq.\eqref{trir}, and the triangle contributions to the 1-loop amplitude is given as
\begin{equation}
M_{\bigtriangleup+ \bigtriangledown} = c_{\bigtriangleup} I_{\bigtriangleup} + c_{\bigtriangledown} I_{\bigtriangledown}\,.
\end{equation}

\paragraph{The classical limit of the triangle coefficients}
$\hphantom{123}$ \newline
\\[-0.3cm]
The triangle coefficients are functions of external kinematics. To extract classical contributions, factors of $\hbar$ are restored through the map
\begin{equation}
m_{a/b} \rightarrow m_{a/b}\,, \quad
p_{1/2} \rightarrow p_{1/2}\,,\quad
q \rightarrow q \hbar\,, \quad 
S_{a/b} \rightarrow \frac{S_{a/b}}{\hbar}\,,
\end{equation}
and only the terms with $\CO (\hbar^0)$ scaling are kept. We expand the triangle coefficient by the following basis of spin operators that scales as $\mathcal{O}(\hbar^0)$:
\begin{equation}\label{CalEDef}
\mathcal{E}_{a/b} 
\,, \quad
q\cdot S_{a/b}
\,, \quad 
q^2 S_i \cdot S_j\,, \quad
q^2 (p_2 \cdot S_a)^2\,, \quad
q^2 (p_2 \cdot S_a) (p_1 \cdot S_b)\,, \quad
q^2 (p_1 \cdot S_b)^2\,.
\end{equation}
We write the amplitude as a sum over various terms with differing powers of the spin operator:
\begin{equation}
M_{\bigtriangleup + \bigtriangledown} = 
- \frac{2\pi^2 G^2}{\sqrt{-q^2}} \sum_{i=0}^{4} \sum_{ \substack{j =0,\\ i + j \leq 4}}^{2} M_{\bigtriangleup + \bigtriangledown}^{(i, j)}\,,
\end{equation}
where $M_{\bigtriangleup + \bigtriangledown}^{(i,j)}$ denotes the amplitude with $S_a^i S_b^j$\,.
\begin{itemize}[leftmargin=*]
\item Scalar coupling
\begin{equation}
M^{(0,0)} = A_{0,0} = 3 \left(1-5 \sigma ^2\right) m_a^2 m_b^2 \left(m_a+m_b\right) \,.
\end{equation}
This coefficient matches with \cite{Bern:2020buy}.
\item $S_a$
\begin{equation}
M^{(1, 0)} = A_{1,0} \mathcal{E}_a = -\frac{i \sigma  \left(5 \sigma ^2-3\right) m_b \left(4 m_a+3 m_b\right)}{\sigma ^2-1}\mathcal{E}_a \,.
\end{equation}
This coefficient matches with \cite{Bern:2020buy}.
\item $S_a S_b$
\begin{subequations}
\begin{equation}
M^{(1, 1)} = A_{1,1}^{(1)} \mathcal{E}_a\mathcal{E}_b + A_{1,1}^{(2)} q^2 (p_2 \cdot S_a)(p_1 \cdot S_b)\,.
\end{equation}
The corresponding coefficeints $A_{1,1}^{i}$ are:
\begin{equation}
\begin{split}
A_{1,1}^{(1)}  &= \frac{\left(20 \sigma ^4-21 \sigma ^2+3\right) \left(m_a+m_b\right)}{\left(\sigma ^2-1\right)^2 m_a m_b} \,, \\
A_{1,1}^{(2)}  &= -\frac{\sigma  \left(5 \sigma ^2-3\right) \left(m_a+m_b\right)}{\left(\sigma ^2-1\right)^2}\,.
\end{split}
\end{equation}
\end{subequations}
These coefficients match with \cite{Bern:2020buy}.
\item $S_a^2$
\begin{subequations}
\begin{equation}
M^{(2, 0)} = A_{2,0}^{(1)} \mathcal{E}_a^2 + A_{2,0}^{(2)} q^2 (p_2 \cdot S_a)^2
\end{equation}
The corresponding coefficeints $A_{2,0}^{i}$ are:
\begin{equation}
\begin{split}
A_{2,0}^{(1)} &= \frac{\left(95 \sigma ^4-102 \sigma ^2+15\right) m_a+4 \left(15 \sigma ^4-15 \sigma ^2+2\right) m_b}{8 \left(\sigma ^2-1\right)^2 m_a^2}\,, \\
A_{2,0}^{(2)} &= \frac{\left(35 \sigma ^4-30 \sigma ^2+3\right) m_a+4 \left(3 \sigma ^2-1\right) m_b}{8 \left(\sigma ^2-1\right)^2}\,.
\end{split}
\end{equation}
\end{subequations}
These coefficiens match with \cite{Kosmopoulos:2021zoq} with $C_{ES^2} = 1$.
\item $S_a^2 S_b$
\begin{subequations}
\begin{equation}
M^{(2, 1)} = 
A_{2,1}^{(1)} \mathcal{E}_a^2\mathcal{E}_b + 
A_{2,1}^{(2)} q^2 (p_2 \cdot S_a)(p_1 \cdot S_b) \mathcal{E}_a+
A_{2,1}^{(3)} q^2 (p_2 \cdot S_a)^2 \mathcal{E}_b\,.
\end{equation}
The corresponding coefficeints $A_{2,1}^{(i)}$ are:
\begin{equation}\label{eq:Tri_Coeff_2_1_res}
\begin{split}
A_{2,1}^{(1)} &= \frac{i \sigma  \left[\left(95 \sigma ^2-51\right) m_a+40 \left(2 \sigma ^2-1\right) m_b\right]}{8 \left(\sigma ^2-1\right)^2 m_a^3 m_b^2}\,,\\
A_{2,1}^{(2)} &= -\frac{i \left(5 \sigma ^2-1\right) \left(3 m_a+4 m_b\right)}{4 \left(\sigma ^2-1\right)^2 m_a^2 m_b}\,, \quad
A_{2,1}^{(3)} = \frac{i \sigma  \left[5 \left(7 \sigma ^2-3\right) m_a+8 m_b\right]}{8 \left(\sigma ^2-1\right)^2 m_a m_b^2}\,.
\end{split}
\end{equation}
\end{subequations}
\item $S_a^3$
\begin{subequations}
\begin{equation}
M^{(3, 0)} = 
A_{3,0}^{(1)} \mathcal{E}_a^3+ 
A_{3,0}^{(2)} q^2 (p_2 \cdot S_a)^2 \mathcal{E}_a\,.
\end{equation}
The corresponding coefficeints $A_{3,0}^{(i)}$ are:
\begin{equation}
\begin{split}
A_{3,0}^{(1)} &= \frac{i \sigma  \left[\left(9 \sigma ^2-5\right) m_a+\left(5 \sigma ^2-2\right) m_b\right]}{2 \left(\sigma ^2-1\right)^2 m_a^4 m_b}\,, \quad
A_{3,0}^{(2)} = \frac{i \sigma  \left[\left(7 \sigma ^2-3\right) m_a+3 m_b\right]}{2 \left(\sigma ^2-1\right)^2 m_a^2 m_b}\,.
\end{split}
\end{equation}
\end{subequations}
\item $S_a^2 S_b^2$
\begin{subequations}
\begin{equation}
\begin{split}
M^{(2, 2)} &= 
A_{2,2}^{(1)} \mathcal{E}_a^2\mathcal{E}_b^2 + 
A_{2,2}^{(2)} q^2 (p_2 \cdot S_a)(p_1 \cdot S_b) \mathcal{E}_a\mathcal{E}_b+
A_{2,2}^{(3)} q^2 (p_1 \cdot S_b)^2 \mathcal{E}_a^2 +
A_{2,2}^{(4)} q^2 (p_2 \cdot S_a)^2 \mathcal{E}_b^2 \\
&\quad +
A_{2,2}^{(5)} q^4 (p_2 \cdot S_a)^2  (p_1 \cdot S_b)^2\,.
\end{split}
\end{equation}
The corresponding coefficeints $A_{2,2}^{(i)}$ are:
\begin{equation}
\begin{split}
A_{2,2}^{(1)} &= -\frac{\left(95 \sigma ^4-95 \sigma ^2+12\right) \left(m_a+m_b\right)}{16 \left(\sigma ^2-1\right)^3 m_a^4 m_b^4},\, \quad
A_{2,2}^{(2)} = \frac{\sigma  \left(15 \sigma ^2-11\right) \left(m_a+m_b\right)}{4 \left(\sigma ^2-1\right)^3 m_a^3 m_b^3}\,,\\
A_{2,2}^{(3)}&= -\frac{\left(7 \sigma ^2-3\right) m_a+\left(35 \sigma ^4-35 \sigma ^2+4\right) m_b}{16 \left(\sigma ^2-1\right)^3 m_a^4 m_b^2}\,, \quad
A_{2,2}^{(4)}= -\frac{\left(35 \sigma ^4-35 \sigma ^2+4\right) m_a+\left(7 \sigma ^2-3\right) m_b}{16 \left(\sigma ^2-1\right)^3 m_a^2 m_b^4}\,,\\
A_{2,2}^{(5)} &= \frac{\left(5 \sigma ^2-1\right) \left(m_a+m_b\right)}{16 \left(\sigma ^2-1\right)^3 m_a^2 m_b^2}\,.
\end{split}
\end{equation}
\end{subequations}
\item $S_a^3 S_b$
\begin{subequations}
\begin{equation}
M^{(3, 1)} = 
A_{3,1}^{(1)} \mathcal{E}_a^3\mathcal{E}_b +  
A_{3,1}^{(2)} q^2 (p_2 \cdot S_a)^2 \mathcal{E}_a \mathcal{E}_b + 
A_{3,1}^{(3)} q^2 (p_2 \cdot S_a)(p_1 \cdot S_b) \mathcal{E}_a^2 + 
A_{3,1}^{(4)} q^4 (p_2 \cdot S_a)^3(p_1 \cdot S_b)\,.
\end{equation}
The corresponding coefficeints $A_{3,1}^{(i)}$ are:
\begin{equation}
\begin{split}
A_{3,1}^{(1)} &= -\frac{3 \left(36 \sigma ^4-37 \sigma ^2+5\right) m_a+4 \left(20 \sigma ^4-19 \sigma ^2+2\right) m_b}{24 \left(\sigma ^2-1\right)^3 m_a^5 m_b^3}\,, \\
A_{3,1}^{(2)} &= \frac{\left(-28 \sigma ^4+27 \sigma ^2-3\right) m_a+\left(4-8 \sigma ^2\right) m_b}{8 \left(\sigma ^2-1\right)^3 m_a^3 m_b^3}\,,\\
A_{3,1}^{(3)} &=\frac{\sigma  \left(\left(13 \sigma ^2-9\right) m_a+4 \left(5 \sigma ^2-4\right) m_b\right)}{8 \left(\sigma ^2-1\right)^3 m_a^4 m_b^2}\,, \quad
A_{3,1}^{(4)} =\frac{\sigma  \left(3-7 \sigma ^2\right) m_a-4 \sigma  m_b}{24 \left(\sigma ^2-1\right)^3 m_a^2 m_b^2}\,.
\end{split}
\end{equation}
\end{subequations}
\item $S_a^4$
\begin{subequations}
\begin{equation}
M^{(4, 0)} = 
A_{4,0}^{(1)} \mathcal{E}_a^4 +  
A_{4,0}^{(2)} q^2 (p_2 \cdot S_a)^2 \mathcal{E}_a^2 + 
A_{4,0}^{(3)} q^4 (p_2 \cdot S_a)^4\,.
\end{equation}
The corresponding coefficeints $A_{4,0}^{(i)}$ are:
\begin{equation}
\begin{split}
A_{4,0}^{(1)} &= \frac{\left(-239 \sigma ^4+250 \sigma ^2-35\right) m_a+24 \left(4-5 \sigma ^2\right) \sigma ^2 m_b}{192 \left(\sigma ^2-1\right)^3 m_a^6 m_b^2}\,,\\
A_{4,0}^{(2)} &=\frac{\left(-49 \sigma ^4+46 \sigma ^2-5\right) m_a+8 \left(2-3 \sigma ^2\right) m_b}{32 \left(\sigma ^2-1\right)^3 m_a^4 m_b^2}\,,\\
A_{4,0}^{(3)} &=\frac{21 \sigma ^4 m_a-14 \sigma ^2 m_a+m_a+8 m_b}{192 \left(\sigma ^2-1\right)^3 m_a^2 m_b^2}\,.
\end{split}
\end{equation}
\end{subequations}
\end{itemize}

\subsubsection{The box and crossed box coefficients}
Two more cut conditions are imposed on eq.\eqref{dbcut} to get the quadruple cut,
\eqa
C_{4, \square} &= (-2l\cdot p_1 ) (2l\cdot p_2 ) C_2\big|_{\ell\cdot p_1=\ell\cdot p_2 = 0}\,, ~
C_{4, \cSquare} &= (2l\cdot p_4 ) (2l\cdot p_2 ) C_2\big|_{\ell\cdot p_4=\ell\cdot p_2 = 0}\,.
\eqae
We only present the computation of $c_{\,\square}$ from $C_{4, \square}$, since the crossed box coefficients can be obtained by a simple substitution rule
\eqa
c_{\cSquare}=c_{\,\square}\,\big|_{p_1\rightarrow -p_4, p_4 \rightarrow -p_1}\,.
\eqae
The quadruple cut conditions
\eqa
\ell^2=(\ell+q)^2=(\ell-p_1)^2-m_a^2=(\ell+p_2)^2-m_b^2=0\,.
\eqae 
are solved by four solutions for the loop momentum~\cite{Bern:2020buy},
\eqa
\ell^\mu_{\pm,1}=-\frac{q^2 \eta^\mu}{2 q\cdot \eta_\pm}\,,~~~~\ell^\mu_{\pm,2}=\frac{N_1 p^\mu_1+N_2 p^\mu_2+N_{q}q^\mu}{\mathcal N}+\frac{q^2 \eta_\pm^\mu}{2 q\cdot \eta_\pm}\,,
\eqae
where
\eqa
\notag &&~~N_1=2m_b(m_b+m_a\sigma)q^2\,,~N_2=-2m_a(m_a+m_b\sigma)q^2\,,\\
&&~~N_q=4m_a^2m_b^2(\sigma^2-1)\,,~~~~\mathcal N=N_q+\frac{N_2-N_1}{2}\,.
\eqae
The null vectors $\eta_\pm$ need to satisfy the properties
\eqa
\eta_\pm^2=\eta_\pm\cdot p_1=\eta_\pm\cdot p_2=0\,.
\eqae
A choice for $\eta_\pm$ is
\eqa
\eta_\pm^{\mu} = \aPs{k_1}{\sigma^{\mu}}{k_2}\,,
\eqae
where
\eqa
(k_1)_\pm^{\mu} = p_1^{\mu} + m_a^2 p_2^{\mu}\zeta_\pm\,, \quad
(k_2)_\pm^{\mu} = p_2^{\mu} + m_b^2  p_1^{\mu}\zeta_\pm\,, \quad
\zeta_{\pm} =  \frac{-\sigma \pm \sqrt{\sigma^2 - 1}}{m_a m_b}\,.
\eqae
Inserting the solutions into the quadruple cut $C_4$ yields the box coefficient as
\begin{equation}
c_{\,\square} = \frac{1}{4}\sum_{i=1,2} \left[C_{4, \square}\left(\ell_{+,i}\right)+C_{4, \square}\left(\ell_{-,i}\right)\right]\,.
\end{equation}
\paragraph{The classical limit of the box and crossed box coefficients}
$\phantom{123123123}$\newline
\\[-0.3mm]
We adopt the normalization:
\begin{equation}
M_{\,\square} = 64\pi^2 G^2 \sum_{i=0}^{4} \sum_{ \substack{j =0,\\ i + j \leq 4}}^{2} c_{\,\square}^{(i, j)} \mathcal{I}_{\,\square}\,.
\end{equation}

\begin{subequations}
\begin{itemize}[leftmargin=*]
\item Scalar coupling
\begin{equation}
c_{\,\square}^{(0,0)} = 4m_a^2 m_b^2(2\sigma^2 - 1)^2\,.
\end{equation}
\item $S_a$
\begin{equation}
c_{\,\square}^{(1,0)} = 8im_a^2 m_b^3 \sigma (2\sigma^2 - 1)\mathcal{E}_a\,.
\end{equation}
\item $S_a S_b$
\begin{equation}
c_{\,\square}^{(1,1)} = 2m_a^3 m_b^3 q\cdot S_a q\cdot S_b - \frac{2m_a m_b(8\sigma^4 - 8\sigma^2 + 1)}{\sigma^2- 1} \mathcal{E}_a \mathcal{E}_b\,.
\end{equation}
\item $S_a^2$
\begin{equation}
c_{\,\square}^{(2,0)} = m_a^2 m_b^4 (q\cdot S_a)^2 - \frac{ m_b^2(8\sigma^4 - 8\sigma^2 + 1)}{\sigma^2- 1} \mathcal{E}_a^2\,.
\end{equation}
\item $S_a^2S_b$
\begin{equation}
c_{\,\square}^{(2,1)} = - \frac{ 4i(2\sigma^2 - 1)\sigma}{m_a(\sigma^2- 1) } \mathcal{E}_a^2 \mathcal{E}_b\,.
\end{equation}
\item $S_a^3$
\begin{equation}
c_{\,\square}^{(3,0)} = - \frac{ 4m_bi(2\sigma^2 - 1) \sigma}{3m_a^2(\sigma^2- 1)} \mathcal{E}_a^3\,.
\end{equation}
\item $S_a^2 S_b^2$
\begin{equation}
c_{\,\square}^{(2,2)} =   \frac{1}{2}m_a^2m_b^2 (q\cdot S_a)^2 (q\cdot S_b)^2 + \frac{8\sigma^4 - 8 \sigma^2 +1}{2m_a^2 m_b^2(\sigma^2 - 1)^2} \mathcal{E}_a^2 \mathcal{E}_b^2\,.
\end{equation}
\item $S_a^3 S_b$
\begin{equation}
c_{\,\square}^{(3,1)} = \frac{1}{3}m_a m_b^3 (q\cdot S_a)^3 (q\cdot S_b) + \frac{8\sigma^4 - 8 \sigma^2 +1}{3m_a^3 m_b(\sigma^2 - 1)^2} \mathcal{E}_a^3 \mathcal{E}_b\,.
\end{equation}
\item $S_a^4$
\begin{equation}
c_{\,\square}^{(4,0)} =   \frac{m_b^4}{12} (q\cdot S_a)^4 + \frac{8\sigma^4 - 8 \sigma^2 +1}{12m_a^4 (\sigma^2 - 1)^2} \mathcal{E}_a^4\,.
\end{equation}
\end{itemize}
\end{subequations}
The crossed box coefficient is obtained by simply replacing $ p_1\rightarrow - p_4$ and $ p_4\rightarrow - p_1$, which is equivalent to $\sigma \rightarrow - \sigma + \mathcal{O}(t)$, $\mathcal{E}_i \rightarrow - \mathcal{E}_i$. The crossed box coefficient turns out to be the same as the box coefficient in the classical limit:
\begin{equation}
c_{\cSquare}
 = c_{\,\square}\,.
\end{equation}
The box the integrals in eq.(\ref{IntResult}) contain a classical piece that is infrared divergent, but combining the box and the crossed box contribution results in a purely imaginary IR divergent piece because they have the same coefficients:
\eq
-\frac{i c_{\,\square}}{16\pi |\vec{q}|^2|\vec{p}|E}\left( \frac{1}{\epsilon} - \log |\vec{q}|^2 \right)\,.
\eqe
This needs to be canceled by the iteration terms from the 1 PM potential, which we compute next. 

\subsection{Iteration terms}\label{subsec:Iteration}
We match the full theory 2 PM classical amplitude to that computed from the EFT to extract the classical potential following refs.\cite{Cheung:2018wkq,Bern:2020buy} (for earlier works at 1 PM see refs.\cite{Vaidya:2014kza, Chung:2020rrz}). For computing iteration terms the on-shell potentials are continued off-shell with a hat to denote the continuation $\hat{V}(\vec{p}^{\,\prime}, \vec{p})$, where $\vec{p}$  $(\vec{p}^{\,\prime})$ is the in-coming (out-going) momentum in the COM frame with $|\vec{p}|^2 \neq |\vec{p}^{\,\prime}|^2$. Spin operators $\vec{S}_n$ satisfying the SU(2) commutation relations are included in the potential for spinning cases, where $n = a, \; b$ is the particle label.
\begin{equation}
[S_n^{i}, S_m^{j}] = i \epsilon^{ijk} \delta_{nm} S^{k}_m\,.
\end{equation}
We begin with the potential expanded in the spin-vector basis~\cite{Bern:2020buy}:
\eqa
\notag \hat{V}_n(\vec{k}^{\,\prime}, \vec{k}) &= 
& \frac{G^n}{|\vec{Q}|^{3-n}} \sum_{\alpha, n_a, n_b} \mathcal{V}^{(\alpha)}_{n_a, n_b}
 \mathbb{O}^{(\alpha)}_{n_a, n_b} \,,\\
\notag \mathbb{O}^{(\alpha)}_{n_a, n_b}&\equiv&(T^{(\alpha)}_{n_a,n_b})^{j_1 \cdots j_{n_a + n_b}, i_1 \cdots i_{n_a}, i_{1} \cdots i_{n_b}} Q^{j_1}\cdots Q^{j_{n_a {+} n_b}} \mathbb{S}^{i_1\dots i_{n_a}}_a \mathbb{S}^{i_1\dots i_{n_b}}_b\,,\\\mathbb{S}_{a/b}^{i_1 \cdots i_n}& \equiv& S_{a/b}^{(i_1}S_{a/b}^{i_2}\cdots S_{a/b}^{i_n)} = \frac{1}{n!} \left[ S_{a/b}^{i_1}\cdots S_{a/b}^{i_n} + \text{all perm.  in $\{i_1 \cdots i_n\}$}\right]\,,
\eqae
where $n_a\,, n_b$ denotes the degree of spin for particles $a,b$ respectively, and $\alpha$ labels distinct tensor structures. The scalar function $\mathcal{V}^{(\alpha)}_{n_a, n_b}$ only depends on $|\vec{P}|^2 = (|\vec{k}'|^2 + |\vec{k}^{\,\prime}|^2)/2$, $\vec{Q}$ is the momentum transfer $\vec{Q} = \vec{k}- \vec{k}'$, and $T^{(\alpha)}_{n_a,n_b}$ denotes a tensorial combination of momentum $\vec{p}$, Kronecker delta, and the 3D Levi-Civita tensor. This is a basis of operators with classical $\hbar$ scaling $\mathcal{O}(\hbar^{-3})$.\footnote{This is the potential in momentum space, and the position space potential gains extra factors of $\hbar^3$ from momentum space measure $\frac{d^3 k}{(2\pi)^3}$.}  The symmetrised product of spin operators $\mathbb{S}$ can be algebraically reduced as
\eqa
S^{i}S^{j} &=& \mathbb{S}^{ij}
+ \frac{i}{2}\epsilon^{ijk}S^{k}\,,\\
S^i S^j S^k 
&=&
\mathbb{S}^{ijk}
+ \frac{i}{2}\left( \epsilon^{jkl}\mathbb{S}^{il} +  \epsilon^{ijl}\mathbb{S}^{kl} +  \epsilon^{ikl}\mathbb{S}^{jl}\right)
-\frac{1}{6}\left(2\delta^{jm}\delta^{ik} - \delta^{im}\delta^{jk}- \delta^{ij}\delta^{km}\right)S^m\,,\nonumber \\
S^i S^j S^k S^l
&=&
\mathbb{S}^{ijkl}
+ \frac{i}{2}
\left(   
\epsilon^{ijn} \mathbb{S}^{kln}+  
\epsilon^{ikn}\mathbb{S}^{jln} +
\epsilon^{iln}\mathbb{S}^{jkn} +
\epsilon^{jkn}\mathbb{S}^{iln} +
\epsilon^{jln}\mathbb{S}^{ikn} +
\epsilon^{kln}\mathbb{S}^{ijn}
\right)\nonumber\\
&&
+
\frac{1}{6}
\left(   
5\delta^{ij}\mathbb{S}^{kl}-
\delta^{ik}\mathbb{S}^{jl}-
7\delta^{il}\mathbb{S}^{jk}-
\delta^{jk}\mathbb{S}^{il} +
5 \delta^{kl}\mathbb{S}^{ij} 
\right)  \nonumber  
\\&&+\frac{i}{6}\left(
\epsilon^{ikn}\delta^{jl} - 
\epsilon^{iln}\delta^{jk} +
\epsilon^{jln}\delta^{ik}
\right) S^{n}- \frac{1}{2}\left(\delta^{ij}\delta^{kl} - \delta^{il}\delta^{jk}\right) \mathbb{S}^{mn}\delta^{mn}\,
\,.\nonumber
\eqae
Each term in the potential can be represented as a contact term in the EFT Lagrangian, coupled to appropriate spin states. The EFT 2 PM amplitude is obtained as
\begin{equation}
M_{\text{EFT}} = 
- \left\langle
\hat{V}_2(\vec{p}^{\,\prime}, \vec{p}) 
+ \int \frac{d^3 \vec{k}}{(2\pi)^3} \frac{\hat{V}_1(\vec{p}^{\,\prime}, \vec{k})\hat{V}_1(\vec{k}, \vec{p})}{E - \sqrt{|\vec{k}|^2 - m_a^2}  - \sqrt{|\vec{k}|^2 - m_b^2}}
\right\rangle \,,
\end{equation}
which can also be understood as solving the Lippman-Schwinger equation to second order in $G$~\cite{Cristofoli:2019neg}. Importantly, the hat denotes that the ``potential" $\hat{V}$ are off-shell potential operators acting on the spin coherent states. Thus the bracket $\AB{\hat{O}}$ denotes sandwiching the operator $\hat{O}$ between the spin coherent state. This momentum space amplitude can then be matched to the classical piece of the QFT computation at 1-loop eq\eqref{1loopb}.
By requiring $M^{\rm 1{-}loop}_{\rm classical}  = M_{\text{EFT}}$ we find the 2PM potential $V_2 = \AB{\hat{V}_{2}}$ is given by: 
\begin{equation}\label{eq:2PM_pot_formula}
V_2 = 
- M^{\rm 1{-}loop}_{\rm classical}
- \left\langle
\int \frac{d^3 \vec{k}}{(2\pi)^3} \frac{\hat{V}_1(\vec{p}^{\,\prime}, \vec{k})\hat{V}_1(\vec{k}, \vec{p})}{E - \sqrt{|\vec{k}|^2 - m_a^2}  - \sqrt{|\vec{k}|^2 - m_b^2}}
\right\rangle\,.
\end{equation}
The second term which stems from contributions from the 1 PM potential is called the iteration terms, and needs to be subtracted from the one-loop amplitude to obtain the 2 PM potential.

As an explicit example, the 1PM potential vertex that goes into the iteration integral is:

\eqa
\label{eq:1PM_Vertex}
&&\hspace{-1cm}\hat{V}_1(\vec{p}_{out}, \vec{p}_{in}) = 
\frac{4\pi G}{|\vec{Q}|^2}
\left[
\mathcal{V}_{0,0} \mathbb{I} + 
\mathcal{V}_{1,0} \vec{p}_{in}\times \vec{Q}\cdot \vec{S}_a + 
\mathcal{V}_{0,1} \vec{p}_{in}\times  \vec{Q}\cdot \vec{S}_b\right. \\
&&\hspace{7cm}\left.+ 
\mathcal{V}_{1,1} \left( \vec{Q}\cdot \vec{S}_a\right)\left(  \vec{Q}\cdot \vec{S}_b\right)
+ \cdots
\right] \,,\notag 
\eqae
where the coefficients $\mathcal{V}_{n_a, n_b}$ can be obtained by expanding the closed form 1PM potential~\cite{Chung:2020rrz}:
\begin{equation}\label{eq:1PM_Ham}
V_{\text{1 PM}} = - \frac{4\pi G}{q^2} \frac{m_a^2 m_b^2}{E_a E_b} \cosh\left[2 \sigma + \frac{ i(m_a^{-1}\mathcal{E}_a + m_b^{-1}\mathcal{E}_b)}{m_a m_b \sinh \sigma}\right]U_a U_b\,.
\end{equation}
and using eq.(\ref{CalEDef}) to rewrite $\mathcal{E}$ in terms of spin vector $\vec{S}$.

\paragraph{The iteration integral term in the 3D integral basis}$\phantom{123}$\newline
\\[-0.3mm]
Evaluation of the full integral is not needed for determining classical contributions, and considering the integrand in the limit $|\vec{k}|^2\sim|\vec{p}|^2$ is enough~\cite{Cheung:2018wkq}. Expanding the integrand of iteration term to $\mathcal{O}(|\vec{k}|^2 - |\vec{p}|^2)^0$ order yields
\eqa\label{eq:Iter_ten_int}
\notag  &&\hspace{-0.2cm}\int \frac{d^3 \vec{k}}{(2\pi)^3} \frac{\hat{V}(\vec{p}^{\,\prime}, \vec{k})\hat{V}(\vec{k}, \vec{p})}{E {-} \sqrt{|\vec{k}|^2 {-} m_a^2}  {-} \sqrt{|\vec{k}|^2 {-} m_b^2}}\Bigg|_{\mathcal{O}(|\vec{k}|^2 - |\vec{p}|^2)} \\
&&= 
\notag \sum_{\substack{n_a,n_b \\ n'_a,n'_b}}\int_{\mathbb{O}}\left(\frac{2\xi E}{(|\vec{k}|^2 {-}|\vec{p}|^2)}{+}\frac{1-3\xi}{2\xi E} + \cdots \right)\left[\mathcal{V}_{n_a,n_b}\mathcal{V}_{n'_a,n'_b} {+} \frac{1}{2} \partial(\mathcal{V}_{n_a,n_b} \mathcal{V}_{n'_a,n'_b})(|\vec{k}|^2{-} |\vec{p}|^2) {+} \cdots\right] \\
\notag &&
=
\sum_{\substack{n_a,n_b \\ n'_a,n'_b}}\int_{\mathbb{O}}\left[
\frac{2\xi E}{|\vec{k}|^2 {-} |\vec{p}|^2}\mathcal{V}_{n_a,n_b}\mathcal{V}_{n'_a,n'_b}+
\frac{1-3\xi}{2\xi E} \mathcal{V}_{n_a,n_b}\mathcal{V}_{n'_a,n'_b}{+} \xi E \partial(\mathcal{V}_{n_a,n_b} \mathcal{V}_{n'_a,n'_b})
\right]
\\
 &&\equiv
- \text{Iter}_{\bigtriangleup} - \text{Iter}_{\cbubble}\,,
\eqae
where we define $\xi=\frac{E_aE_b}{(E_a{+}E_b)^2}$ and
\eqa
\notag \int_{\mathbb O}\equiv-\int \frac{d^3 \vec{k}}{(2\pi)^3}\mathbb{O}_{n'_a,n'_b}(\vec{k})\mathbb{O}_{n_a,n_b}(\vec{k}) \frac{16\pi^2 G^2}{|\vec{k} {-}\vec{p}|^2|\vec{k} {-} \vec{p} {+} \vec{q}|^2}\,.
\eqae
In the first equality $\mathcal{V}_{n_a,n_b}$ has been expanded at $|\vec k|=|\vec p|$ so that $\mathcal{V}_{n_a,n_b}$ becomes a function of $|\vec p|$, and the following identity was used.
\begin{equation}
\begin{split}
&\frac{\partial}{\partial |\vec{k}|^2} \left[\mathcal{V}_{i,j}(|\vec{P}|^2)\mathcal{V}_{i',j'}(|\vec{P}|^2)\right]\Big|_{|\vec{k}|^2 = |\vec{p}|^2}
= \frac{1}{2}\frac{\partial}{\partial |\vec{P}|^2} \left[\mathcal{V}_{i,j} i(|\vec{P}|^2)\mathcal{V}_{i',j'}(|\vec{P}|^2)\right]\Big|_{|\vec{P}|^2 = |\vec{p}|^2} \\
=& \frac{1}{2}\frac{\partial}{\partial |\vec{p}|^2} \left[\mathcal{V}_{i,j}(|\vec{p}|^2)\mathcal{V}_{i',j'}(|\vec{p}|^2)\right]
\equiv
\frac{1}{2}\partial(\mathcal{V}_{i,j} \mathcal{V}_{i',j'})\,.
\end{split}
\end{equation}
$\text{Iter}_{\bigtriangleup}$ corresponds to the first term in the penultimate line of \eqc{eq:Iter_ten_int} having three propagators, and $ \text{Iter}_{\cbubble}$ corresponds to the remaining terms with two propagators. Note that the operators $\mathbb{O}_{i}(\vec{k})$ are tensors of $\vec{k}$ contracted to other vectors, therefore the last line of eq.\eqref{eq:Iter_ten_int} is a tensor integral of $\vec{k}$. We expand the tensor integrals by scalar intgrals
\begin{equation}\label{eq:3D_ten_int_to_sca_int}
\begin{split}
\int \frac{d^3 \vec{k}}{(2\pi)^3} \frac{k^{i_1} \cdots k^{i_n}}{|\vec{k} - \vec{p}|^2 |\vec{k} - \vec{p} + \vec{q}|^2} & = c_2^{i_1 \cdots i_n}(|\vec{p}|^2, |\vec{q}|^2) I + \cdots,\\
\int \frac{d^3 \vec{k}}{(2\pi)^3} \frac{k^{i_1} \cdots k^{i_n}}{|\vec{k} - \vec{p}|^2 |\vec{k} - \vec{p} + \vec{q}|^2(|\vec{k}|^2 - |\vec{p}|^2)} & = 
c_2^{i_1 \cdots i_n}(|\vec{p}|^2, |\vec{q}|^2) I
+ 
c_3^{i_1 \cdots i_n}(|\vec{p}|^2, |\vec{q}|^2) J
+ \cdots\,,
\end{split}
\end{equation}
where $c_2^{i_1 \cdots i_n}$ and $c_3^{i_1 \cdots i_n}$ are the tensors constructed from 3D Kronecker deltas and a scalar function depending on $|\vec{p}|^2, |\vec{q}|^2$. Ellipsis denotes terms irrelevant for classical physics. The 3D scalar integrals $I, \; J$ are given as:
\begin{equation}
\begin{split}
I &= \int \frac{d^3 \vec{k}}{(2\pi)^3}\frac{1}{|\vec{k} - \vec{p}|^2 |\vec{k} - \vec{p} + \vec{q}|^2} = \frac{1}{8|\vec{q}|},\\
J &= \int \frac{d^3 \vec{k}}{(2\pi)^3}\frac{1}{|\vec{k} - \vec{p}|^2 |\vec{k} - \vec{p} + \vec{q}|^2 (|\vec{k}|^2 - |\vec{p}|^2)}  = \frac{1}{8\pi} \frac{1}{|\vec{p}||\vec{q}|^2} \left( \frac{1}{\epsilon} - \log|\vec{q}|^2\right)\,.
\end{split}
\end{equation}
The full theory 1-loop amplitude of section \ref{subsec:Int_Coeff} should also be recast using three-dimensional variables. The integral coefficients are converted using eq.\eqref{eq:Kinematics} and eq.\eqref{eq:Spin_Vec_3D}. The 4D scalar integrals are reduced to the 3D scalar integrals as
\begin{equation}
\begin{split}
\mathcal{I}_{\bigtriangleup,\bigtriangledown} &= -\frac{i}{4m_{a,b}} \int \frac{d^3 \vec{k}}{(2\pi)^3} \frac{1}{|\vec{k} - \vec{p}|^2  |\vec{k} - \vec{p} + \vec{q}|^2} + \cdots= -\frac{i}{4m_{a,b}}I+ \cdots,\\
\text{Im}(\mathcal{I}_{\,\square}) &= \frac{i}{2E} \int \frac{d^3 \vec{k}}{(2\pi)^3} \frac{1}{|\vec{k} - \vec{p}|^2  |\vec{k} - \vec{p} + \vec{q}|^2(|\vec{k}|^2 - |\vec{p}|^2)} + \cdots= \frac{i}{2E}J+ \cdots\,.
\end{split}
\end{equation}
The 4D triangle integral is proportional to the 3D scalar bubble integral $I$, and the imaginary part of the 4D box integral is proportional to the 3D scalar triangle integral $J$.

\subsubsection{The 3D bubble coefficient}\label{subsubsec:3D_bubble_coeff_result}
We present the 3D bubble part of the iteration integrals up to quartic order in spin based on the discussions of section \ref{subsec:Iteration}.
\begin{equation}
\text{Iter}_{\cbubble} = \frac{2\pi^2 G^2}{|\vec{q}|} \sum_{n_a = 0}^{4}\sum_{\substack{n_b =0,\\ n_a + n_b \leq 4}}^{2} \tilde{I}_{n_a, n_b}\,.
\end{equation}
The variables are defined as
\begin{equation}
\begin{split}
\IterCoeff{i}{j}{i'}{j'} &= \mathcal{V}_{i,j}\mathcal{V}_{i',j'}\,,\\
\IterCoeff{i}{j}{i'}{j'}^{(k)} &= \left[ (1-3\xi) + \frac{k \xi^2 E^2}{\Pv{2}} + 2\xi^2 E^2 \partial \right]\left( \mathcal{V}_{i,j}\mathcal{V}_{i',j'} \right)\,,
\end{split}
\end{equation}
where $\mathcal{V}_{i,j}$ are the 1PM coefficients defined in eq.\eqref{eq:1PM_Vertex}.
\begin{itemize}[leftmargin=*]
\item Scalar coupling
\begin{equation}
\tilde{I}_{0, 0} = \frac{1}{2\xi E} \IterCoeff{0}{0}{0}{0}^{(0)}\,.
\end{equation}
This result is consistent with the one given in eq.(6.24) of \cite{Bern:2020buy}

\item $S_a$
\begin{equation}
\tilde{I}_{1, 0} = \frac{1}{2\xi E} \IterCoeff{1}{0}{0}{0}^{(2)}\,.
\end{equation}
This result is consistent with the one given in eq.(6.24) of \cite{Bern:2020buy}

\item $S_a S_b$
\begin{subequations}
\begin{equation}
\tilde{I}_{1, 1} = 
\tilde{I}_{1, 1}^{(1)} \; |\vec{q}|^2 \vec{p}\cdot \vec{S}_a \vec{p}\cdot \vec{S}_b +
\tilde{I}_{1, 1}^{(2)} \; |\vec{q}|^2 \vec{S}_a \cdot \vec{S}_b +
\tilde{I}_{1, 1}^{(3)} \; \vec{q}\cdot \vec{S}_a \vec{q}\cdot \vec{S}_b \,.
\end{equation}
The corresponding coefficients are:
\begin{align}
\tilde{I}_{1,1}^{(1)} &= 
\frac{E \xi}{\Pv{4}} \IterCoeff{1}{1}{0}{0} 
+ \frac{1}{4 E \xi} \IterCoeff{1}{0}{0}{1}^{(4)} 
- \frac{E \xi}{2 \Pv{2}} ( \IterCoeff{1}{1}{0}{1} + \IterCoeff{1}{1}{1}{0} )\,,  \nonumber \\
\tilde{I}_{1,1}^{(2)} &= 
-\frac{1}{8E \xi} \IterCoeff{1}{1}{0}{0}^{(4)} 
- \frac{\Pv{2}}{4E \xi} \IterCoeff{1}{0}{0}{1}^{(4)} 
+ \frac{E \xi}{8} ( \IterCoeff{1}{1}{0}{1} + \IterCoeff{1}{1}{1}{0} )\,,   \\
\tilde{I}_{1,1}^{(3)} &= 
\frac{3}{8E \xi} \IterCoeff{1}{1}{0}{0}^{(\frac{4}{3})} 
+ \frac{\Pv{2}}{8E \xi} \IterCoeff{1}{0}{0}{1}^{(4)} 
+ \frac{E \xi}{8} ( \IterCoeff{1}{1}{0}{1} + \IterCoeff{1}{1}{1}{0} )\,.\nonumber
\end{align}
\end{subequations}
This result is consistent with the one given in eq.(6.24) of \cite{Bern:2020buy}

\item $S_a^2$
\begin{equation}
\tilde{I}_{2, 0} =
\tilde{I}_{2, 0}^{(1)} \; |\vec{q}|^2 (\vec{p}\cdot \vec{S}_a)^2  +
\tilde{I}_{2, 0}^{(2)} \; |\vec{q}|^2 |\vec{S}_a|^2 +
\tilde{I}_{2, 0}^{(3)} \; (\vec{q}\cdot \vec{S}_a)^2\,.
\end{equation}
The corresponding coefficients are:
\begin{align}
\tilde{I}_{2, 0}^{(1)} &= 
\frac{E \xi}{\Pv{4}}\IterCoeff{2}{0}{0}{0} 
+ \frac{1}{8\xi E} \IterCoeff{1}{0}{1}{0} 
- \frac{E \xi}{\Pv{2}}\IterCoeff{2}{0}{1}{0} \nonumber\,, \\
\tilde{I}_{2, 0}^{(2)} &= 
-\frac{1}{8 E \xi}\IterCoeff{2}{0}{0}{0}^{(4)} 
- \frac{\Pv{2}}{8\xi E} \IterCoeff{1}{0}{1}{0} 
+ \frac{E \xi}{4}\IterCoeff{2}{0}{1}{0} \,,\\
\tilde{I}_{2, 0}^{(3)} &= 
\frac{3}{8 E \xi}\IterCoeff{2}{0}{0}{0}^{(\frac{4}{3})} 
+ \frac{\Pv{2}}{16\xi E} \IterCoeff{1}{0}{1}{0}^{(4)} 
+ \frac{E \xi}{4}\IterCoeff{2}{0}{1}{0}\,. \nonumber
\end{align}
This result is consistent with the one given in eq.(4.11) of \cite{Kosmopoulos:2021zoq}

\item $S_a^2S_b$
\begin{equation}
\begin{split}
\tilde{I}_{2,1} &= 
i\left[ \tilde{I}_{2,1}^{(1)} |\vec{q}|^2 (\vec{p} \cdot \vec{S}_a)^2 (\vec{p} \times \vec{q}) \cdot \vec{S}_b +
 \tilde{I}_{2,1}^{(2)} |\vec{q}|^2 (\vec{p} \cdot \vec{S}_a)(\vec{p} \cdot \vec{S}_b) (\vec{p} \times \vec{q}) \cdot \vec{S}_a \right.\\
&\quad +
\tilde{I}_{2,1}^{(3)} |\vec{q}|^2 \vec{S}_a \cdot \vec{S}_b (\vec{p} \times \vec{q}) \cdot \vec{S}_a + 
\tilde{I}_{2,1}^{(4)}  \vec{q}\cdot\vec{S}_a \vec{q}\cdot \vec{S}_b (\vec{p} \times \vec{q}) \cdot \vec{S}_a \\
&\quad 
\left. 
+
\tilde{I}_{2,1}^{(5)} |\vec{q}|^2 |\vec{S}_a|^2 (\vec{p} \times \vec{q}) \cdot \vec{S}_b +
\tilde{I}_{2,1}^{(6)} (\vec{q}\cdot \vec{S}_a)^2 (\vec{p} \times \vec{q}) \cdot \vec{S}_b \right].
\end{split}
\end{equation}
Note that this basis is not linearly independent, since different tensor structures may be related by the Schouten identity:
\begin{equation}\label{eq: Sa2_Sb_Identity}
\begin{split}
&|\vec{q}|^2 \vec{p} \cdot \vec{S}_a 
\left[- (\vec{p}\times \vec{q}\cdot \vec{S}_a) \vec{p} \cdot \vec{S}_b + 
(\vec{p}\times \vec{q}\cdot \vec{S}_b) \vec{p} \cdot \vec{S}_a \right] \\
-&
|\vec{p}|^2
\left[|\vec{q}|^2 |\vec{S}_a|^2 - (\vec{q}\cdot \vec{S}_a)^2\right] (\vec{S}_b \cdot \vec{p} \times \vec{q})
+|\vec{p}|^2
\left[|\vec{q}|^2 \vec{S}_a \cdot \vec{S}_b - (\vec{q}\cdot \vec{S}_a)(\vec{q}\cdot \vec{S}_b)\right](\vec{S}_a \cdot \vec{p} \times \vec{q})\\
=&
\vec{p} \cdot \vec{q}
\left[ 
|\vec{q}|^2 (\vec{p}\cdot\vec{S}_a) (\vec{p}\cdot \vec{S}_a \times \vec{S}_b)
-
|\vec{p}|^2 (\vec{q}\cdot\vec{S}_a) (\vec{q}\cdot \vec{S}_a \times \vec{S}_b)
\right]\\
=&
\frac{|\vec{q}|^2}{2}
\left[ 
|\vec{q}|^2 (\vec{p}\cdot\vec{S}_a) (\vec{p}\cdot \vec{S}_a \times \vec{S}_b)
-
|\vec{p}|^2 (\vec{q}\cdot\vec{S}_a) (\vec{q}\cdot \vec{S}_a \times \vec{S}_b)
\right],
\end{split}
\end{equation}
where we used the on-shell condition $\vec{p}\cdot \vec{q} = \frac{|\vec{q}|^2}{2}$ on the last line. When restoring the $\hbar$ factors, the last line of eq.\eqref{eq: Sa2_Sb_Identity} carries one more power of $\hbar$, so it can be neglected. The coorsponding coefficients we derived are:
{
\small
\begin{align} \label{eq:Iter_2_1_res}
\small
\tilde{I}_{2,1}^{(1)} &= \frac{3 E \xi}{4\Pv{4}} (\IterCoeff{2}{1}{0}{0} + \IterCoeff{2}{0}{0}{1}) - \frac{E \xi}{4 \Pv{2}} \IterCoeff{2}{1}{0}{1} - \frac{E \xi}{8 \Pv{4}}\IterCoeff{1}{1}{1}{1}, \nonumber\,,\\
\tilde{I}_{2,1}^{(2)} &=\frac{3 E \xi}{4\Pv{4}} \IterCoeff{1}{1}{1}{0} - \frac{E \xi}{2\Pv{4}} (\IterCoeff{2}{1}{1}{0}\Pv{2}+\IterCoeff{2}{0}{1}{1}) \nonumber\,,\\
\tilde{I}_{2,1}^{(3)} &=
\frac{E \xi}{4\Pv{2}} 
(\IterCoeff{2}{1}{0}{0}
+\IterCoeff{2}{0}{0}{1} 
+ 2 \IterCoeff{1}{1}{2}{0}) 
- \frac{1}{16 E \xi} \IterCoeff{1}{1}{1}{0}^{(8)}  \nn \\ &\phantom{=asdf}
+ \frac{E \xi}{2}\IterCoeff{1}{0}{2}{1} 
- \frac{E \xi}{4}\IterCoeff{0}{1}{2}{1}
- \frac{E \xi}{8 \Pv{2}}\IterCoeff{1}{1}{1}{1}
\nonumber\,,\\
\tilde{I}_{2,1}^{(4)} &=
\frac{1}{8E \xi} (\IterCoeff{2}{1}{0}{0}^{(2)} 
+ \IterCoeff{1}{1}{1}{0}^{(4)} 
- \IterCoeff{0}{1}{2}{0}^{(6)}) 
- \frac{E \xi}{2} \IterCoeff{2}{1}{1}{0} 
+ \frac{E \xi}{4} \IterCoeff{2}{1}{0}{1} 
+ \frac{E \xi}{8 \Pv{2}}\IterCoeff{1}{1}{1}{1}\,,
\nonumber
\\
\tilde{I}_{2,1}^{(5)} &=
-\frac{1}{16E \xi} (\IterCoeff{2}{1}{0}{0}^{(10)} + \IterCoeff{2}{0}{0}{1}^{(10)}) 
+ \frac{E \xi}{8 \Pv{2}}\IterCoeff{1}{1}{1}{0} 
- \frac{3E \xi}{8} \IterCoeff{2}{1}{1}{0}  \nn \\ &\phantom{=asdf}
+ \frac{E \xi}{4}\IterCoeff{2}{1}{0}{1} 
- \frac{E \xi}{4\Pv{2}}\IterCoeff{1}{1}{2}{0} 
+ \frac{3E \xi}{16\Pv{2}}\IterCoeff{1}{1}{1}{1}
\nonumber\,,\\
\tilde{I}_{2,1}^{(6)} &=
\frac{3}{16E \xi} 
(\IterCoeff{2}{1}{0}{0}^{(\frac{14}{3})} 
- \frac{1}{3}\IterCoeff{1}{1}{1}{0}^{(6)} 
+ \IterCoeff{2}{0}{0}{1}^{(\frac{14}{3})}) 
+ \frac{5E\xi}{8}\IterCoeff{2}{1}{1}{0} \nn \\ &\phantom{=asdf}
- \frac{E \xi}{4}\IterCoeff{2}{1}{0}{1} 
+ \frac{E \xi}{4 \Pv{2}}\IterCoeff{1}{1}{2}{0} 
- \frac{E \xi}{16 \Pv{2}}\IterCoeff{1}{1}{1}{1}\,.
\end{align}
}
\item $S_a^3$
\begin{equation}
\begin{split}
\tilde{I}_{3,0} = 
i
&\left[ \tilde{I}_{3,0}^{(1)} |\vec{q}|^2 (\vec{p} \cdot \vec{S}_a)^2 (\vec{p} \times \vec{q}) \cdot \vec{S}_a +
 \tilde{I}_{3,0}^{(2)} |\vec{q}|^2 |\vec{S}_a|^2 (\vec{p} \times \vec{q}) \cdot \vec{S}_a \right.\\
&\quad 
\left.
+
\tilde{I}_{3,0}^{(3)} (\vec{q}\cdot\vec{S}_a)^2 (\vec{p} \times \vec{q}) \cdot \vec{S}_a \right]\,.
\end{split}
\end{equation}
The corresponding coefficients $\tilde{I}_{3,0}^{(\alpha)}$ are:
\begin{align}
\tilde{I}_{3,0}^{(1)} &= \frac{3E\xi}{4 \Pv{4}}(\IterCoeff{3}{0}{0}{0} + \IterCoeff{2}{0}{1}{0} ) - \frac{E \xi}{2 \Pv{4}}\IterCoeff{2}{0}{2}{0} - \frac{3E \xi}{4 \Pv{2}}\IterCoeff{3}{0}{1}{0} 
\nonumber\,,\\
\tilde{I}_{3,0}^{(2)} &= -\frac{1}{16 E\xi}(\IterCoeff{3}{0}{0}{0}^{(6)} + \IterCoeff{2}{0}{1}{0}^{(6)} ) + \frac{E \xi}{4 \Pv{2}}\IterCoeff{2}{0}{2}{0} + \frac{E \xi}{8}\IterCoeff{3}{0}{1}{0}\,,\\
\tilde{I}_{3,0}^{(3)} &= 
\frac{5}{16 E\xi}\IterCoeff{3}{0}{0}{0}^{(\frac{18}{5})} 
+ \frac{1}{16 E\xi}\IterCoeff{2}{0}{1}{0}^{(2)}   + \frac{E \xi}{4 \Pv{2}}\IterCoeff{2}{0}{2}{0} + \frac{E \xi}{8}\IterCoeff{3}{0}{1}{0}
\nonumber\,.
\end{align}

\item Quartic Spin Order: $S_a^2 S_b^2$, $S_a^3 S_b$, $S_a^4$ \newline
The expressions are too lengthy to be shown here. The explicit results can be found in the ancillary Mathematica file ``3D Integral Coefficients.wl".

\end{itemize}

\subsubsection{The 3D triangle coefficient: Cancellation of IR divergence}
IR divergences should not appear in the classical potential. This requires cancellation between the 3D triangle iteration integral and the sum of 4D box and crossed box integrals.
\begin{equation}
\text{Iter}_{\triangle} = \frac{M_{\,\square + \cSquare}}{4E_a E_b}U_a U_b\,.
\end{equation}
Decomposing $\text{Iter}_{\triangle}$ into a sum over different power of spin:
\begin{equation}
\text{Iter}_{\triangle} = J \sum_{n_a, n_b}\tilde{J}_{n_a, n_b}\,,
\end{equation}
where $\tilde{J}_{n_a, n_b}$ are the 3D triangle coefficients for various powers of spin, we obtain the following consistency condition:
\begin{equation}
\tilde{J}_{n_a, n_b} = \frac{1}{8\xi E^3} \hat{c}_{B}^{(n_a, n_b)}\,,
\end{equation}
where $\hat{c}_{B}$ denotes the 4D box coefficient with rotation factors $U_a U_b$ incorporated. This is the primary consistency check of our computations.
\begin{itemize}[leftmargin=*]
\item Scalar coupling
\begin{equation}
\tilde{J}_{0, 0} = 2 E \xi  \IterCoeff{0}{0}{0}{0}\,.
\end{equation}
\item $S_a$
\begin{equation}
\begin{split}
\tilde{J}_{1, 0} &= 
i\tilde{J}_{1, 0}^{(1)} \vec{p} \times \vec{q} \cdot \vec{S}_a  = 2i E \xi  \IterCoeff{1}{0}{0}{0}\vec{p} \times \vec{q} \cdot \vec{S}_a\,.
\end{split}
\end{equation}

This result is consistent with the one presented in \cite{Bern:2020buy}.

\item $S_a S_b$
\begin{subequations}
\begin{equation}
\tilde{J}_{1, 1} = 
\tilde{J}_{1, 1}^{(1)} |\vec{q}|^2 (\vec{p} \cdot \vec{S}_a)  (\vec{p} \cdot \vec{S}_b) +
\tilde{J}_{1, 1}^{(2)} |\vec{q}|^2 (\vec{S}_a \cdot \vec{S}_b) +
\tilde{J}_{1, 1}^{(3)} (\vec{q} \cdot \vec{S}_a)(\vec{q} \cdot \vec{S}_b)\,.
\end{equation}
The coorsponding coefficients $\tilde{J}_{1,1}^{(\alpha)}$ are:
\begin{align}
\tilde{J}_{1, 1}^{(1)} &=  \frac{E \xi}{\Pv{2}} \IterCoeff{1}{1}{0}{0} + E \xi \IterCoeff{1}{0}{0}{1}\,, \nonumber\\
\tilde{J}_{1, 1}^{(2)} &= -E \xi \IterCoeff{1}{1}{0}{0} - E \xi \Pv{2} \IterCoeff{1}{0}{0}{1}\,, \\
\tilde{J}_{1, 1}^{(3)} &= 2 E \xi \IterCoeff{1}{1}{0}{0}\,.\nonumber
\end{align}
\end{subequations}
This result is consistent with the one presented in \cite{Bern:2020buy}.

\item $S_a^2$
\begin{subequations}
\begin{equation}
\tilde{J}_{2, 0} = 
\tilde{J}_{2, 0}^{(1)} |\vec{q}|^2 (\vec{p} \cdot \vec{S}_a)^2+
\tilde{J}_{2, 0}^{(2)} |\vec{q}|^2 |\vec{S}_a|^2 +
\tilde{J}_{2, 0}^{(3)} (\vec{q} \cdot \vec{S}_a)^2\,.
\end{equation}
The coorsponding coefficients $\tilde{J}_{2,0}^{(\alpha)}$ are:
\begin{align}
\tilde{J}_{2, 0}^{(1)} &=  \frac{E \xi}{\Pv{2}} \IterCoeff{2}{0}{0}{0} + \frac{E \xi}{2} \IterCoeff{1}{0}{1}{0}\,, \nonumber\\
\tilde{J}_{2, 0}^{(2)} &= - E \xi \IterCoeff{2}{0}{0}{0} - \frac{E \xi \Pv{2}}{2} \IterCoeff{1}{0}{1}{0}\,, \\
\tilde{J}_{2, 0}^{(3)} &= 2 E \xi \IterCoeff{2}{0}{0}{0}\,.\nonumber
\end{align}
\end{subequations}

\item $S_a^2S_b$
\begin{subequations}
\begin{equation}
\begin{split}
\tilde{J}_{2,1} 
&=
i \left[ \tilde{J}_{2,1}^{(1)} |\vec{q}|^2 ( \vec{p}\cdot \vec{S}_a)^2 \vec{p}\times \vec{q}\cdot \vec{S}_b \right.
+
\tilde{J}_{2,1}^{(2)} |\vec{q}|^2 \vec{p}\cdot \vec{S}_a \vec{p}\cdot \vec{S}_b \vec{p}\times \vec{q}\cdot \vec{S}_a \\ & \phantom{=} \qquad
+
\tilde{J}_{2,1}^{(3)} |\vec{q}|^2  \vec{S}_a\cdot \vec{S}_b \vec{p}\times \vec{q}\cdot \vec{S}_a
+
\tilde{J}_{2,1}^{(4)} |\vec{q}|^2 |\vec{S}_a|^2  \vec{p}\times \vec{q}\cdot \vec{S}_b \\ & \phantom{=asd} \qquad
\left.
+
\tilde{J}_{2,1}^{(5)} |\vec{q}|^2  \vec{q}\cdot \vec{S}_a \vec{p}\cdot \vec{S}_a\times \vec{S}_b
+
\tilde{J}_{2,1}^{(6)} (\vec{q}\cdot \vec{S}_a)^2 \vec{p}\times \vec{q}\cdot \vec{S}_b \right]\,.
\end{split}
\end{equation}
The corresponding coefficients $\tilde{J}_{2,1}^{(\alpha)}$ are:
\begin{align}
\tilde{J}_{2,1}^{(1)} &=  \frac{E \xi}{2 \Pv{2}} (\IterCoeff{2}{1}{0}{0} + \IterCoeff{2}{0}{0}{1} )\,, \quad
\tilde{J}_{2,1}^{(2)} = \frac{E \xi}{2 \Pv{2}} \IterCoeff{1}{0}{1}{1}\,, 
\nonumber
\\
\tilde{J}_{2,1}^{(3)} &= -\frac{E \xi}{2 } \IterCoeff{1}{1}{1}{0}\,, \quad
\tilde{J}_{2,1}^{(4)} = -\frac{E \xi}{2} (\IterCoeff{2}{1}{0}{0} + \IterCoeff{2}{0}{0}{1} )\,,
\\
\tilde{J}_{2,1}^{(5)} &= -\frac{E \xi}{2 } (2\IterCoeff{2}{1}{0}{0} + \IterCoeff{1}{1}{1}{0} - 2\IterCoeff{2}{0}{0}{1})\,, \quad
\tilde{J}_{2,1}^{(6)} = 2E \xi \IterCoeff{2}{1}{0}{0}\,.
\nonumber
\end{align}
\end{subequations}

\item $S_a^3$
\begin{subequations}
\begin{equation}
\tilde{J}_{3,0} = i(\vec{p}\times \vec{q} \cdot \vec{S}_a)\left[
\tilde{J}_{3, 0}^{(1)} |\vec{q}|^2 (\vec{p} \cdot \vec{S}_a)^2+
\tilde{J}_{3, 0}^{(2)} |\vec{q}|^2 |\vec{S}_a|^2 +
\tilde{J}_{3, 0}^{(3)} (\vec{q} \cdot \vec{S}_a)^2
\right]\,.
\end{equation}
The coorsponding coefficients $\tilde{J}_{3,0}^{(\alpha)}$ are:
\begin{align}
\tilde{J}_{3, 0}^{(1)} &= \frac{E \xi}{2\Pv{2}} \IterCoeff{2}{0}{1}{0} + \frac{E \xi }{2 \Pv{2}} \IterCoeff{0}{0}{3}{0},\, \nonumber\\
\tilde{J}_{3, 0}^{(2)} &= -\frac{E \xi}{2} \IterCoeff{2}{0}{1}{0} - \frac{E \xi}{2} \IterCoeff{3}{0}{0}{0}\,, \\
\tilde{J}_{3, 0}^{(3)} &= 2 E \xi \IterCoeff{3}{0}{0}{0}\,.\nonumber
\end{align}

\end{subequations}

\item Quartic Spin Order: $S_a^2 S_b^2$, $S_a^3 S_b$, $S_a^4$ \newline
These coefficients also cancel with their corresponding 4D box coefficients. The expressions are too lengthy to be shown here and they can be found in the ancillary Mathematica file ``3D Integral Coefficients.wl".

\end{itemize}

\subsection{Conservative spin Hamiltonian}\label{subsec: Hamiltonian}
The full spinning Hamiltonian is:
\begin{equation}
H = \sqrt{|\vec{p}|^2 + m_{a\vphantom{b}}^2} + \sqrt{|\vec{p}|^2 + m_b^2} + H_{1\text{PM}} + H_{2\text{PM}} + \mathcal{O}(G^3)\,,
\end{equation}
where $H_{1\text{PM}}$ is the Fourier transform of \eqref{eq:1PM_Ham}. The components of the 2 PM Hamiltonian can be constructed from the components presented in sections \ref{subsec:Int_Coeff} and \ref{subsec:Iteration}. Only the 4D triangle integrals and the 3D bubble integrals contribute to eq.\eqref{eq:2PM_formula} and the momentum space Hamiltonian is
\begin{equation}\label{eq:Hq_Simp}
\tilde{H}_{\text{2PM}}(\vec{q}) = -\frac{M_{\bigtriangleup + \bigtriangledown}}{4E_a E_b}U_{a} U_{b} + \text{Iter}_{\cbubble}\,,
\end{equation}
where $M_{\bigtriangleup + \bigtriangledown}$ was given in section \ref{subsubsec:Triangle_Coefficient_Results} and $\text{Iter}_{\cbubble}$ was given in section \ref{subsubsec:3D_bubble_coeff_result}. 

We report the full momentum dependence in the ancillary file ``2PM Hamiltonian.wl" and only present the Hamiltonian to the first few orders in the $|\vec{p}|$ expansion. The cancellation between $1/|\vec{p}|^n$ poles in the iteration term and the triangle part of the 1-loop amplitude provides another consistency check of our results.

\begin{itemize}[leftmargin=*]
\item Scalar coupling
\begin{equation}
\small
\begin{split}
\tilde{H}_{0,0} 
&= \frac{2\pi^2 G^2}{|\vec{q}|}\tilde{H}_{0,0}^{(1)}\\
&= \frac{\pi ^2 G^2m_a m_b \left( m_a^2+3 m_a m_b+m_b^2\right)}{m_a+m_b} \\
&\qquad + \frac{\pi^2 G^2  \left(10 m_a^4+67 m_a^3 m_b+117 m_a^2 m_b^2+67 m_a m_b^3+10 m_b^4\right)}{2 m_a m_b \left(m_a+m_b\right)} |\vec{p}|^2 + \mathcal{O}(|\vec{p}|^4)\,.
\end{split}
\end{equation}
This matches with the results in \cite{Bern:2020buy}.
\item $S_a$
{
\begin{equation}
\begin{split}
\tilde{H}_{1,0} &= 
\frac{2i \pi^2 G^2 }{|\vec{q}|} \tilde{H}^{(1)}_{1,0} \vec{p} \times \vec{q} \cdot \vec{S}_a
\\
&
=
\frac{i \pi^2 G^2 }{|\vec{q}|} 
\left[
\frac{12 m_a^3 + 45m_a^2 m_b +41m_a m_b^2 + 10 m_b^3}{2m_a(m_a + m_b)} \right.\\
&\qquad\qquad
+
\left.
\frac{3(64 m_a^4 + 182m_a^3 m_b + 137m_a^2 m_b^2 + 13 m_a m_b^3 - 9m_b^4)}{8m_a^3(m_a + m_b)} |\vec{p}|^2
+
\mathcal{O}(|\vec{p}|^4)
\right] \vec{p} \times \vec{q} \cdot \vec{S}_a\,.
\end{split}
\end{equation}
}
This matches with the results in \cite{Bern:2020buy}.
\item $S_a S_b$
\begin{equation}
\tilde{H}_{1,1} = 
\frac{2 \pi^2 G^2 }{|\vec{q}|} \left[
\tilde{H}_{1,1}^{(1)} (\vec{q} \cdot \vec{S}_a)(\vec{q} \cdot \vec{S}_b)
+
\tilde{H}_{1,1}^{(2)} |\vec{q}|^2 (\vec{S}_a \cdot \vec{S}_b)
+
\tilde{H}_{1,1}^{(3)} |\vec{q}|^2 (\vec{p} \cdot \vec{S}_a) (\vec{p} \cdot \vec{S}_b) 
\right]\,.
\end{equation}
The corresponding coefficients $\tilde{H}_{1,1}^{(\alpha)}$ are:
{
\small
\begin{align}
\tilde{H}_{1,1}^{(1)} &= 
\frac{21m_a^2+53 m_a m_b+21 m_b^2}{16 \left(m_a+m_b\right)}
+ \frac{\left(63 m_a^4+708 m_a^3 m_b+1366 m_a^2 m_b^2+708 m_a m_b^3+63 m_b^4\right)}{64 m_a^2 m_b^2 \left(m_a+m_b\right)} |\vec{p}|^2 + \mathcal{O}(|\vec{p}|^4) \nonumber\,,\\
\tilde{H}_{1,1}^{(2)} &=
-\frac{19m_a^2+41 m_a m_b+19 m_b^2}{16 \left(m_a+m_b\right)}
-\frac{\left(21 m_a^4+465 m_a^3 m_b+920 m_a^2 m_b^2+465 m_a m_b^3+21 m_b^4\right)}{64 m_a^2 m_b^2 \left(m_a+m_b\right)} |\vec{p}|^2+ \mathcal{O}(|\vec{p}|^4)\,,\\
\tilde{H}_{1,1}^{(3)} &=
\frac{105 m_a^2+214 m_a m_b+105 m_b^2}{32 m_a^2 m_b+32 m_a m_b^2} + \mathcal{O}(|\vec{p}|^2) \,.\nonumber
\end{align}
}
This matches with the results in \cite{Bern:2020buy}.
\item $S_a^2$
\begin{equation}
\tilde{H}_{2,0} = 
\frac{2 \pi^2 G^2 }{|\vec{q}|} \left[
\tilde{H}_{2,0}^{(1)} (\vec{q} \cdot \vec{S}_a)^2
+
\tilde{H}_{2,0}^{(2)} |\vec{q}|^2 |\vec{S}_a|^2
+
\tilde{H}_{2,0}^{(3)} |\vec{q}|^2 (\vec{p} \cdot \vec{S}_a)^2 
\right]\,.
\end{equation}
The corresponding coefficients $\tilde{H}_{2,0}^{(\alpha)}$ are:
{
\small
\begin{align}
\tilde{H}_{2,0}^{(1)} &= \frac{m_b \left(18 m_a^2+58 m_am_b+27 m_b^2\right)}{32 m_a \left(m_a+m_b\right)}
+
\frac{\left(92 m_a^4 + 542 m_a^3 m_b+701 m_a^2 m_b^2+209 m_a m_b^3-26 m_b^4\right)}{64 m_a^3 m_b \left(m_a+m_b\right)}|\vec{p}|^2 + \mathcal{O}(|\vec{p}|^4)\,, \nonumber \\
\tilde{H}_{2,0}^{(2)} &= -\frac{m_b \left(22 m_a^2+46 m_am_b+19 m_b^2\right)}{32 m_a \left(m_a+m_b\right)}
-
\frac{\left(32 m_a^4 + 193 m_a^3 m_b+235 m_a^2 m_b^2+59 m_a m_b^3-14 m_b^4\right)}{32 m_a^3 m_b \left(m_a+m_b\right)} |\vec{p}|^2+ \mathcal{O}(|\vec{p}|^4)\,,\\
\tilde{H}_{2,0}^{(3)} &=\frac{14 m_a^4 + 60 m_a^3 m_b+51 m_a^2 m_b^2-m_a m_b^3-8 m_b^4}{32 m_a^3 m_b \left(m_a+m_b\right)} + \mathcal{O}(|\vec{p}|^2)\,.
\nonumber
\end{align}
}
This matches with the results in \cite{Kosmopoulos:2021zoq}.
\item $S_a^2S_b$
\begin{equation}
\begin{split}
\tilde{H}_{2,1} = 
\frac{2i\pi^2G^2}{|\vec{q}|} 
&
\left[ 
\tilde{H}_{2,1}^{(1)} |\vec{q}|^2 (\vec{p} \cdot \vec{S}_a)^2 (\vec{p} \times \vec{q}) \cdot \vec{S}_b 
+
\tilde{H}_{2,1}^{(2)} |\vec{q}|^2 (\vec{p} \cdot \vec{S}_a)(\vec{p} \cdot \vec{S}_b) (\vec{p} \times \vec{q}) \cdot \vec{S}_a \right.\\
&\quad +
\tilde{H}_{2,1}^{(3)} |\vec{q}|^2 \vec{S}_a \cdot \vec{S}_b (\vec{p} \times \vec{q}) \cdot \vec{S}_a + 
\tilde{H}_{2,1}^{(4)}  \vec{q}\cdot\vec{S}_a \vec{q}\cdot \vec{S}_b (\vec{p} \times \vec{q}) \cdot \vec{S}_a \\
&\quad 
\left. 
+
\tilde{H}_{2,1}^{(5)} |\vec{q}|^2 |\vec{S}_a|^2 (\vec{p} \times \vec{q}) \cdot \vec{S}_b +
\tilde{H}_{2,1}^{(6)} (\vec{q}\cdot \vec{S}_a)^2 (\vec{p} \times \vec{q}) \cdot \vec{S}_b \right]\,.
\end{split}
\end{equation}
Putting eq.\eqref{eq:Tri_Coeff_2_1_res} and eq.\eqref{eq:Iter_2_1_res} into eq.\eqref{eq:Hq_Simp}, one would get:
\begin{equation}\label{eq:Pole_UnRemoved_21}
\small
\begin{split}
\frac{|\vec{q}|\tilde{H}_{2,1}}{2\pi^2i G^2} = 
&\left[
\frac{m_b(6m_a + m_b)}{8(m_a + m_b)\Pv{4}} +
\frac{30m_a^3 + 78m_a^2 m_b + 29 m_a m_b^2 -3m_b^3}{32 m_a^2 m_b(m_a +m_b)|\vec{p}|^2} \right] \\
&\hphantom{123123123123123123123123123123}
\times|\vec{q}|^2(\vec{p} \cdot \vec{S}_a) \left[ (\vec{p}\cdot \vec{S}_b) \vec{p}\times \vec{q}\cdot \vec{S}_a - (\vec{p}\cdot \vec{S}_a) \vec{p}\times \vec{q}\cdot \vec{S}_b \right]\\
&
-\frac{m_b(6m_a + m_b)}{8(m_a + m_b)\Pv{2}}
\left[ \vec{p}\times \vec{q} \cdot \vec{S}_a \left(|\vec{q}|^2 \vec{S}_a \cdot \vec{S}_b - (\vec{q} \cdot \vec{S}_a)(\vec{q} \cdot \vec{S}_b)\right)
-
\vec{p}\times \vec{q} \cdot \vec{S}_b \left(|\vec{q}|^2 |\vec{S}_a|^2 - (\vec{q} \cdot \vec{S}_a)^2\right)
 \right]\\
&
+ 
\mathcal{O}(|\vec{p}|^0)\,.
\end{split}
\end{equation}
The superficial $1/|\vec{p}|^n$ poles are removed by applying eq.\eqref{eq: Sa2_Sb_Identity} so that they pushed to higher order in $\hbar$, leaving only:\footnote{In the ancillary file ``2PM Hamiltonian.wl", we introduce a parameter $\delta$ to turn on and off the Schouten identity. When $\delta = 0$, it gives the $\tilde{H}_{2,1}^{\alpha}$ in eq.\eqref{eq:Pole_UnRemoved_21}. When $\delta = 1$, it gives the $\tilde{H}_{2,1}^{(\alpha)}$ coefficients eq.\eqref{eq:Pole_Removed_21}.}
\begin{equation}\label{eq:Pole_Removed_21}
\begin{split}
\tilde{H}_{2,1}^{(1)} &= - \frac{-24m_a^5 + 42m_a^4 m_b + 171 m_a^3 m_b^2 + 10 m_a^2 m_b^3 - 36m_a m_b^4 + 6m_b^5}{128 m_a^4 m_b^3(m_a + m_b)} + \mathcal{O}(|\vec{p}|^2)\,,\\
\tilde{H}_{2,1}^{(2)} &= \frac{-24 m_a^5 + 82 m_a^4 m_b+243 m_a^3 m_b^2+34 m_a^2 m_b^3-44 m_a m_b^4+6 m_b^5}{128 m_a^4 m_b^3 \left(m_a+m_b\right)} + \mathcal{O}(|\vec{p}|^2)\,,\\
\tilde{H}_{2,1}^{(3)} &= - \frac{3(4m_a^2 - 6m_a m_b - 3m_b^2)}{64m_a^2 m_b}+\mathcal{O}(|\vec{p}|^2)\,,\\
\tilde{H}_{2,1}^{(4)} &= \frac{14m_a^3 + 52m_a^2 m_b + 45 m_a m_b^2 + 14 m_b^3}{32m_a^2m_b(m_a + m_b)} +\mathcal{O}(|\vec{p}|^2)\,,\\
\tilde{H}_{2,1}^{(5)} &= -\frac{14 m_a^3+98 m_a^2 m_b+100 m_a m_b^2+17 m_b^3}{64 m_a^2 m_b \left(m_a+m_b\right)} + \mathcal{O}(|\vec{p}|^2)\,,\\
\tilde{H}_{2,1}^{(6)} &= \frac{54 m_a^3 + 220 m_a^2 m_b + 199 m_a m_b^2 + 36 m_b^3}{64m_a^2m_b(m_a + m_b)}+\mathcal{O}(|\vec{p}|^2)\,.
\end{split}
\end{equation}
Note that these numbers are not unique due to the Schouten identity eq.\eqref{eq: Sa2_Sb_Identity}, but physical observables are not affected by the identity.

\item $S_a^3$
\begin{subequations}
\begin{equation}
\tilde{H}_{3, 0} = \frac{2i\pi^2 G^2}{|\vec{q}|} \vec{p} \times \vec{q}\cdot \vec{S}_a 
\left[
\tilde{H}_{3,0}^{(1)}|\vec{q}|^2(\vec{p} \cdot \vec{S}_a)^2 + 
\tilde{H}_{3,0}^{(2)}|\vec{q}|^2|\vec{S}_a|^2+
\tilde{H}_{3,0}^{(3)}(\vec{q} \cdot \vec{S}_a)^2
\right]\,.
\end{equation}
The corresponding coefficients $H_{3,0}^{\alpha}$ are:
{
\footnotesize
\begin{equation}
\begin{split}
\tilde{H}_{3,0}^{(1)} &= 
-\frac{3 m_a^2+4 m_a m_b+m_b^2}{16 \left(m_a^4 m_b\right)}
+
\frac{\left(m_a+m_b\right) \left(8 m_a^3-6 m_a^2 m_b+8 m_a m_b^2+5 m_b^3\right)}{64 m_a^6 m_b^3} |\vec{p}|^2+O\left(|\vec{p}|^4\right)\,,\\
\tilde{H}_{3,0}^{(2)} &= 
\frac{-130 m_a^2 m_b-70
   m_a m_b^2-40 m_a^3+m_b^3}{192 m_a^3 \left(m_a+m_b\right)}
+
\frac{ \left(-446 m_a^3 m_b+24 m_a^2 m_b^2+160 m_a m_b^3-270 m_a^4+3 m_b^4\right)}{384 m_a^5 m_b \left(m_a+m_b\right)}|\vec{p}|^2 
+O\left(|\vec{p}|^4\right)\,,\\
\tilde{H}_{3,0}^{(3)} &=
\frac{104 m_a^3+398 m_a^2
   m_b+266 m_a m_b^2+31 m_b^3}{192 m_a^3 \left(m_a+m_b\right)} + 
\frac{\left(618 m_a^4 + 1114 m_a^3 m_b-60 m_a^2 m_b^2-500 m_a m_b^3-81 m_b^4\right)}{384 m_a^5 m_b \left(m_a+m_b\right)}|\vec{p}|^2
+O\left(|\vec{p}|^4\right)\,.
\end{split}
\end{equation}
}
\end{subequations}

\item $S_a^2 S_b^2$
\begin{equation}
\begin{split}
\tilde{H}_{2,2} =
\frac{2\pi^2 G^2}{|\vec{q}|}
&\left[
\tilde{H}_{2,2}^{(1)} |\vec{q}|^4  ( \vec{p}\cdot \vec{S}_a)^2 (\vec{p}\cdot \vec{S}_b )^2
+
\tilde{H}_{2,2}^{(2)} \vec{q}^4 \vec{S}_b^2  ( \vec{p}\cdot \vec{S}_a)^2
+
\tilde{H}_{2,2}^{(3)} |\vec{q}|^2 (\vec{p}\cdot \vec{S}_a )^2 (\vec{q}\cdot \vec{S}_b )^2
\right.
\\
&
+
\tilde{H}_{2,2}^{(4)} |\vec{q}|^4 \vec{S}_a\cdot \vec{S}_b \vec{p}\cdot \vec{S}_a \vec{p}\cdot \vec{S}_b
+
\tilde{H}_{2,2}^{(5)} |\vec{q}|^2  \vec{p}\cdot \vec{S}_a \vec{q}\cdot \vec{S}_a \vec{p}\cdot \vec{S}_b \vec{q}\cdot \vec{S}_b
+
\tilde{H}_{2,2}^{(6)} |\vec{q}|^4 |\vec{S}_a|^2 (\vec{p}\cdot \vec{S}_b )^2\\
&
+
\tilde{H}_{2,2}^{(7)} |\vec{q}|^2  (\vec{q}\cdot \vec{S}_a )^2 (\vec{p}\cdot \vec{S}_b)^2
+
\tilde{H}_{2,2}^{(8)} |\vec{q}|^4 |\vec{S}_a|^2 |\vec{S}_b|^2 
+
\tilde{H}_{2,2}^{(9)} |\vec{q}|^4 (\vec{S}_a\cdot \vec{S}_b)^2\\
&
+
\tilde{H}_{2,2}^{(10)} |\vec{q}|^2 |\vec{S}_b|^2  (\vec{q}\cdot \vec{S}_a)^2
+
\tilde{H}_{2,2}^{(11)} |\vec{q}|^2  \vec{S}_a\cdot \vec{S}_b \vec{q}\cdot \vec{S}_a \vec{q}\cdot \vec{S}_b
+
\tilde{H}_{2,2}^{(12)} \vec{q}^2 \vec{S}_a^2  (\vec{q}\cdot \vec{S}_b)^2\\
&
\left.
+
\tilde{H}_{2,2}^{(13)} (\vec{q}\cdot \vec{S}_a)^2 (\vec{q}\cdot \vec{S}_b)^2
\right]\,.
\end{split}
\end{equation}
Note that this basis is also not linearly independent since different contractions can be related by the Schouten identity:
\begin{equation}\label{eq: Sa2_Sb2_Identity}
\begin{split}
\left| (\vec{p} \cdot \vec{S}_a)(\vec{q} \times \vec{S}_b) - (\vec{p} \cdot \vec{S}_b)(\vec{q} \times \vec{S}_a) \right|^2 &=
\left|-\vec{p} (\vec{q}\cdot \vec{S}_a \times \vec{S}_b) + (\vec{p} \cdot \vec{q}) (\vec{S}_a \times \vec{S}_b)
\right|^2 \\
&= 
|\vec{p}|^2 (\vec{q}\cdot \vec{S}_a \times \vec{S}_b)^2 + \mathcal{O}(\hbar)\,, 
\end{split}
\end{equation}
where we use the on-shell condition $\vec{p} \cdot \vec{q} = \frac{|\vec{q}|^2}{2}$ on the second line. Application of eq.\eqref{eq:Hq_Simp} results in the expression
\begin{equation}\label{eq:Pole_UnRemoved_22}
\begin{split}
\tilde{H}_{2,2} =
\frac{2\pi^2 G^2}{|\vec{q}|}
&\left[
\frac{m_a m_b|\vec{q}|^2}{8(m_a + m_b)|\vec{p}|^4}\left| (\vec{p} \cdot \vec{S}_a)(\vec{q} \times \vec{S}) - (\vec{p} \cdot \vec{S}_b)(\vec{q} \times \vec{S}_b) \right|^2
\right.
\\
&
\quad
- \frac{(m_a^2 - 4m_am_b + m_b^2)|\vec{q}|^2}{32m_a m_b(m_a + m_b)|\vec{p}|^2}\left| (\vec{p} \cdot \vec{S}_a)(\vec{q} \times \vec{S}) - (\vec{p} \cdot \vec{S}_b)(\vec{q} \times \vec{S}_b) \right|^2 \\
&
\quad
\left.
-
\frac{m_am_b|\vec{q}|^2}{8(m_a + m_b)|\vec{p}|^2}(\vec{q} \cdot \vec{S}_a \times \vec{S}_b)^2\right] + \mathcal{O}(|\vec{p}|^0)\,, 
\end{split}
\end{equation}
which contains superficial $1/|\vec{p}|^n$ poles which are removed by eq.\eqref{eq: Sa2_Sb2_Identity}.\footnote{In the ancillary file ``2PM Hamiltonian.wl", we introduce a parameter $\delta$ to turn on and off the Schouten identity. When $\delta = 0$, it gives the $\tilde{H}_{2,2}^{\alpha}$ in eq.\eqref{eq:Pole_UnRemoved_22}. When $\delta = 1$, it gives the $\tilde{H}_{2,2}^{(\alpha)}$ coefficients eq.\eqref{eq:Pole_Removed_22}.}
\begingroup
\allowdisplaybreaks
\begin{align}\label{eq:Pole_Removed_22}
\tilde{H}_{2,2}^{(1)} &= -\frac{\left(m_a+m_b\right) \left(2 m_a^2-5 m_a m_b+2 m_b^2\right)}{64 m_a^4 m_b^4} + \mathcal{O}(|\vec{p}|^2)\,, \nonumber\\
\tilde{H}_{2,2}^{(2)} &= \frac{34m_a^4-27 m_a^3 m_b+34 m_a^2 m_b^2+71 m_a m_b^3+46 m_b^4}{512 m_a^3 m_b^3 \left(m_a+m_b\right)} + \mathcal{O}(|\vec{p}|^2)\,,
\nonumber\\
\tilde{H}_{2,2}^{(3)} &= \frac{-88 m_a^4 + 134 m_a^3 m_b+157 m_a^2 m_b^2-72 m_a m_b^3-132 m_b^4}{1024 m_a^3 m_b^3\left(m_a+m_b\right)} + \mathcal{O}(|\vec{p}|^2)\,,
\nonumber\\
\tilde{H}_{2,2}^{(4)} &= -\frac{26 m_a^4+47 m_a^3 m_b+109 m_a^2 m_b^2+47 m_a m_b^3+26 m_b^4}{256 m_a^3 m_b^3 \left(m_a+m_b\right)} + \mathcal{O}(|\vec{p}|^2)\,,
\nonumber\\
\tilde{H}_{2,2}^{(5)} &= \frac{26 m_a^4+47 m_a^3 m_b+109 m_a^2 m_b^2+47 m_a m_b^3+26 m_b^4}{256 m_a^3 m_b^3 \left(m_a+m_b\right)} + \mathcal{O}(|\vec{p}|^2)\,,
\nonumber\\
\tilde{H}_{2,2}^{(6)} &= \frac{46m_a^4+71 m_a^3 m_b+34 m_a^2 m_b^2-27 m_a m_b^3+34 m_b^4}{512 m_a^3 m_b^3 \left(m_a+m_b\right)} + \mathcal{O}(|\vec{p}|^2)\,,
\nonumber\\
\tilde{H}_{2,2}^{(7)} &= \frac{-132 m_a^4-72 m_a^3 m_b+157 m_a^2 m_b^2+134 m_a m_b^3-88 m_b^4}{1024 m_a^3 m_b^3\left(m_a+m_b\right)} + \mathcal{O}(|\vec{p}|^2)\,,\\
\tilde{H}_{2,2}^{(8)} &= \frac{3 \left(25 m_a^2+76 m_a m_b+25 m_b^2\right)}{512 m_a m_b \left(m_a+m_b\right)} + \mathcal{O}(|\vec{p}|^2)\,,
\nonumber\\
\tilde{H}_{2,2}^{(9)} &= -\frac{17 m_a^2+76 m_a m_b+17m_b^2}{256 m_a m_b \left(m_a+m_b\right)} + \mathcal{O}(|\vec{p}|^2)\,,
\nonumber\\
\tilde{H}_{2,2}^{(10)} &= -\frac{143 m_a^2+348 m_a m_b+131 m_b^2}{512 m_a m_b \left(m_a+m_b\right)} + \mathcal{O}(|\vec{p}|^2)\,,
\nonumber\\
\tilde{H}_{2,2}^{(11)} &= \frac{9 m_a^2+28 m_a m_b+9m_b^2}{128 m_a m_b \left(m_a+m_b\right)} + \mathcal{O}(|\vec{p}|^2)\,,
\nonumber\\
\tilde{H}_{2,2}^{(12)} &= -\frac{131 m_a^2+348 m_a m_b+143 m_b^2}{512 m_a m_b \left(m_a+m_b\right)} + \mathcal{O}(|\vec{p}|^2)\,,
\nonumber\\
\tilde{H}_{2,2}^{(13)} &= \frac{213m_a^2+572 m_a m_b+213 m_b^2}{512 m_a m_b \left(m_a+m_b\right)} + \mathcal{O}(|\vec{p}|^2)\,.
\nonumber
\end{align}
\endgroup
\item $S_a^3 S_b$

\begin{subequations}
\begin{equation}
\begin{split}
\tilde{H}_{3,1} 
=
\frac{2\pi^2 G^2}{|\vec{q}|}
& 
\left[
\tilde{H}_{3,1}^{(1)} |\vec{q}|^4  ( \vec{p}\cdot \vec{S}_a)^3 \vec{p}\cdot \vec{S}_b+
\tilde{H}_{3,1}^{(2)} |\vec{q}|^4  \vec{S}_a\cdot \vec{S}_b (\vec{p}\cdot \vec{S}_a )^2
+
\tilde{H}_{3,1}^{(3)} |\vec{q}|^2 ( \vec{p}\cdot \vec{S}_a)^2 \vec{q}\cdot \vec{S}_a \vec{q}\cdot \vec{S}_b \right.
\\
&
+
\tilde{H}_{3,1}^{(4)} |\vec{q}|^4 \vec{S}_a^2 \vec{p}\cdot \vec{S}_a \vec{p}\cdot \vec{S}_b
+
\tilde{H}_{3,1}^{(5)}  |\vec{q}|^2 \vec{p}\cdot \vec{S}_a ( \vec{q}\cdot \vec{S}_a )^2 \vec{p}\cdot \vec{S}_b
+
\tilde{H}_{3,1}^{(6)} |\vec{q}|^4 |\vec{S}_a|^2  \vec{S}_a\cdot \vec{S}_b\\
&
\left.
+
\tilde{H}_{3,1}^{(7)} |\vec{q}|^2 \vec{S}_a\cdot \vec{S}_b (\vec{q}\cdot \vec{S}_a)^2
+
\tilde{H}_{3,1}^{(8)} |\vec{q}|^2 |\vec{S}_a|^2  \vec{q}\cdot \vec{S}_a \vec{q}\cdot \vec{S}_b
+
\tilde{H}_{3,1}^{(9)} (\vec{q}\cdot \vec{S}_a)^3 \vec{q}\cdot \vec{S}_b
\right]\,.
\end{split}
\end{equation}
The corresponding coefficients $H_{3,1}^{(\alpha)}$ are:
{
\footnotesize
\begin{align}
\tilde{H}_{3,1}^{(1)} &= \frac{\left(m_a+m_b\right) \left(
7 m_a^2-3 m_a m_b-m_b^2\right)}{96 m_a^5 m_b^3} -\frac{5 m_a^3 m_b^2+12 m_a^2 m_b^3-9 m_a m_b^4+12 m_a^5-4 m_b^5}{192 m_a^7 m_b^5}|\vec{p}|^2 + \mathcal{O}(|\vec{p}|^4) \nonumber\,,\\
\tilde{H}_{3,1}^{(2)} &=\frac{32 m_a^3+48 m_a^2 m_b+18 m_a m_b^2+m_b^3}{96 m_a^4 m_b \left(m_a+m_b\right)} +\mathcal{O}(|\vec{p}|^2) \nonumber\,,\\
\tilde{H}_{3,1}^{(3)} &=-\frac{-8 m_a^4 + 367 m_a^3 m_b+644 m_a^2 m_b^2+303 m_a m_b^3+68 m_b^4}{1536 m_a^4 m_b^2 \left(m_a+m_b\right)}+ \mathcal{O}(|\vec{p}|^2) \nonumber\,,\\
\tilde{H}_{3,1}^{(4)}&= \frac{4 m_a^4-327 m_a^3 m_b-460 m_a^2 m_b^2-99 m_a m_b^3+24 m_b^4}{768 m_a^4 m_b^2 \left(m_a+m_b\right)}+ \mathcal{O}(|\vec{p}|^2) \nonumber\,,\\
\tilde{H}_{3,1}^{(5)}&= \frac{4 m_a^4 + 445 m_a^3 m_b+584 m_a^2 m_b^2+75 m_a m_b^3-40 m_b^4}{768 m_a^4 m_b^2 \left(m_a+m_b\right)}+ \mathcal{O}(|\vec{p}|^2) \nonumber \,,\\
\tilde{H}_{3,1}^{(6)}&=\frac{11 m_a^2+15 m_a m_b+3 m_b^2}{128 m_a^2 \left(m_a+m_b\right)}+
\frac{94 m_a^4 + 1229 m_a^3 m_b+1306 m_a^2 m_b^2-145 m_a m_b^3-220 m_b^4}{3072 m_a^4 m_b^2 \left(m_a+m_b\right)}|\vec{p}|^2 + \mathcal{O}(|\vec{p}|^4)\nonumber\,, \\
\tilde{H}_{3,1}^{(7)}&=-\frac{27 m_a^2+67 m_a m_b+27 m_b^2}{128 m_a^2 \left(m_a+m_b\right)}+\frac{-158 m_a^4-2777 m_a^3 m_b-3626 m_a^2 m_b^2+101 m_a m_b^3+684 m_b^4}{3072 m_a^4 m_b^2 \left(m_a+m_b\right)}|\vec{p}|^2 +\mathcal{O}(|\vec{p}|^4) \nonumber\,,\\
\tilde{H}_{3,1}^{(8)}&=-\frac{ 11 m_a^2+33 m_a m_b+17 m_b^2}{128 m_a^2 \left(m_a+m_b\right)} + \frac{-51 m_a^4-852 m_a^3 m_b-1075 m_a^2 m_b^2+53 m_a m_b^3+292 m_b^4}{1536 m_a^4 m_b^2 \left(m_a+m_b\right)}|\vec{p}|^2+ \mathcal{O}(|\vec{p}|^4) \nonumber\,,\\
\tilde{H}_{3,1}^{(9)}&=\frac{89 m_a^2+291 m_a m_b+127 m_b^2}{384 m_a^2 \left(m_a+m_b\right)}
+
\frac{251 m_a^4 + 2356 m_a^3 m_b+3035 m_a^2 m_b^2+91 m_a m_b^3-532 m_b^4}{1536 m_a^4 m_b^2 \left(m_a+m_b\right)} |\vec{p}|^2 + \mathcal{O}(|\vec{p}|^4)\,.
\end{align}
}
\end{subequations}

\item $S_a^4$
\begin{subequations}
\begin{equation}
\begin{split}
\tilde{H}_{4,0} &= 
\tilde{H}_{4,0}^{(1)} |\vec{q}|^4 ( \vec{p}\cdot \vec{S}_a)^4
+
\tilde{H}_{4,0}^{(2)} |\vec{q}|^4 |\vec{S}_a|^2 ( \vec{p}\cdot \vec{S}_a )^2
+
\tilde{H}_{4,0}^{(3)} |\vec{q}|^2 ( \vec{p}\cdot \vec{S}_a)^2 (\vec{q}\cdot \vec{S}_a)^2\\
&
+
\tilde{H}_{4,0}^{(4)} |\vec{q}|^4 |\vec{S}_a|^4 
+
\tilde{H}_{4,0}^{(5)} |\vec{q}|^2 |\vec{S}_a|^2 ( \vec{q}\cdot \vec{S}_a)^2
+
\tilde{H}_{4,0}^{(6)} ( \vec{q}\cdot \vec{S}_a)^4\,.
\end{split}
\end{equation}
The corresponding coefficients $H_{4,0}^{(\alpha)}$ are:
{
\footnotesize
\begin{align}
\tilde{H}_{4,0}^{(1)} &= -\frac{\left(m_a+m_b\right) \left(14 m_a^2 + 3 m_a m_b-2 m_b^2\right)}{384 m_a^6 m_b^2} + \frac{\left(m_a+m_b\right) \left(55 m_a^4 + 5 m_a^3 m_b+76 m_a^2 m_b^2+33 m_a m_b^3-9 m_b^4\right)}{1536 m_a^8 m_b^4} + \mathcal{O}(|\vec{p}|^4) \nonumber\,,\\
\tilde{H}_{4,0}^{(2)} &= \frac{-24 m_a^4-48 m_a^3 m_b+70 m_a^2 m_b^2+86 m_a m_b^3+11 m_b^4}{1536 m_a^5 m_b \left(m_a+m_b\right)} + \mathcal{O}(|\vec{p}|^2) \nonumber\,,\\
\tilde{H}_{4,0}^{(3)} &= \frac{144 m_a^4 + 384 m_a^3 m_b-62 m_a^2 m_b^2-286 m_a m_b^3-55 m_b^4}{3072 m_a^5 m_b \left(m_a+m_b\right)} + \mathcal{O}(|\vec{p}|^2)\,,\\
\tilde{H}_{4,0}^{(4)}&=\frac{m_b \left(34 m_a^2+66 m_a m_b+11 m_b^2\right)}{1536 m_a^3 \left(m_a+m_b\right)} + \frac{63 m_a^4 + 289 m_a^3 m_b+146 m_a^2 m_b^2-86 m_a m_b^3-25 m_b^4}{1536 m_a^5 m_b \left(m_a+m_b\right)}|\vec{p}|^2 + \mathcal{O}(|\vec{p}|^4)\nonumber\,,\\
\tilde{H}_{4,0}^{(5)}&= -\frac{m_b \left(22 m_a^2+54 m_a m_b+23 m_b^2\right)}{256 m_a^3 \left(m_a+m_b\right)}
+
\frac{-792 m_a^4-4200 m_a^3 m_b-2426 m_a^2 m_b^2+1286 m_a m_b^3+779 m_b^4}{6144 m_a^5 m_b \left(m_a+m_b\right)}|\vec{p}|^2+ \mathcal{O}(|\vec{p}|^4)\,,
\nonumber\\
\tilde{H}_{4,0}^{(6)}&=\frac{m_b \left(82 m_a^2+306 m_a m_b+143 m_b^2\right)}{1536 m_a^3 \left(m_a+m_b\right)} +
\frac{764 m_a^4 + 4036 m_a^3 m_b+2618 m_a^2 m_b^2-1062 m_a m_b^3-739 m_b^4}{6144 m_a^5 m_b \left(m_a+m_b\right)}|\vec{p}|^2 + \mathcal{O}(|\vec{p}|^4)\,.
\nonumber
\end{align}
}
\end{subequations}
\end{itemize}
~\\[-1cm]
\paragraph{Position Space Hamiltonian}
$\hphantom{123123123}$\newline
\\[-0.3mm]
The position space Hamiltonian is obtained by a Fourier transform on the momentum space Hamiltonian $\tilde{H}_{n_a, n_b}$, and in the process the different tensor structures $\tilde{H}_{n_a,n_b}^{(\alpha)}$ become redistributed among the different tensor structures of the position space Hamiltonian.
\begin{itemize}[leftmargin = *]
\item Scalar
\begin{equation}
H_{0, 0} = \frac{G^2}{r^2} \tilde{H}_{0,0}\,.
\end{equation}
\item $S_a$
\begin{equation}
H_{1, 0} = \frac{2G^2}{r^4} \tilde{H}_{1,0}^{(1)} \vec{p}\times \vec{r} \cdot \vec{S}_a\,.
\end{equation}
\item $S_a S_b$
\begin{subequations}
\begin{equation}
H_{1,1} = \frac{G^2}{r^6}
\left[ 
H^{(1)}_{1,1} \vec{r}\cdot \vec{S}_a \vec{r}\cdot \vec{S}_b +
H^{(2)}_{1,1} r^2 \vec{S}_a \cdot \vec{S}_b +
H^{(3)}_{1,1} r^2 \vec{p} \cdot \vec{S}_a \vec{p} \cdot \vec{S}_b
\right]\,.
\end{equation}
The $H_{1,1}^{(\alpha)}$ coefficients are related to the $\tilde{H}_{1,1}^{(\alpha)}$ by the following relation:
\begin{equation}
\begin{split}
H_{1,1}^{(1)} &= -8 \tilde{H}_{1,1}^{(1)} \,,\\
H_{1,1}^{(2)} &= 2 \tilde{H}_{1,1}^{(1)} -  2 \tilde{H}_{1,1}^{(2)}\,,\\
H_{1,1}^{(3)} &= -2 \tilde{H}_{1,1}^{(3)} \,.
\end{split}
\end{equation}
\end{subequations}

\item $S_a^2$
\begin{subequations}
\begin{equation}
H_{2,0} = \frac{G^2}{r^6}
\left[ 
H^{(1)}_{2,0} (\vec{r}\cdot \vec{S}_a)^2  +
H^{(2)}_{2,0} r^2 |\vec{S}_a|^2 +
H^{(3)}_{2,0} r^2 (\vec{p} \cdot \vec{S}_a)^2
\right]\,.
\end{equation}
The $H_{2,0}^{(\alpha)}$ coefficients are related to the $\tilde{H}_{2,0}^{(\alpha)}$ by the following relation:
\begin{equation}
\begin{split}
H_{2,0}^{(1)} &= -8 \tilde{H}_{2,0}^{(1)}\,, \\
H_{2,0}^{(2)} &= 2 \tilde{H}_{2,0}^{(1)} -  2 \tilde{H}_{2,0}^{(2)}\,,\\
H_{2,0}^{(3)} &= -2 \tilde{H}_{2,0}^{(3)}\,.
\end{split}
\end{equation}
\end{subequations}
\item $S_a^2 S_b$
\begin{subequations}
\begin{equation}
\begin{split}
H_{2,1} = \frac{G^2}{r^8}
&\left[ 
H_{2,1}^{(1)} r^2 (\vec{p}\cdot \vec{S}_a)^2 \vec{p}\times \vec{r}\cdot \vec{S}_b+
H_{2,1}^{(2)} r^2  \vec{p}\cdot \vec{S}_a \vec{P}\cdot \vec{S}_b \vec{p}\times \vec{r}\cdot \vec{S}_a+
H_{2,1}^{(3)} r^2 \vec{r}\cdot \vec{S}_a \vec{S}_a\times \vec{S}_b\cdot \vec{p} \right.\\
&+
H_{2,1}^{(4)} r^2 |\vec{S}_a|^2 \vec{p}\times \vec{r}\cdot \vec{S}_b+
H_{2,1}^{(5)} r^2 \vec{S}_a\cdot \vec{S}_b \vec{p}\times \vec{r}\cdot \vec{S}_a+
H_{2,1}^{(6)} (\vec{r}\cdot \vec{S}_a)^2 \vec{p}\times \vec{r}\cdot \vec{S}_b\\ &+
\left.
H_{2,1}^{(7)} \vec{r}\cdot \vec{S}_a \vec{r}\cdot \vec{S}_b \vec{p}\times \vec{r}\cdot \vec{S}_a
\right]\,.
\end{split}
\end{equation}The $H_{2,1}^{(\alpha)}$ coefficients are related to the $\tilde{H}_{2,1}^{(\alpha)}$ by the following relation:
\begin{equation}
\begin{alignedat}{3}
& H_{2,1}^{(1)} = 8 \tilde{H}_{2,1}^{(1)}\,,\qquad
&& H_{2,1}^{(2)} = 8 \tilde{H}_{2,1}^{(2)}\,, \qquad
&& H_{2,1}^{(3)} = 8 \tilde{H}_{2,1}^{(4)} - 16 \tilde{H}_{2,1}^{(6)}\,,\\
& H_{2,1}^{(4)} = 8 \tilde{H}_{2,1}^{(5)} - 8 \tilde{H}_{2,1}^{(6)}\,,\qquad
&& H_{2,1}^{(5)} = 8 \tilde{H}_{2,1}^{(3)} - 8\tilde{H}_{2,1}^{(4)}\,, \qquad
&& H_{2,1}^{(6)} = 48 \tilde{H}_{2,1}^{(6)}\,,\\
& H_{2,1}^{(7)} = 48 \tilde{H}_{2,1}^{(4)}\,.
\end{alignedat}
\end{equation}
\end{subequations}

\item $S_a^3$
\begin{subequations}
\begin{equation}
H_{3,0} = \frac{G^2}{r^8} \vec{p}\times\vec{r}\cdot\vec{S}_a
\left[ 
H^{(1)}_{3,0} r^2 (\vec{p} \cdot \vec{S}_a)^2 +
H^{(2)}_{3,0} r^2 |\vec{S}_a|^2 +
H^{(3)}_{3,0} (\vec{r}\cdot \vec{S}_a)^2 
\right]\,.
\end{equation}
The $H_{3,0}^{(\alpha)}$ coefficients are related to the $\tilde{H}_{3,0}^{(\alpha)}$ by the following relation:
\begin{equation}
\begin{split}
H_{3,0}^{(1)} &= 8 \tilde{H}_{3,0}^{(1)}\,, \\
H_{3,0}^{(2)} &= 8 \tilde{H}_{3,0}^{(2)} - 8 \tilde{H}_{2,0}^{(3)}\,,\\
H_{3,0}^{(3)} &= 48 \tilde{H}_{3,0}^{(3)}\,.
\end{split}
\end{equation}
\end{subequations}

\item $S_a^2 S_b^2$
\begin{subequations}
\begin{equation}
\begin{split}
H_{2,2} = 
\frac{G^2}{|\vec{q}|}
&
\left[
H_{2,2}^{(1)} r^4 (\vec{p}\cdot \vec{S}_a)^2 (\vec{p}\cdot \vec{S}_b)^2
+
H_{2,2}^{(2)} r^4 \vec{S}_b^2(\vec{p}\cdot \vec{S}_a)^2
+
H_{2,2}^{(3)} r^2 (\vec{p}\cdot \vec{S}_a)^2 (r\cdot \vec{S}_b)^2\right.\\
&
+ 
H_{2,2}^{(4)} r^4 \vec{S}_a\cdot \vec{S}_b \vec{p}\cdot \vec{S}_a \vec{p}\cdot \vec{S}_b
+
H_{2,2}^{(5)} r^2 \vec{p}\cdot \vec{S}_a \vec{r}\cdot \vec{S}_a \vec{p}\cdot \vec{S}_b \vec{r}\cdot \vec{S}_b
+
H_{2,2}^{(6)} r^4 |\vec{S}_a|^2 (\vec{p}\cdot \vec{S}_b)^2\\
&
+
H_{2,2}^{(7)} r^2 (\vec{r}\cdot \vec{S}_a)^2 (\vec{p}\cdot \vec{S}_b)^2
+
H_{2,2}^{(8)} r^4 |\vec{S}_a|^2 |\vec{S}_b|^2 
+
H_{2,2}^{(9)} r^4 (\vec{S}_a\cdot\vec{S}_b)^2\\
&
+
H_{2,2}^{(10)} r^2 |\vec{S}_b|^2 (\vec{r}\cdot \vec{S}_a)^2
+
H_{2,2}^{(11)} r^2 \vec{S}_a\cdot \vec{S}_b \vec{r}\cdot \vec{S}_a r\cdot \vec{S}_b
+
H_{2,2}^{(12)} r^2 |\vec{S}_a|^2 (\vec{r}\cdot\vec{S}_b)^2\\
&
\left.
+
H_{2,2}^{(13)} (\vec{r}\cdot \vec{S}_a)^2 (\vec{r}\cdot \vec{S}_b)^2 \right].
\end{split}
\end{equation}
The $H_{2,2}^{(\alpha)}$ coefficients are related to the $\tilde{H}_{2,2}^{(\alpha)}$ by the following relation:
{\footnotesize
\begin{equation}
\begin{alignedat}{3}
&H_{2,2}^{(1)} = 24 \tilde{H}_{2,2}^{(1)}\,, \qquad
&& H_{2,2}^{(2)} = 24 \tilde{H}_{2,2}^{(2)}-8 \tilde{H}_{2,2}^{(3)}\,, \qquad
&& H_{2,2}^{(3)} = 48 \tilde{H}_{2,2}^{(3)}\,, \\
& H_{2,2}^{(4)} = 24 \tilde{H}_{2,2}^{(4)}-8 \tilde{H}_{2,2}^{(5)}\,, \qquad
&& H_{2,2}^{(5)} = 48\tilde{H}_{2,2}^{(5)}\,, \qquad
&& H_{2,2}^{(6)} = 24 \tilde{H}_{2,2}^{(6)}-8 \tilde{H}_{2,2}^{(7)}\,, \\
& H_{2,2}^{(7)} = 48 \tilde{H}_{2,2}^{(7)}\,, \qquad
&& H_{2,2}^{(8)} = 24 \tilde{H}_{2,2}^{(8)}-8 \tilde{H}_{2,2}^{(10)}-8 \tilde{H}_{2,2}^{(12)}+8\tilde{H}_{2,2}^{(13)}\,, \qquad
&& H_{2,2}^{(9)} = 24 \tilde{H}_{2,2}^{(9)}-8 \tilde{H}_{2,2}^{(11)}+16 \tilde{H}_{2,2}^{(13)}\,, \\
& H_{2,2}^{(10)} = 48 \tilde{H}_{2,2}^{(10)}-48 \tilde{H}_{2,2}^{(13)}\,, \qquad
&& H_{2,2}^{(11)} = 48\tilde{H}_{2,2}^{(11)}-192 \tilde{H}_{2,2}^{(13)}\,, \qquad
&& H_{2,2}^{(12)} = 48 \tilde{H}_{2,2}^{(12)}-48 \tilde{H}_{2,2}^{(13)}\,, \\
& H_{2,2}^{(13)} = 384 \tilde{H}_{2,2}^{(13)}\,.
\end{alignedat}
\end{equation}
}
\end{subequations}

\item $S_a^3 S_b$
\begin{subequations}
\begin{equation}
\begin{split}
H_{3,1} = \frac{G^2}{r^{10}}
&
\left[
H_{3,1}^{(1)} r^4 (\vec{p}\cdot \vec{S}_a)^3 \vec{p}\cdot \vec{S}_b
+
H_{3,1}^{(2)} r^4 \vec{S}_a\cdot \vec{S}_b (\vec{p}\cdot \vec{S}_a )^2
+
H_{3,1}^{(3)} r^2 ( \vec{p}\cdot \vec{S}_a )^2 \vec{r}\cdot \vec{S}_a \vec{r}\cdot \vec{S}_b \right.
\\
&
+
H_{3,1}^{(4)} r^4 |\vec{S}_a|^2 \vec{p}\cdot \vec{S}_a \vec{p}\cdot \vec{S}_b
+
H_{3,1}^{(5)} r^2 \vec{p}\cdot \vec{S}_a ( \vec{r}\cdot \vec{S}_a )^2 \vec{p}\cdot \vec{S}_b
+
H_{3,1}^{6} r^4 |\vec{S}_a|^2 \vec{S}_a\cdot \vec{S}_b\\
&
\left.
+
H_{3,1}^{(7)} r^2 \vec{S}_a\cdot \vec{S}_b ( \vec{r}\cdot \vec{S}_a )
+
H_{3,1}^{(8)} r^2 |\vec{S}_a|^2 \vec{r}\cdot \vec{S}_a \vec{r}\cdot \vec{S}_b
+
H_{3,1}^{(9)} (\vec{r}\cdot \vec{S}_a )^3 \vec{r}\cdot \vec{S}_b \right]\,.
\end{split}
\end{equation}
The $H_{3,1}^{(\alpha)}$ coefficients are related to the $\tilde{H}_{3,1}^{(\alpha)}$ by the following relation:
{\small
\begin{equation}
\begin{alignedat}{3}
&H_{3,1}^{(1)} = 24 \tilde{H}_{3,1}^{(1)}\,,  \qquad
&&H_{3,1}^{(2)} = 24 \tilde{H}_{3,1}^{(2)}-8 \tilde{H}_{3,1}^{(3)}\,,  \qquad
&&H_{3,1}^{(3)} = 48 \tilde{H}_{3,1}^{(3)}\,, \\
&H_{3,1}^{(4)} = 24 \tilde{H}_{3,1}^{(4)} - 8 \tilde{H}_{3,1}^{(5)}\,, \qquad
&&H_{3,1}^{(5)} = 48\tilde{H}_{3,1}^{(5)}\,, \qquad
&&H_{3,1}^{(6)} = 24 \tilde{H}_{3,1}^{(6)} - 8 \tilde{H}_{3,1}^{(7)} - 8 \tilde{H}_{3,1}^{(8)}+24 \tilde{H}_{3,1}^{(9)}\,, \\ 
&H_{3,1}^{(7)} = 48 \tilde{H}_{3,1}^{(7)}-144 \tilde{H}_{3,1}^{(9)}\,, \qquad
&&H_{3,1}^{(8)} = 48 \tilde{H}_{3,1}^{(8)} - 144 \tilde{H}_{3,1}^{(9)}\,, \qquad
&&H_{3,1}^{(9)} = 384 \tilde{H}_{3,1}^{(9)}\,.
\end{alignedat}
\end{equation}
}
\end{subequations}
\item $S_a^4$
\begin{subequations}
\begin{equation}
\begin{split}
H_{4,0} = \frac{G^2}{r^{10}}
&
\left[ 
H_{4,0}^{(1)} r^4  ( \vec{p}\cdot \vec{S}_a )^4
+
H_{4,0}^{(2)} r^4 |\vec{S}_a|^2 (\vec{p} \cdot \vec{S}_a)^2
+
H_{4,0}^{(3)} r^2 (\vec{p} \cdot \vec{S}_a)^2 (\vec{r}\cdot \vec{S}_a)^2 \right.\\
&
\left.
+
H_{4,0}^{(4)} r^4 |\vec{S}_a|^4
+
H_{4,0}^{(5)} r^2 |\vec{S}_a|^2 (\vec{r} \cdot \vec{S}_a)^2
+
H_{4,0}^{(6)} (\vec{r}\cdot \vec{S}_a)^4
\right]\,.
\end{split}
\end{equation}
The $H_{4,0}^{(\alpha)}$ coefficients are related to the $\tilde{H}_{4,0}^{(\alpha)}$ by the following relation:
\begin{equation}
\begin{alignedat}{3}
&
H_{4,0}^{(1)} = 24 \tilde{H}_{4,0}^{(1)}\,, \qquad
&&
H_{4,0}^{(2)} = 24 \tilde{H}_{4,0}^{(2)}-8 \tilde{H}_{4,0}^{(3)}\,, \qquad
&&
H_{4,0}^{(3)} = 48 \tilde{H}_{4,0}^{(3)}\,, \\
&
H_{4,0}^{(4)} = 24 \tilde{H}_{4,0}^{(4)}-8 \tilde{H}_{4,0}^{(5)}+24\tilde{H}_{4,0}^{(6)}\,, \qquad
&&
H_{4,0}^{(5)} = 48 \tilde{H}_{4,0}^{(5)}-288 \tilde{H}_{4,0}^{(6)}\,, \qquad
&&
H_{4,0}^{(6)} = 384 \tilde{H}_{4,0}^{(6)}\,.
\end{alignedat}
\end{equation}

\end{subequations}
\end{itemize}

\subsection{Matching with other results up to $\mathcal{O}(S^2)$}
The Hamiltonian can come in different forms, but the different expressions are equivalent when one can find a canonical transform that relates them
\begin{equation}\label{eq:Gen_Can}
\Delta H = \{H, g\}\,, 
\end{equation}
where $g$ is the generator of the canonical transformation. Such equivalence between amplitude methods and the EFT formalisms \cite{Levi:2015msa, Levi:2014sba, Levi:2014gsa} was established up to the NNLO $\mathcal{O}(G)$ quadratic in spin sector in \cite{Chung:2020rrz}. We further show that the leading order in $\vec{p}$ contribution of the 2PM Hamiltonian obtained from amplitudes are equivalent to the NLO $\mathcal{O}(G^2)$ contributions computed from the EFT formalism up to quadratic order in spin.

In general, the canonical transformation is given by eq.\eqref{eq:Gen_Can}. However at NLO, $\Delta H_{n_a, n_b}^{\text{NLO}}$ takes the form:
\begin{equation}
\Delta H_{n_a, n_b}^{\text{NLO}} = \{H_N^{\text{EFT}}, g^{\text{NLO}}_{n_a, n_b}\} + \{H^{\text{EFT}}_{LO, (n_a, n_b)}, g_{0,0}^{\text{NLO}}\}\,, 
\end{equation}
where $H_N$ is the Newtonian Hamiltonian:
\begin{equation}
H_N = \left( \frac{|\vec{p}|^2}{2m_a} + \frac{|\vec{p}|^2}{2m_b} \right) - \frac{Gm_a m_b}{r}\,, 
\end{equation}
$g^{\text{NLO}}_{n_a, n_b}$ is the generator of order $\mathcal{O}(G S_a^{n_a} S_b^{n_b} \vec{p}^1)$ and $H_{LO, (n_a, n_b)}^{\text{EFT}}$ are the 1PM spinning potential given by eq.\eqref{eq:1PM_Ham} with $|\vec{p}| = 0$.

Starting from the scalar sector, one needs the generator:
\begin{equation}\label{eq:g00}
g_{0,0}^{(\text{NLO})} = \frac{1}{2}\frac{m_a}{(1+ \zeta^{-1})} \frac{G}{r} \vec{r}\cdot \vec{p}\,, \qquad \zeta = \frac{m_b}{m_a}\,.
\end{equation}
to show the equivalence of our leading order 2PM scalar Hamiltonian with that from ref.\cite{Levi:2014sba}. For the spinning sectors, the NLO $\mathcal{O}(G)$ generators $g^{\text{NLO}}_{n_a, n_b}$ are already fixed by the NLO $\mathcal{O}(G)$ Hamiltonian matching in \cite{Chung:2020rrz}, and $g^{\text{NLO}}_{0, 0} $ is fixed by eq.\eqref{eq:g00}, leaving no more room for additional generators at NLO. Indeed, combining the results in \cite{Chung:2020rrz} and eq.\eqref{eq:g00}, we match the results from EFT formalism up to NLO quadratic in spin order.

\section{Observables}

We compute gauge invariant observables from Hamiltonian equations of motion (EOM) in the COM frame, focussing on the impulse $\Delta \vec{p}$ and the spin kick $\Delta \vec{S}_{a/b}$. Alternative method for computing these observables is to use the eikonal phase~\cite{Kosower:2018adc, Maybee:2019jus, Bern:2020buy}, or in the  test body limit to use the Mathisson-Papapetrou-Dixon (MPD) equations. Matching with these results provide further consistency checks of our Hamiltonian.

\subsection{Observales in the COM frame}
We compute the impulse and spin kick in the COM frame using two methods; by integrating the Hamiltonian EOM iteratively to $\mathcal{O}(G^2)$, and by applying differential operators to 2 PM eikonal phase~\cite{Bern:2020buy}. The results were shown to be consistent for all computed sectors.

\subsubsection{Observables from the Hamiltonian EOM}\label{subsubsec:Ham_Obs}
The Hamiltonian EOM of an observable $O^i$ is 
\begin{equation}
\dot{O}^i = \{O^i, H\} 
= 
\left( \frac{\partial O^i}{\partial r^j}\frac{\partial H}{\partial p^j} - \frac{\partial O^i}{\partial p^j}\frac{\partial H}{\partial r^j} \right) + 
\sum_{n = a,b} \left( \epsilon^{jkl} \frac{\partial O^i}{\partial S_n^j} \frac{\partial H}{\partial S_n^k} S_n^l \right)\,.
\end{equation}
The EOM for the observables of our interest are
\begin{equation}
\frac{d\vec{r}}{dt} = \frac{\partial H}{\partial \vec{p}}, \quad
\frac{d\vec{p} }{dt}= -\frac{\partial H}{\partial \vec{r}}, \quad
\frac{d\vec{S}_n }{dt} = -\vec{S}_n \times \frac{\partial H}{\partial \vec{S}_n}\,.
\end{equation}
We write $O^i$ as a series expansion in $G$, i.e.
\begin{equation}
O_i = O_{i}^{(0)} + G O_{i}^{(1)} + G^2 O_{i}^{(2)} + \cdots\,.
\end{equation}
The scattering variables are integrals of the EOM from the past to future infinity, 
\begin{equation}
\Delta O_i = \int_{-\infty}^{\infty} \dot{O}^i dt\,,
\end{equation}
which often suffers from IR divergences that can be avoided by a suitable change of integration variable;\footnote{We acknowledge Chia-Hsien Shen for discussion of the following method.} we divide $\dot{O}_i$ by $\dot{z}$ and use $dz$ as integration variable instead
\begin{equation}
\frac{dO_i}{dz} = \frac{\dot{O_i}}{\dot{z}}, \quad \dot{z} = \frac{\partial H}{\partial p^z} \,, \quad \Delta O_i = \int_{-\infty}^{\infty} \frac{dO_i}{dz} dz \,.
\end{equation}
The suitable boundary conditions are:
\begin{equation}
\begin{split}
x^{(0)}(z {\rightarrow-}\infty) &{=} b\,, \quad 
y^{(0)}(z {\rightarrow-} \infty) {=} 0\,, 
\quad x^{(n {>}0)}(z {\rightarrow-} \infty) {=} y^{({n}>0)}(z {\rightarrow-} \infty) {=} 0\,, \\
\vec{p}^{(0)}(z {\rightarrow-} \infty) &{=} (0, 0, p_{\infty})\,, \quad \vec{p}^{(n{>}0)}(z {\rightarrow-} \infty) {=} \vec{0}\,, \\
\vec{S}_{a/b}^{(0)}(z {\rightarrow-} \infty) &= (\vec{S}_{a/b,x}, \vec{S}_{a/b,y}, \vec{S}_{a/b,z})\,, \quad \vec{S}_{a/b}^{(n{>}0)}(z {\rightarrow-} \infty) {=} \vec{0}\,.
\end{split}
\end{equation}
We compute the 2PM impulse $\Delta \vec{p}$ and spin kicks $\Delta\vec{S}_{a/b}$ for general initial spin directions to $\mathcal{O}(S_a^0S_b^0)$, $\mathcal{O}(S_a)$, $\mathcal{O}(S_aS_b)$, $\mathcal{O}(S_a^2)$, $\mathcal{O}(S_a^2 S_b)$, $\mathcal{O}(S_a^3)$, $\mathcal{O}(S_a^4)$.

\subsubsection{Observables from the eikonal phase}\label{subsubsection:Eikonal_Obs}
The 1PM and 2PM eikonal phases are defined as \cite{Bern:2020buy}:
\begin{equation}
\begin{split}
\chi_{1\text{PM}} &= \frac{1}{4m_a m_b\sqrt{\sigma^2 - 1}} \int \frac{d^2 \vec{q}}{(2\pi)^2}e^{-i \vec{q} \cdot \vec{b}}M_{\text{tree}}, \\
\chi_{2\text{PM}} &= \frac{1}{4m_a m_b\sqrt{\sigma^2 - 1}} \int \frac{d^2 \vec{q}}{(2\pi)^2}e^{-i \vec{q} \cdot \vec{b}}M_{\bigtriangleup + \bigtriangledown}\,,
\end{split}
\end{equation}
where $\vec{b}$ is the impact parameter. It was shown that the impact parameter and the spin-kick can be obtained as suitable derivatives of eikonal phase at $\mathcal{O}(G)$ order~\cite{Kosower:2018adc, Maybee:2019jus}, which was generalized to $\mathcal{O}(G^2)$ in ref.\cite{Bern:2020buy}. The latter is the version we utilise:
\begin{equation}\label{eq:Eikonal_Obs_Formula}
\begin{split}
\Delta \vec{p}_{\perp} &= -\{ \vec{p}_{\perp}, \chi\} - \frac{1}{2} \{\chi, \{ \vec{p}_{\perp}, \chi \}\} - \mathcal{D}_{SL}\left( \chi, \{\vec{p}_{\perp}, \chi\} \right) + \frac{1}{2}\{\vec{p}_{\perp}, \mathcal{D}_{SL}(\chi, \chi)\}, \\
\Delta \vec{S}_{a/b} &= -\{ \vec{S}_{a/b}, \chi\} - \frac{1}{2} \{\chi, \{ \vec{S}_{a/b}, \chi \}\} - \mathcal{D}_{SL}\left( \chi, \{\vec{S}_{a/b}, \chi\} \right) + \frac{1}{2}\{\vec{S}_{a/b}, \mathcal{D}_{SL}(\chi, \chi)\}\,,
\end{split}
\end{equation}
where $\chi = \chi_{1\text{PM}} + \chi_{2\text{PM}} + \cdots$ is the full eikonal phase, and the Poisson bracket and $\mathcal{D}_{SL}$ are defined by:
\begin{equation}
\begin{split}
\{O^i, \chi\} 
&= 
\left( \frac{\partial O^i}{\partial b^j}\frac{\partial H}{\partial p_{\perp}^j} - \frac{\partial O^i}{\partial p_{\perp}^j}\frac{\partial H}{\partial b^j} \right) + 
\sum_{n = a,b} \left( \epsilon^{jkl} \frac{\partial O^i}{\partial S_n^j} \frac{\partial \chi}{\partial S_n^k} S_n^k \right),\\
\mathcal{D}_{SL}(f, g) &= \frac{1}{p_{\infty}} \sum_{n = a,b}\left( \frac{\partial f}{\partial S_n^{i}} \frac{\partial g}{\partial b^i} \vec{S}_n \cdot \vec{p} - p^i \frac{\partial f}{\partial S_n^i} \vec{S}_n \cdot \nabla_{\vec{b}} g\right)\,.
\end{split}
\end{equation}
$\vec{p}_{\perp}$ denotes the momentum components perpendicular to the incoming momentum $\vec{p}^{(0)} = (0, 0, p_{\infty})$, and corresponds to $x$ and $y$ components in our set-up. The $z$ component of the impulse $\Delta p_z$ is determined by the energy conservation condition
\begin{equation}
2 p_{\infty} \Delta p_z + |\Delta \vec{p}_{\perp}|^2 + \Delta{p}_z^2 = 0\,.
\end{equation}
We present our impulse $\Delta \vec{p}$ and the spin kicks $\Delta \vec{S}_a$, $\Delta \vec{S}_b$ from eq.\eqref{eq:Eikonal_Obs_Formula} up to quartic order in spin in the ancillary file ``Observables.wl". We have verified that the results match with that from Hamiltonian EOM for all computed sectors, to all orders in momentum $|\vec{p}|$.

As an additional check we compared and matched with the results of ref.\cite{Guevara:2018wpp} where the 2 PM aligned-spin scattering angle was computed up to quartic order in spin. The aligned-spin kinematics imposes the condition $\vec{b} \cdot \vec{S} = \vec{p} \cdot \vec{S} = 0$ and the scattering angle simplifies to \cite{Kosmopoulos:2021zoq}:
\begin{equation}\label{eq:Scattering_Ang formula}
\theta = - \frac{E}{m_a m_b \sqrt{\sigma^2 - 1}} \partial_b \bar{\chi}\,,
\end{equation}
where $\bar{\chi}$ is the eikonal phase without the rotation factors $U_{a,b}$. The rotation factors are dropped because ref.\cite{Guevara:2018wpp} uses the proper impact parameter, which is different from the canonical impact parameter~\cite{Vines:2017hyw,Liu:2021zxr} corresponding to the canonical (or Pryce-Newton-Wigner) SSC~\cite{Pryce:1935ibt,Pryce:1948pf,Newton:1949cq} suited for Hamiltonian dynamics;\footnote{A review on consequences for different choices of SSCs can be found in ref.\cite{Steinhoff:2011sya}.} see ref.\cite{Bern:2020buy} for the rotation factor's connection to SSC choice. The difference between covariant and canonical impact parameter will be important when comparing with test black hole scattering results in the next section.

\subsection{The background-probe limit}
In the extreme mass ratio limit the dynamics of the binary system reduces to that of a test spinning black hole moving on a Kerr background. Scattering observables in this set-up can be calculated using the Mathisson-Papapetrou-Dixon (MPD) equations~\cite{Mathisson:1937zz,Papapetrou:1951pa,doi:10.1098/rspa.1970.0020}, which describes a test body moving on a curved background. We use the pole-dipole approximation
\bl
\bld
\frac{Dp^\m}{D \t} &= - \half R^{\m}_{~\n\a\b} u^\n S^{\a\b} \,,
\\ \frac{DS^{\m\n}}{D \t} &= p^\m u^\n - p^\n u^\m \,,
\eld \nn
\el
which describes a test particle endowed with a monopole moment (momentum) $p^\m$ and a dipole moment (spin tensor) $S^{\m\n}$. The equations can be solved iteratively to compute scattering variables to linear order in test-body spin, which we provide the details in appendix \ref{app:MPD}. This method provides an independent computation that can be used as consistency checks of our results.

We consider scattering on a Kerr background with initial data
\bl
\bld
b^\m = \left( 0 \,, b \,, 0 \,, 0 \right) ,\, u_0^\m = \left( \frac{1}{\sqrt{1-v^2}} \,, 0 \,, \frac{v}{\sqrt{1-v^2}} \,, 0 \right) ,\, s_0^\m (\t_0) = \left( 0 \,, s_0 \,, 0 \,, 0 \right) \,,
\eld
\el
where $b^\m$ is the impact parameter, $u_0^\m$ is the initial 4-velocity, and $s_0^\m$ is the mass-rescaled initial Pauli-Lubanski spin vector $s_0^\m = {S_0^\m}/{m}$. The black hole spin is directed along the $z$-axis and covariant SSC is implied. We choose this configuration to test genericity of our results beyond well-studied aligned-spin cases while keeping some simplicity. The full results are presented in appendix \ref{app:MPDexample1}.

To make contact with the observables computed in section \ref{subsubsec:Ham_Obs} and \ref{subsubsection:Eikonal_Obs} we take the test body limit, by keeping only the leading contributions in the large mass ratio $\frac{M}{m} \rightarrow \infty$ expansion, where we identify $m_a = M$ as the black hole's mass and $m_b = m$ as the spinning test body's mass. The difference between canonical and covariant variables need to be accounted for; Hamiltonian EOM uses canonical variables, while MPD EOM uses covariant variables. This would require us to do a shift to the impact parameter \cite{Vines:2017hyw}
\begin{equation}
\vec{b}_{\text{covariant}} = \vec{b}_{\text{canonical}} - \frac{\vec{p} \times \vec{S}_b}{m_b(m_b + E_b)}\,.
\end{equation}
For the spin kick, we compare the spin 4-vector since it is free of ``frame dependence'' as demonstrated in ref.\cite{Aoude:2021oqj}. In practice, the spin kick $\Delta S_b$ computed from the Hamiltonian EOM is boosted as
\begin{equation}
m \Delta s^{\mu}_0 = \Delta S^{\mu}_b = \left( 
\frac{\vec{p}' \cdot \vec{S}_b' }{m_b},
 \vec{S}_b'+ \frac{ \vec{p}'  \cdot \vec{S}_b'}{m_b(m_b + E_b)} \vec{p}'
\right) - 
 \left( 
\frac{\vec{p} \cdot \vec{S}_b }{m_b},
 \vec{S}_b+ \frac{ \vec{p} \cdot \vec{S}_b}{m_b(m_b + E_b)} \vec{p}
\right)\,,
\end{equation}
where $\vec{p}' = \vec{p} + \Delta \vec{p}$, $\vec{S}_b = \vec{S} + \Delta \vec{S}_b$\,. Up to the explained technicalities, the observables computed using three different methods were confirmed to be consistent with each other.

For the simpler case of aligned-spins, the scattering angle for test black hole scattering was computed in ref.\cite{Vines:2018gqi} to $G^2$ order with quadrupole moment $J^{\m\n\l\s}$ included in the analysis. The computations amount to quadratic-in-spin order computation for the scattering angle, which was shown to match with amplitude results eq.(1.12) of ref.\cite{Guevara:2018wpp}, which was also compared with quartic-in-spin order results for the test black hole in ref.\cite{Siemonsen:2019dsu}. Therefore matching with these results would be redundant, since the scattering angle of aligned spin case presented in ref.\cite{Guevara:2018wpp} was already compared in section \ref{subsubsection:Eikonal_Obs}. 

\section{Conclusions and outlook}
In this paper, we've computed the 2 PM conservative Hamiltonian up to quadratic order in total spin for binary spinning black holes without resorting to Feynman diagrams. As NLO corrections to the effective action that are cubic and quartic in spin PN expansion are given in~\cite{Levi:2019kgk,Levi:2020lfn}, it will be interesting to compare at the observable level. It will certainly be interesting to extend this to general spinning two bodies, which has been explored to quadratic order in spin~\cite{Kosmopoulos:2021zoq,Liu:2021zxr, Jakobsen:2021zvh} based on Feynman diagrams. As discussed in appendix~\ref{app:genWC}, direct implementation of the BCFW recursion for non-minimal coupling results in an expression for the Compton amplitudes with spurious poles, obstructing application of on-shell methods to the gravitational two-body problem. It would be interesting to develop a systematic way of cancelling the spurious poles in the Compton amplitude, and to compare with that based on EFT Feynman rules. 

Certainly the most interesting theoretical question is the Kerr Compton amplitude beyond spin-2. Recently a new spin-$\frac{5}{2}$ amplitude was proposed based on double copy structure~\cite{Chiodaroli:2021eug}. It would be interesting to see if there are any further properties that singles out this amplitude. For example: When compared to the BCFW result, is the difference minimal in the basis of local contact terms constructed along the lines of ref.\cite{Durieux:2020gip}? Does it have the best Regge behaviour at high energies?

On the other hand, it may very well be that these contact term ambiguities, which by definition is of order $G$, will not contribute classically. Indeed tidal operators for scalars, which appear as contact terms in the Compton amplitude, leads to modification to the triangle coefficients that has distinct $q$-scalings~\cite{Cheung:2020sdj,Haddad:2020que}. Assuming the same behaviour carries over to spinning cases, unless the contact terms are allowed to have extra inverse powers of $\hbar$ they will lead to quantum effects and become irrelevant for classical observables. If this is the case, then one only needs a systematic, and preferably economic, way of removing the spurious poles.

\vskip 1cm 

\acknowledgments 

We are grateful for Mich\`ele Levi, Chia-Hsien Shen, Dimitrios Kosmopoulos, Andres Luna, and Andreas Brandhuber for discussions. We also would like to thank Mich\`ele Levi and Roger M. Espasa for sharing unpublished results for initial comparison~\cite{RogerThesis}.
WMC is supported by in part by JSPS KAKENHI Grant Number 21F21317.
MZC and YTH is supported by MoST Grant No. 109-2112-M-002 -020 -MY3. 
JWK was supported by the Science and Technology Facilities Council (STFC) Consolidated Grant ST/T000686/1 \textit{``Amplitudes, Strings and Duality”}.

\appendix

\section{Spin factor composition} \label{app:spinprod}
The main objectives of this appendix are to prove representation independence of the expression \eqc{eq:minampexp} and to evaluate composition of the spin factors in amplitude products of the form $M_3 \times M_3$, which should be evaluated as operator products.

The following relations are satisfied by 3pt on-shell kinematics $p_1^2 = p_2^2 = m^2$, $k_3^2 = 0$, and $p_1 + p_2 + k_3 = 0$.
\bl
\la \bf{21} \ra &= \sbra{\bf{2}} \left(\iden + \frac{\sket{3} \sbra{3}}{m x} \right) \sket{\bf{1}} = \sbra{\bf{2}} \exp \left( \frac{\sket{3} \sbra{3}}{m x} \right) \sket{\bf{1}} \,,
\\ \frac{\sket{3} \sbra{3}}{m x} &= \half\frac{\sqrt{2} \la 3 \eta \ra}{\sbra{3} p_1 \ket{\eta}} \sket{3} \bra{3}  \left( \sqrt{2} \frac{\ket{\eta} \sbra{3}}{\la 3 \eta \ra} \right) = -i \frac{k_3^\m \ve_{3+}^\n (J_{\m\n})^{\dot\a}_{~\dot\b}}{p_1 \cdot \ve_{3+}} \,.
\el
Considering $k_3^\m \ve_{3+}^\n (J_{\m\n})_{\a}^{~\b} = 0$ with the above relations, we can write
\bl
\la \mathbf{21} \ra &= \bra{\bf{2}} \exp \left( -i \frac{k_3^\m \ve_{3+}^\n (J_{\m\n})_{\a}^{~\b}}{p_1 \cdot \ve_{3+}} \right) \ket{\bf{1}} = \sbra{\bf{2}} \exp \left( -i \frac{k_3^\m \ve_{3+}^\n (J_{\m\n})^{\dot\a}_{~\dot\b}}{p_1 \cdot \ve_{3+}} \right) \sket{\bf{1}} \,,
\el
because $e^{0} = \iden$. In other words,
\bl
\la \mathbf{21} \ra &= \left[ \ve_{2} \cdot \exp \left( -i \frac{k_3^\m \ve_{3+}^\n J_{\m\n} }{p_1 \cdot \ve_{3+}} \right) \cdot \ve_{1} \right]_{\text{Rep}(1/2)} \stackrel{\cdot}{=} \exp \left( -i \frac{k_3^\m \ve_{3+}^\n J_{\m\n} }{p_1 \cdot \ve_{3+}} \right) \,, \label{eq:anglebkexp}
\el
where $\text{Rep}(1/2)$ stands for either of chiral or anti-chiral spin-$\half$ representation. The right equality given by $\stackrel{\cdot}{=}$ follows because the two exhausts all spin-$\half$ representations. The relation \eqc{eq:sqrbkexp} follows from similar considerations.
\bl
\bld
{}[ \mathbf{21} ] &= \sbra{\bf{2}} \exp \left( -i \frac{k_3^\m \ve_{3-}^\n (J_{\m\n})^{\dot\a}_{~\dot\b}}{p_1 \cdot \ve_{3-}} \right) \sket{\bf{1}} = \bra{\bf{2}} \exp \left( -i \frac{k_3^\m \ve_{3-}^\n (J_{\m\n})_{\a}^{~\b}}{p_1 \cdot \ve_{3-}} \right) \ket{\bf{1}}
\\ &= \left[ \ve_{2} \cdot \exp \left( -i \frac{k_3^\m \ve_{3-}^\n J_{\m\n} }{p_1 \cdot \ve_{3-}} \right) \cdot \ve_{1} \right]_{\text{Rep}(1/2)} \stackrel{\cdot}{=} \exp \left( -i \frac{k_3^\m \ve_{3-}^\n J_{\m\n} }{p_1 \cdot \ve_{3-}} \right) \,. \label{eq:sqrbkexp}
\eld
\el
The two relations \eqc{eq:anglebkexp} and \eqc{eq:sqrbkexp} establish \eqc{eq:minampexp}, because all finite spin representations are constructed from tensor products of spin-$\half$ representations.
\bfig
\centering
\begin{tikzpicture}[line width=1. pt, scale=2,
sines/.style={
        line width=1pt,
        line join=round, 
        draw=black, 
        decorate, 
        decoration={complete sines, number of sines=4, amplitude=4pt}
    }
]

\node[scale=1] at (-0.6,-0.6) {$p_1$};
\node[scale=1] at (-0.6,0.6) {$k_2$};
\node[scale=1] at (1.6,0.6) {$k_3$};
\node[scale=1] at (1.6,-0.6) {$p_4$};
\node[scale=1] at (0.5,-0.2) {$P$};
\draw [-stealth](0.49,0) -- (0.48,0);
\draw [-stealth](-0.25,-0.25) -- (-0.249,-0.249);
\draw [-stealth](1.25,-0.25) -- (1.249,-0.249);
\draw[white,postaction={sines}] (0,0) -- (-0.5,0.5);
\draw[white,postaction={sines}] (1,0) -- (1.5,0.5);
\draw[black, line width=1.pt] (-0.005,0) -- (1.005,0);
\draw[black, line width=1.pt] (0,0) -- (-0.5,-0.5);
\draw[black, line width=1.pt] (1,0) -- (1.5,-0.5);
\draw [-stealth](1.2,0.4) -- (1,0.2);
\draw [-stealth](-0.2,0.4) -- (0,0.2);
\end{tikzpicture}

\centering
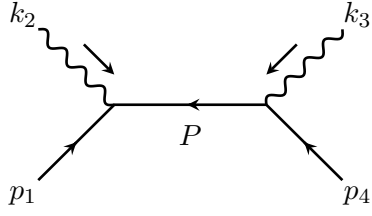
\caption{Generic $A_3 \times A_3$ gluing configuration with all edges on-shell} \label{fig:GF}
\efig

Consider the gluing configuration as depicted in \fig{fig:GF}, where all lines are considered on-shell. We attach auxiliary variables $z_i$ to the spin factor of \eqc{eq:minampexp}, bearing in mind the 3pt amplitude representation \eqc{eq:3ptresint}. The composition of spin factors is an \emph{operator product}
\bl
\exp \left( -i \frac{k_3^\m \ve_{3}^\n J_{\m\n}}{(- P \cdot \ve_{3}) z_4} \right) \exp \left( -i \frac{k_2^\m \ve_{2}^\n J_{\m\n}}{(p_1 \cdot \ve_{2}) z_1} \right) &= \exp \left( i \frac{k_3^\m \ve_{3}^\n J_{\m\n}}{(P \cdot \ve_{3}) z_4} \right) \exp \left( i \frac{k_2^\m \ve_{2}^\n J_{\m\n}}{(P \cdot \ve_{2}) z_1} \right) \,, \label{eq:spinprodgen}
\el
which we work out explicitly in this section. The composition can be used in the BCFW construction of 4pt amplitudes and in triple/quadruple cut computation of the spin factor for amplitude products of the form $M_3 \times M_3$.

\subsection{Opposite helicity case}
\subsubsection{Helicity configuration $2^+ 3^-$}
Without loss of generality we may set the reference spinors of $\ve_{2+}$($\ve_{3-}$) as $\ket{3}$($\sket{2}$). The factor \eqc{eq:spinprodgen} can be worked out explicitly in chiral and aniti-chiral spin-$\half$ representations.
\bl
\exp \left( i \frac{k_3^\m \ve_{3-}^\n J_{\m\n}}{(P \cdot \ve_{3-}) z_4} \right) \exp \left( i \frac{k_2^\m \ve_{2+}^\n J_{\m\n}}{(P \cdot \ve_{2+}) z_1} \right) &= \left\{
\begin{aligned}
\exp \left( - \frac{\ket{3}[32]\bra{3}}{\sbra{2}P\ket{3} z_4} \right) && &~~(\text{chiral})
\\ \exp \left( - \frac{\sket{2} \la 23 \ra \sbra{2}}{\sbra{2}P\ket{3} z_1} \right) && &(\text{anti-chiral})
\end{aligned}\,.
\right. \label{eq:2p3mSF}
\el
As emphasised in the beginning of this appendix, the LHS is a product of operators and the RHS are $2 \times 2$ matrix expressions for the operator in chiral and aniti-chiral spin-$\half$ representations. The task is to find the $b_{\m\n}$ that solves the operator relation
\bl
\exp (-\frac{i}{2} b_{\m\n} J^{\m\n}) &= \left\{
\begin{aligned}
\exp \left( - \frac{\ket{3}[32]\bra{3}}{\sbra{2}P\ket{3} z_4} \right) && &~~(\text{chiral})
\\ \exp \left( - \frac{\sket{2} \la 23 \ra \sbra{2}}{\sbra{2}P\ket{3} z_1} \right) && &(\text{anti-chiral})
\end{aligned}\,.
\right.
\el
in both spin-$\half$ representations. The solution to the two matrix equations is
\bg
\bgd
\exp \left( i \frac{k_3^\m \ve_{3-}^\n J_{\m\n}}{(P \cdot \ve_{3-}) z_4} \right) \exp \left( i \frac{k_2^\m \ve_{2+}^\n J_{\m\n}}{(P \cdot \ve_{2+}) z_1} \right) = \exp \left( i K^\m L^\n J_{\m\n} \right)\,,
\\ K^\m = \frac{k_2^\m}{z_1} + \frac{k_3^\m}{z_4} \,,~~~~~~~\, L^\m = \frac{\bra{3} \s^\m \sket{2}}{\bra{3} P \sket{2}} \,.
\egd \label{eq:SFprodPM}
\eg
Setting $z_1 = z_4 = 1$, the above formula interpolates the two expressions of (2.35) in ref.\cite{Guevara:2018wpp} to arbitrary spin representations.

\subsubsection{Helicity configuration $2^- 3^+$}
Without loss of generality we may set the reference spinors of $\ve_{2-}$($\ve_{3+}$) as $\sket{3}$($\ket{2}$).
\bl
\exp \left( i \frac{k_3^\m \ve_{3+}^\n J_{\m\n}}{(P \cdot \ve_{3+}) z_4} \right) \exp \left( i \frac{k_2^\m \ve_{2-}^\n J_{\m\n}}{(P \cdot \ve_{2-}) z_1} \right) &= \left\{
\begin{aligned}
\exp \left( - \frac{\ket{2}[23]\bra{2}}{\sbra{3}P\ket{2} z_1} \right) && &~~(\text{chiral})
\\ \exp \left( - \frac{\sket{3} \la 32 \ra \sbra{3}}{\sbra{3}P\ket{2} z_4} \right) && &(\text{anti-chiral})
\end{aligned}\,.
\right.
\el
The rest is straightforward; we only exchange the roles of angles and squares.
\bg
\bgd
\exp \left( i \frac{k_3^\m \ve_{3+}^\n J_{\m\n}}{(P \cdot \ve_{3+}) z_4} \right) \exp \left( i \frac{k_2^\m \ve_{2-}^\n J_{\m\n}}{(P \cdot \ve_{2-}) z_1} \right) = \exp \left( i K^\m L^\n J_{\m\n} \right)\,,
\\ K^\m = \frac{k_2^\m}{z_1} + \frac{k_3^\m}{z_4} \,,~~~~~~~\, L^\m = \frac{\bra{2} \s^\m \sket{3}}{\bra{2} P \sket{3}} \,.
\egd \label{eq:SFprodMP}
\eg

\subsection{Equal helicity case}
\subsubsection{Helicity configuration $2^+ 3^+$}
Without loss of generality we may set the reference spinors of $\ve_{2+}$($\ve_{3+}$) as $\ket{3}$($\ket{2}$). Working out the factor \eqc{eq:spinprodgen} in chiral and aniti-chiral spin-$\half$ representations yields
\bl
\exp \left( i \frac{k_3^\m \ve_{3+}^\n J_{\m\n}}{(P \cdot \ve_{3+}) z_4} \right) \exp \left( i \frac{k_2^\m \ve_{2+}^\n J_{\m\n}}{(P \cdot \ve_{2+}) z_1} \right) &= \left\{
\begin{aligned}
\iden~~~~~~~~~~~~~~~~~~~~~~~~~~~~~~~~~~~~~~~~~~~ && &~~(\text{chiral})
\\ \exp \left( - \frac{\sket{3} \la 32 \ra \sbra{3}}{\sbra{3}P\ket{2} z_4} \right) \exp \left( - \frac{\sket{2} \la 23 \ra \sbra{2}}{\sbra{2}P\ket{3} z_1} \right) && &(\text{anti-chiral})
\end{aligned}
\right.
\el
First we find a basis for $b_{\m\n}$ that solves the relation
\bl
\exp (-\frac{i}{2} b_{\m\n} J^{\m\n}) &= \exp \left( i \frac{k_3^\m \ve_{3+}^\n J_{\m\n}}{(P \cdot \ve_{3+}) z_4} \right) \exp \left( i \frac{k_2^\m \ve_{2+}^\n J_{\m\n}}{(P \cdot \ve_{2+}) z_1} \right) \,,
\el
for the two matrix equations. This is just a pair of $2 \times 2$ matrix equations, which is simple enough to be solved without the Baker-Campbell-Hausdorff formula. After some algebra, one finds the ansatz
\bg
\bgd
b^{\m\n} = a_1 \left( k_2^{[\m} k_3^{\n]} + l^{[\m} \bar{l}^{\n]} \right) + a_2 k_2^{[\m} \bar{l}^{\n]} + a_3k_3^{[\m} l^{\n]} \,,
\\ l^\m = \frac{\bra{2} \s^\m \sket{3}}{2} \,,\, \bar{l}^\m = \frac{\bra{3} \s^\m \sket{2}}{2} \,,
\egd \label{eq:sameheltent1}
\eg
where the coefficients $a_i$ are \emph{not} simple (rational) functions of kinematics. A formal series expansion in $z_i^{-1}$ yields the solution
\bl
\bld
\exp (-\frac{i}{2} b_{\m\n} J^{\m\n}) &= \exp \left( i \frac{k_3^\m \ve_{3+}^\n J_{\m\n}}{(P \cdot \ve_{3+}) z_4} \right) \exp \left( i \frac{k_2^\m \ve_{2+}^\n J_{\m\n}}{(P \cdot \ve_{2+}) z_1} \right) \,,
\\ b^{\m\n} &= 2 f \left( \frac{2(k_2 \cdot k_3)}{m^2 z_1 z_4} \right) \left[ \frac{k_2^{[\m} k_3^{\n]} + l^{[\m} \bar{l}^{\n]}}{m^2 z_1 z_4} - \frac{2 k_2^{[\m} \bar{l}^{\n]}}{\sbra{2}P\ket{3} z_1} - \frac{2 k_3^{[\m} l^{\n]}}{\sbra{3} P \ket{2} z_4} \right] \,,
\eld \label{eq:sameheltent3}
\el
where the function $f(x)$ is defined after \eqc{eq:BCFWsamehelSFschanmin}, reproduced below.
\bl
\bld
f(x) &= \sum_{n=0}^{\infty} \frac{(n!)^2 x^n}{(2n+1)!} = \frac{4 \sin^{-1} (\sqrt{x}/2)}{\sqrt{x(4-x)}} = \frac{-4 i}{\sqrt{x(4-x)}} \log \left[ \frac{\sqrt{4-x}+i\sqrt{x}}{2} \right] 
\,. 
\eld \label{eq:ffactor}
\el
Restoring $\hbar$ to massless momenta, the function $f(x)$ can be simplified to $f(x) = 1 + \CO(\hbar^2)$ for computing the classical limit.

\subsubsection{Helicity configuration $2^- 3^-$}
The results can be obtained by exchanging angle and square spinors from helicity configuration $2^+ 3^+$;
\bl
\bld
\exp (-\frac{i}{2} b_{\m\n} J^{\m\n}) &= \exp \left( i \frac{k_3^\m \ve_{3-}^\n J_{\m\n}}{(P \cdot \ve_{3-}) z_4} \right) \exp \left( i \frac{k_2^\m \ve_{2-}^\n J_{\m\n}}{(P \cdot \ve_{2-}) z_1} \right) \,,
\\ b^{\m\n} &= 2 f \left( \frac{2(k_2 \cdot k_3)}{m^2 z_1 z_4} \right) \left[ \frac{k_2^{[\m} k_3^{\n]} - l^{[\m} \bar{l}^{\n]}}{m^2 z_1 z_4} - \frac{2 k_2^{[\m} {l}^{\n]}}{\bra{2}P\sket{3} z_1} - \frac{2 k_3^{[\m} \bar{l}^{\n]}}{\bra{3} P \sket{2} z_4} \right] \,.
\eld \label{eq:sameheltent4}
\el

\section{Classical spinning Compton amplitude for generic Wilson coefficients} \label{app:genWC}
We construct a generalisation of matter-graviton coupling \eqc{eq:minampexp} which maintains the exponential structure while allowing for non-minimal Wilson coefficients. The generalisation is used to construct Compton amplitudes and to compute scalar integral coefficients for non-minimal Wilson coefficients.

\subsection{The residue integral representation and the spin representation of spinning matter} \label{app:resint3pt}
Motivated by the residue integral representation introduced in ref.\cite{Chung:2019duq}, we deform the 3pt amplitude \eqc{eq:minampexp0} to incorporate the Wilson coefficients $C_{\text{S}^n}$ of ref.\cite{Levi:2015msa} parametrising how different the coupling is from black holes. We use the notations of ref.\cite{Chung:2018kqs} for the Wilson coefficients where $C_{\text{S}^n} = 1$ is the Kerr black hole limit.
\bl
\bld
M^{s,\eta \abs{h}}_{3} &= M^{0,\eta \abs{h}}_{3} \oint \frac{dz}{2\pi i z} \left( \sum C_{\text{S}^n} z^n \right) \left[ \ve_{2} \cdot \exp \left( -i \frac{k_3^\m \ve_{3\eta}^\n J_{\m\n}}{(p_1 \cdot \ve_{3\eta}) z} \right) \cdot \ve_{1} \right]_{\text{Rep}(s)}
\\ &= M^{0,\eta \abs{h}}_{3} \sum \frac{C_{\text{S}^n}}{n!} \left[ \ve_{2} \cdot \left( - \eta \frac{k_3 \cdot S}{m} \right)^n \cdot \ve_{1} \right]_{\text{Rep}(s)} \,.
\eld \label{eq:3ptresint}
\el
The parameter $\eta$ denotes the helicity of the massless boson, and the scalar factor $M^{0,\eta \abs{h}}_{3}$ denotes the coupling to spinless matter; $g_0 x^{\abs{h}}$ or $g_0 x^{-\abs{h}}$. The integration variable $z$ is a bookkeeping device where the integration contour encircles the origin CCW, and for minimal coupling the \emph{Wilson polynomial} $\sum C_{\text{S}^n} z^n$ resums to $\sum z^n = \frac{1}{1-z}$, localising the variable to $z=1$ and yielding \eqc{eq:minampexp}. To get to the second line, we substitute the Lorentz generator $J^{\m\n}$ by the spin generator $S^{\m\n} = - \frac{1}{m} \e^{\m\n\l\s} p_{1\l} S_\s$ and use 3pt kinematics to write
\bl
\e_{\a\b\g\m} k_{3}^{\a} \ve_{3\pm}^\b p_1^\g &= \pm i (\ve_{3 \pm} \cdot p_1) k_{3\m} && \Rightarrow && -i \frac{k_3^\m \ve_{3\pm}^\n J_{\m\n}}{p_1 \cdot \ve_{3\pm}} = \mp \frac{k_3 \cdot S}{m} \,.
\el
The second line of \eqc{eq:3ptresint} was used in ref.\cite{Chung:2018kqs} to argue that Kerr black holes couple minimally to gravitons. We remark that while ref.\cite{Jakobsen:2021zvh} considers the $\CO(S^2)$ order operator corresponding to deviations from the black hole limit to be distinct from the $\CO(S^2)$ order operator corresponding to the black hole limit, they seem to be on an equal footing for \eqc{eq:3ptresint} and ref.\cite{Kosmopoulos:2021zoq}.

We are interested in the classical spin limit, $s \to \infty$ with $S = s \hbar$ fixed, of the expression \eqc{eq:3ptresint}. Unlike \eqc{eq:minampexp}, the amplitude \eqc{eq:3ptresint} depends on the spin representation of the polarisation tensor for the spinning matter. Therefore the classical spin limit depends on how the limit is approached, and we have restored polarisation tensors and spin representation in \eqc{eq:3ptresint}. For example, consider the following two extreme cases for Rep$(s)$ when coupling to a positive helicity massless particle.
\bl
\left[ \ve_{2} \cdot\exp \left( -i \frac{k_3^\m \ve_{3+}^\n J_{\m\n}}{(p_1 \cdot \ve_{3+}) z} \right) \cdot \ve_{1} \right]_{\text{Rep}(s)} &= \left\{
\begin{aligned}
&\la \mathbf{21} \ra^{2s} && (\text{totally chiral})\phantom{-anti}
\\ &\left( [\mathbf{21}] + \frac{[\mathbf{2}3][3\mathbf{1}]}{mxz}\right)^{2s} && (\text{totally anti-chiral})
\end{aligned}
\right. \label{eq:spinrep}\,.
\el
It is clear that unless $z$ is localised to $z=1$ (which is the case for minimal coupling) the representation \eqc{eq:3ptresint} will lead to very different results, depending on the spin representation used. For definiteness, we fix the sequence of spin representations approaching the classical limit to be the chirally averaged representations \eqc{eq:CArep}, which we define shortly.

On the other hand, we expect the specifics of the representations to be insignificant in the classical limit, if the sequence of representations satisfy
the following properties:
\bn
\item The representation is symmetric in treating the chiral and anti-chiral representations.
\item The representation is scalable, i.e. symmetrising the product of spin-$s_1$ representation and spin-$s_2$ representation yields spin-$(s_1 + s_2)$ representation.
\en
One of the representations that satisfies the above criteria is the one implicitly used in ref.\cite{Chung:2019duq}, which we refer to as the \emph{chirally averaged} representation. In the chirally averaged representation different spin-$s$ representations are weighted by the binomial coefficient.
\bl
\bld
\left[ \ve_{2} \cdot \exp \left( -i \frac{k_3^\m \ve_{3+}^\n J_{\m\n}}{(p_1 \cdot \ve_{3+}) z} \right) \cdot \ve_{1} \right]_{\text{Rep}(s)=\text{CA}} &:= \left\{ \frac{\la \mathbf{21} \ra}{2} + \half \left( [\mathbf{21}] + \frac{[\mathbf{2}3][3\mathbf{1}]}{mxz}\right) \right\}^{2s} \,,
\\ \left[ \ve_{2} \cdot \exp \left( -i \frac{k_3^\m \ve_{3-}^\n J_{\m\n}}{(p_1 \cdot \ve_{3-}) z} \right) \cdot \ve_{1} \right]_{\text{Rep}(s)=\text{CA}} &:= \left\{ \half \left( \la \mathbf{21} \ra + \frac{x \la \mathbf{2}3 \ra \la 3\mathbf{1} \ra}{mz}\right) + \frac{[ \mathbf{21} ]}{2} \right\}^{2s} \,.
\eld \label{eq:CArep}
\el
Another example is polarisation tensors with Lorentz indices, which are usually used for integer spin particles. This is the sequence of representations used in refs.\cite{Bern:2020buy,Kosmopoulos:2021zoq}. In spinor-helicity variables the representation would translate into the expressions
\bl
\bld
\left[ \ve_{2} \cdot \exp \left( -i \frac{k_3^\m \ve_{3+}^\n J_{\m\n}}{(p_1 \cdot \ve_{3+}) z} \right) \cdot \ve_{1} \right]_{\text{Rep}(s)=\text{PT}} &:= \left\{ \la \mathbf{21} \ra \left( [\mathbf{21}] + \frac{[\mathbf{2}3][3\mathbf{1}]}{mxz}\right) \right\}^{s} \,,
\\ \left[ \ve_{2} \cdot \exp \left( -i \frac{k_3^\m \ve_{3-}^\n J_{\m\n}}{(p_1 \cdot \ve_{3-}) z} \right) \cdot \ve_{1} \right]_{\text{Rep}(s)=\text{PT}} &:= \left\{ \left( \la \mathbf{21} \ra + \frac{x \la \mathbf{2}3 \ra \la 3\mathbf{1} \ra}{mz}\right) [\mathbf{21}] \right\}^{s} \,.
\eld \label{eq:PTrep}
\el
As can be seen from \eqc{eq:CArep} and \eqc{eq:PTrep}, the representations lead to different expressions for the amplitude.
However, the amplitude \eqc{eq:3ptresint} for the two representations are the same in the $s \to \infty$ limit as expected, in the sense that the $g_i$ coefficients of the standard expansion for coupling to a positive helicity massless particle~\cite{Arkani-Hamed:2017jhn}
\bl
M_3^{+\abs{h}} &= x^{+\abs{h}} \left[ g_0 \la \mathbf{21} \ra^{2s} + g_1 \frac{x \la \mathbf{2}3 \ra \la 3\mathbf{1} \ra}{m} \la \mathbf{21} \ra^{2s-1} + \cdots + g_{2s} \left( \frac{x \la \mathbf{2}3 \ra \la 3\mathbf{1} \ra}{m} \right)^{2s} \right]\,,
\el
computed using \eqc{eq:3ptresint} has the same asymptotic behaviour. The $g_i$ coefficients for the chirally averaged representation \eqc{eq:CArep} are given as
\bl
\bld
\left[ \ve_{2} \cdot \exp \left( -i \frac{k_3^\m \ve_{3+}^\n J_{\m\n}}{(p_1 \cdot \ve_{3+}) z} \right) \cdot \ve_{1} \right]_{\text{Rep}(s)=\text{CA}} &= \left( \la \mathbf{21} \ra + \frac{z-1}{2z} \frac{x \la \mathbf{2}3 \ra \la 3\mathbf{1} \ra}{m} \right)^{2s} \,,
\\ \therefore g_{i,\text{CA}} &= \frac{1}{2^i} {2s \choose i} \sum_{n=0}^{\infty} (-1)^n {i \choose n} C_{\text{S}^n} \,,
\eld
\el
where the first line was inserted into \eqc{eq:3ptresint} to obtain the second line. For the polarisation tensors \eqc{eq:PTrep} the same computation leads to
\bl
\bld
\left[ \ve_{2} \cdot \exp \left( -i \frac{k_3^\m \ve_{3+}^\n J_{\m\n}}{(p_1 \cdot \ve_{3+}) z} \right) \cdot \ve_{1} \right]_{\text{Rep}(s)=\text{PT}} &= \la \mathbf{21} \ra^s \left( \la \mathbf{21} \ra + \frac{z-1}{z} \frac{x \la \mathbf{2}3 \ra \la 3\mathbf{1} \ra}{m} \right)^{s} \,,
\\ \therefore g_{i,\text{PT}} &= {s \choose i} \sum_{n=0}^{\infty} (-1)^n {i \choose n} C_{\text{S}^n} \,.
\eld
\el
While the coefficients $g_{i,\text{CA}}$ and $g_{i,\text{PT}}$ are distinct, their ratios converge to unity in the limit $s \to \infty$.
\bl
\frac{g_{i,\text{CA}}}{g_{i,\text{PT}}} &= \frac{(s-1/2)(s-1)\cdots(s-(i-1)/2)}{(s-1)(s-2)\cdots(s-i+1)} = 1 + \frac{i(i-1)}{4s} + \CO (s^{-2}) \,.
\el
Therefore the two sequences of reprentations can be considered equivalent in the classical limit.
We use the equality $\stackrel{\cdot}{=}$ for the amplitudes in the following sections and omit unnecessary details.

\subsection{BCFW construction of Compton amplitude (opposite helicity)} \label{app:mixedComp}
We use the set-up described in section \ref{sec:mixhelComp} to construct the Compton amplitude. For the $s$-channel, the product of spin factors is given by inserting shifted variables into \eqc{eq:SFprodPM}
\bg
\bgd
\exp \left( i \frac{\hat{k}_3^\m \hat{\ve}_{3-}^\n J_{\m\n}}{(\hat{P} \cdot \hat{\ve}_{3-}) z_4} \right) \exp \left( i \frac{\hat{k}_2^\m \hat{\ve}_{2+}^\n J_{\m\n}}{(\hat{P} \cdot \hat{\ve}_{2+}) z_1} \right) = \exp \left( - i K^\m L^\n J_{\m\n} \right)\,,
\\ K^\m = \frac{k_2^\m}{z_1} + \frac{k_3^\m}{z_4} \,,~~~~~~~\, L^\m = \frac{\bra{3} \s^\m \sket{2}}{\bra{3} p_1 \sket{2}} \,.
\egd \label{eq:BCFWspinFgenWC}
\eg
The bookkeeping parameter $z_1$ is associated with leg 1 and $z_4$ is associated with leg 4. The $s$-channel propagator pole is absorbed by the scalar factor as in section \ref{sec:mixhelComp}. Similar to the minimal coupling case, the $u$-channel spin factor turns out to be the same.
\bg
\bgd
\exp \left( - i \frac{\hat{k'}_2^\m \hat{\ve'}_{2+}^\n J_{\m\n}}{(\hat{P'} \cdot \hat{\ve'}_{2+}) z_1} \right) \exp \left( - i \frac{\hat{k'}_3^\m \hat{\ve'}_{3-}^\n J_{\m\n}}{(\hat{P'} \cdot \hat{\ve'}_{3-}) z_4} \right) = \exp \left( - i K^\m L^\n J_{\m\n} \right)\,.
\egd
\eg
Note that the bookkeeping parameters $z_1$ and $z_4$ have been exchanged; $z_1$ is the parameter associated with leg 4 and $z_4$ is the parameter associated with leg 1. Adding up all contributions, the 4pt amplitude $M_4(p_1,k_2^+,k_3^-,p_4)$ is given as
\bl
M_4^{s} &\stackrel{\cdot}{=} M_4^{0} \oint \prod_{i=1,4} \frac{dz_i}{2\pi i z_i} \left( \sum_m C_{\text{S}^m} z_i^m \right) \exp \left( -i K^\m L^\n J_{\m\n} \right) \,, \nn
\\ M_4^{0} &= \left\{ \begin{aligned}
- \frac{\sbra{2}p_1\ket{3}^2}{(s-m^2)(u-m^2)}\,, && \abs{h} = 1\,, \\
\frac{\sbra{2}p_1\ket{3}^4}{(s-m^2)t(u-m^2)}\,, && \abs{h} = 2\,,
\end{aligned} \right. \label{eq:BCFWmixComp}
\el
with the relevant definitions in \eqc{eq:BCFWspinFgenWC}. This expression should be understood as a sum over symmetric polynomials of $J$ with coefficients determined by residue integral of bookkeeping parameters $z_1$ and $z_4$. In other words, the exponential factor containing $J$ should be expanded before performing the auxiliary residue integrals. Note that for minimal coupling the residue integrals localise to $z_1 = z_4 = 1$, yielding \eqc{eq:BCFWmixCompmin} upon resummation. 

An important feature of the amplitude \eqc{eq:BCFWmixComp} is that all physical channel residues including the $t$-channel for graviton coupling are correctly captured to all orders in $J$, just like the minimally coupled case \eqc{eq:BCFWmixCompmin}. Also, the order of $J$ at which spurious poles start to appear is the same for \eqc{eq:BCFWmixComp} and \eqc{eq:BCFWmixCompmin}.

\subsection{BCFW construction of Compton amplitude (same helicity)} \label{app:sameComp}
We construct the 4pt amplitude $M_4(p_1,k_2^+,k_3^+,p_4)$ using BCFW recursion in two ways; i) the $\la 23 ]$ shift used in section \ref{sec:mixhelComp}, and ii) the other viable shift, $\sket{2} \to \sket{2} + w\sket{3}$ and $\ket{3} \to \ket{3}-w\ket{2}$ (the $[23\ra$ shift), which we only report the final outcome. The reason we include both shifts is because they lead to \emph{different} expressions, in contrast to minimal coupling case.

For the $\la 23 ]$ shift, the $s$-channel spin factor is given as
\bl
\bgd
\exp \left( i \frac{\hat{k}_3^\m \hat{\ve}_{3+}^\n J_{\m\n}}{(\hat{P} \cdot \hat{\ve}_{3+}) z_4} \right) \exp \left( i \frac{\hat{k}_2^\m \hat{\ve}_{2+}^\n J_{\m\n}}{(\hat{P} \cdot \hat{\ve}_{2+}) z_1} \right) = \exp \left( - \frac{i}{2} b_s^{\m\n} J_{\m\n} \right)\,,
\\ b_s^{\m\n} = 4 f \left( \frac{t/m^2}{z_1 z_4} \right) \left[ a_1 \left( k_2^{[\m} k_3^{\n]} + l^{[\m} \bar{l}^{\n]} \right) + a_2 k_2^{[\m} \bar{l}^{\n]} + a_3k_3^{[\m} l^{\n]} \right] \,,\,
\\ l^\m = \frac{\bra{2} \s^\m \sket{3}}{2} \,,\, \bar{l}^\m = \frac{\bra{3} \s^\m \sket{2}}{2} \,,
\\ a_1 = \frac{s-m^2}{t m^2 z_4} + \frac{1}{2m^2 z_1 z_4} \,,\, a_3 = -\frac{\sbra{2}p_1\ket{3}}{t m^2 z_4} \,,
\\ a_2 = \frac{1}{\sbra{2}p_1\ket{3} z_1} + \frac{(s-m^2)^2}{t \sbra{2}p_1\ket{3} m^2 z_4} + \frac{s-m^2}{\sbra{2}p_1\ket{3} m^2 z_1 z_4} \,,
\egd \label{eq:BCFWsamehelSFschan}
\el
where the function $f(x)$ is given in \eqc{eq:ffactor}.  
For minimal coupling the bookkeeping parameters localise to $z_1 = z_4 = 1$ and reduces to \eqc{eq:BCFWsamehelSFschanmin}. The spin factor composition \eqc{eq:BCFWsamehelSFschan} follows from \eqc{eq:sameheltent3} with shifted momenta $\hat{k}_2$ and $\hat{k}_3$, where $z_i^{-1}$ are used as formal expansion parameters\footnote{Although the integration contours for $z_i$ are small circles encircling the origin, the formal expansion is justified because residue integrals in $z_i$ are bookkeeping devices and have no physical content.}. The $u$-channel spin factor is given as
\bl
\bgd
\exp \left( - i \frac{\hat{k'}_2^\m \hat{\ve'}_{2+}^\n J_{\m\n}}{(\hat{P'} \cdot \hat{\ve'}_{2+}) z_1} \right) \exp \left( - i \frac{\hat{k'}_3^\m \hat{\ve'}_{3+}^\n J_{\m\n}}{(\hat{P'} \cdot \hat{\ve'}_{3+}) z_4} \right) = \exp \left( - \frac{i}{2} b_u^{\m\n} J_{\m\n} \right)\,,
\\ b_u^{\m\n} = 4 f\left( \frac{t/m^2}{z_1 z_4} \right) \left[ a_4 \left( k_2^{[\m} k_3^{\n]} + l^{[\m} \bar{l}^{\n]} \right) + a_5 k_2^{[\m} \bar{l}^{\n]} + a_6 k_3^{[\m} l^{\n]} \right] \,,\,
\\ a_4 = - \frac{u-m^2}{t m^2 z_4} - \frac{1}{2m^2 z_1 z_4} \,,\, a_6 = -\frac{\sbra{2}p_1\ket{3}}{t m^2 z_4} \,,
\\ a_5 = \frac{1}{\sbra{2}p_1\ket{3} z_1} + \frac{(u-m^2)^2}{t \sbra{2}p_1\ket{3} m^2 z_4} + \frac{u-m^2}{\sbra{2}p_1\ket{3} m^2 z_1 z_4} \,,
\egd
\el
which is \emph{different} from the $s$-channel spin factor \eqc{eq:BCFWsamehelSFschan}, unless we impose the minimal coupling condition $z_1 = z_4 = 1$. This difference does not allow a complete factorisation of the amplitude into the scalar factor and the spin factor as in \eqc{eq:samehelBCFWmin} or \eqc{eq:BCFWmixComp}.

It also turns out that the $[23\ra$ shift leads to \emph{different} spin factors, unless we impose the minimal coupling condition. Summing up and reorganising the expression to look as symmetric as possible, we get the following expression for $M_4(p_1,k_2^+,k_3^+,p_4)$ using BCFW construction.
\bl
\bld
M_4^s &\stackrel{\cdot}{=} - \left(- \frac{ [23]^2 }{ t } \right)^{\abs{h}} \oint \prod_{i=1,4} \frac{dz_i}{2\pi i z_i} \left( \sum_m C_{\text{S}^m} z_i^m \right) \left[ \frac{e^{- \frac{i}{2} b_s^{\m\n} J_{\m\n}}}{s-m^2} + \frac{e^{- \frac{i}{2} b_u^{\m\n} J_{\m\n}}}{u-m^2} \right] \,,
\\ b_s^{\m\n} &= 4 f\left( \frac{t/m^2}{z_1 z_4} \right) \left[ a_1 \left( k_2^{[\m} k_3^{\n]} + l^{[\m} \bar{l}^{\n]} \right) + a_2 k_2^{[\m} \bar{l}^{\n]} + a_3k_3^{[\m} l^{\n]} \right] \,,
\\ b_u^{\m\n} &= 4 f\left( \frac{t/m^2}{z_1 z_4} \right) \left[ a_4 \left( k_2^{[\m} k_3^{\n]} + l^{[\m} \bar{l}^{\n]} \right) + a_5 k_2^{[\m} \bar{l}^{\n]} + a_6 k_3^{[\m} l^{\n]} \right] \,,
\eld \label{eq:samehelBCFW}
\el
where $l^\m$ and $\bar{l}^\m$ are given in \eqc{eq:BCFWsamehelSFschan}, $f(x)$ is given in \eqc{eq:ffactor}, and $a_i$ coefficients are
\bl
\bld
a_1 &= \frac{1}{m^2} \left[ \frac{s-m^2}{t} \left( \frac{x_1}{z_4} + \frac{x_2}{z_1} \right) + \frac{1}{2 z_1 z_4} \right] \,,
\\ a_2 &= - \frac{\sbra{3}p_1\ket{2}}{m^2 t} \left( \frac{x_1}{z_4} + \frac{x_2}{z_1} \right) + \frac{x_1}{\sbra{2}p_1\ket{3}} \left( \frac{s-m^2}{m^2 z_1 z_4} + \frac{1}{z_1} - \frac{s}{m^2} \frac{1}{z_4} \right) \,,
\\ a_3 &= - \frac{\sbra{2}p_1\ket{3}}{m^2 t} \left( \frac{x_1}{z_4} + \frac{x_2}{z_1} \right) + \frac{x_2}{\sbra{3}p_1\ket{2}} \left( \frac{s-m^2}{m^2 z_1 z_4} + \frac{1}{z_4} - \frac{s}{m^2} \frac{1}{z_1} \right) \,,
\\ a_4 &= - \frac{1}{m^2} \left[ \frac{u-m^2}{t} \left( \frac{x_1}{z_4} + \frac{x_2}{z_1} \right) + \frac{1}{2 z_1 z_4} \right] \,,
\\ a_5 &= - \frac{\sbra{3}p_1\ket{2}}{m^2 t} \left( \frac{x_1}{z_4} + \frac{x_2}{z_1} \right) + \frac{x_1}{\sbra{2}p_1\ket{3}} \left( \frac{u-m^2}{m^2 z_1 z_4} + \frac{1}{z_1} - \frac{u}{m^2} \frac{1}{z_4} \right) \,,
\\ a_6 &= - \frac{\sbra{2}p_1\ket{3}}{m^2 t} \left( \frac{x_1}{z_4} + \frac{x_2}{z_1} \right) + \frac{x_2}{\sbra{3}p_1\ket{2}} \left( \frac{u-m^2}{m^2 z_1 z_4} + \frac{1}{z_4} - \frac{u}{m^2} \frac{1}{z_1} \right) \,,
\eld \label{eq:samehelBCFWconj2}
\el
where parameter choice $x_1 = 1$ and $x_2 = 0$ corresponds to the $\la 23 ]$ shift, while the choice $x_1 = 0$ and $x_2 = 1$ corresponds to the $[23\ra$ shift. Similar to the opposite helicity case \eqc{eq:BCFWmixComp}, the exponential factor containing $J$ should be expanded before performing the auxiliary residue integrals. Note that for minimal coupling the residue integrals localise to $z_1 = z_4 = 1$ and the expression resums to \eqc{eq:samehelBCFWmin}. We make the following remarks regarding the expression \eqc{eq:samehelBCFW}:
\bn
\item All physical channel residues including the $t$-channel for graviton coupling are correctly captured by this expression to all orders in $J$, subject to the BCFW condition $x_1 + x_2 = 1$.
\item Unlike minimal coupling case \eqc{eq:samehelBCFWmin} nor the opposite helicity case \eqc{eq:BCFWmixComp}, the expression \eqc{eq:samehelBCFW} does not factorise into scalar and spin factors.
\item Unlike the opposite helicity case \eqc{eq:BCFWmixComp}, the expression \eqc{eq:samehelBCFW} develops spurious poles even at linear order in $J$ (quadratic order when considering the condition $C_{\text{S}^1} = 1$), \emph{unless} we impose the minimal coupling condition $z_1 = z_4 = 1$ which removes all spurious poles as in \eqc{eq:samehelBCFWmin}.
\item Similar to the Kerr case, the full form of $f(x)$ factor is only relevant for the classical spin limit \eqc{eq:cslimdef} and the factor can be simplified to $f(x) = 1 + \CO(\hbar^2)$ when computing the classical limit of the integral coefficients. This limit was used at the integrand level when computing the coefficients given in appedix \ref{app:trigcoeff}.
\item When computing triangle integral coefficients, the classical contributions seem to vanish when $(J)^{n \ge 2}$-order expansion term of \eqc{eq:samehelBCFW} is involved, subject to the condition $x_1 + x_2 = 1$. This behaviour was observed in computation of $c_{\bigtriangleup, \pm\pm}^{S_1^0 S_2^2}$ and $c_{\bigtriangleup, \pm\pm}^{S_1^1 S_2^2}$, the notations following that of appendix \ref{app:trigcoeff}.
\en
Based on the last observation, we expect spurious pole cancellation terms to play an important role for classical scattering beyond binary black holes. Systematic removal of spurious poles will be treated in a future work.

\subsection{Partial results for integral coefficients in the classical limit}
The results presented in this section do not contain the Hilbert space matching factor. The integral coefficients were evaluated using the cut parametrisations of ref.\cite{Bern:2020buy} summarised in section \ref{sec:IntCoeff}. The notation for scalar integrals follow that of section \ref{sec:IntCoeff}.

\subsubsection{Box integral coefficients}
The integral coefficients are normalised by $i M = (16 \pi G)^2 c_{\,\square} \CI_{\,\square} + \cdots$ where elipsis denotes triangle and quantum contributions. The coefficients are evaluated to the leading $\hbar$ order, which is sufficient for checking IR divergence cancellation and eikonal exponentiation. The orthogonal vector $n^\m$ is defined as $n^\m = \e^{\a\b\g\m}p_{1\a} p_{2\b} q_{\g}$.
\bl
c_{\square}^{S_1^0 S_2^0} &= m_1^4 m_2^4 \left(1-2 \sigma ^2\right)^2\,,
\\ c_{\square}^{S_1^1 S_2^0} &= - 2 i m_1^2 m_2^3 \sigma  \left(2 \sigma ^2-1\right) n\cdot S_1\,,
\\ c_{\square}^{S_1^2 S_2^0} &= \frac{1}{4} m_2^2 \left(m_1^2 m_2^2 \left(\left(1-2 \sigma ^2\right)^2 C_{\text{S}_1^2}-4 \sigma
   ^4+4 \sigma ^2\right) \left(q\cdot S_1\right){}^2 \phantom{+ \frac{\left(-\left(\sigma
   ^2\right)^2 \right)}{\sigma
   ^2-1}}\right. \nn
   \\ &\phantom{=} \left.\phantom{asdfasdfasdfasdf}+\frac{\left(-\left(1-2 \sigma
   ^2\right)^2 C_{\text{S}_1^2}-4 \sigma ^4+4 \sigma ^2\right) \left(n\cdot S_1\right){}^2}{\sigma
   ^2-1}\right)\,,
\\ c_{\square}^{S_1^1 S_2^1} &= \frac{1}{2} m_1 m_2 \left(m_1^2 m_2^2 q\cdot S_1 q\cdot S_2+\frac{\left(-8 \sigma ^4+8
   \sigma ^2-1\right) n\cdot S_1 n\cdot S_2}{\sigma ^2-1}\right)\,,
\\ c_{\square}^{S_1^3 S_2^0} &= \frac{1}{4} i m_2^3 \sigma  \left(2 \sigma ^2-1\right) \left(C_{\text{S}_1^2}-C_{\text{S}_1^3}\right)
   n\cdot S_1 \left(q\cdot S_1\right){}^2 \nn
   \\ &\phantom{=asdfasdf} + \frac{i m_2 \sigma  \left(2 \sigma ^2-1\right) \left(3 C_{\text{S}_1^2}+C_{\text{S}_1^3}\right)
   \left(n\cdot S_1\right){}^3}{12 m_1^2 \left(\sigma ^2-1\right)}\,,
\\ c_{\square}^{S_1^2 S_2^1} &= -\frac{1}{2} i m_1 m_2^2 \sigma  \left(2 \sigma ^2-1\right) \left(C_{\text{S}_1^2}-1\right) n\cdot
   S_2 \left(q\cdot S_1\right){}^2 \nn
   \\ &\phantom{=asdfasdf} + \frac{i \sigma  \left(2 \sigma ^2-1\right) \left(C_{\text{S}_1^2}+1\right) \left(n\cdot
   S_1\right){}^2 n\cdot S_2}{2 m_1 \left(\sigma ^2-1\right)}\,,
\\ c_{\square}^{S_1^4 S_2^0} &= \frac{1}{192} m_2^4 \left(-16 \left(\sigma ^2-1\right) \sigma ^2 C_{\text{S}_1^3}+3 \left(1-2
   \sigma ^2\right)^2 C_{\text{S}_1^2}^2+\left(1-2 \sigma ^2\right)^2 C_{\text{S}_1^4}\right) \left(q\cdot
   S_1\right){}^4 \nn
   \\ &\phantom{=as} + \frac{\left(4 \left(\sigma ^2-1\right) \sigma ^2 \left(4 C_{\text{S}_1^3}+C_{\text{S}_1^4}\right)+3
   \left(1-2 \sigma ^2\right)^2 C_{\text{S}_1^2}^2+C_{\text{S}_1^4}\right) \left(n\cdot
   S_1\right){}^4}{192 m_1^4 \left(\sigma ^2-1\right)^2} \nn
   \\ &\phantom{=asdfas} + \frac{m_2^2 \left(1-2 \sigma ^2\right)^2 \left(C_{\text{S}_1^2}^2-C_{\text{S}_1^4}\right) \left(n\cdot
   S_1\right){}^2 \left(q\cdot S_1\right){}^2}{32 m_1^2 \left(\sigma ^2-1\right)}\,,
\\ c_{\square}^{S_1^3 S_2^1} &= \frac{1}{48} m_1 m_2^3 \left(3 C_{\text{S}_1^2}+C_{\text{S}_1^3}\right) \left(q\cdot S_1\right){}^3
   q\cdot S_2  + \frac{m_2 \left(C_{\text{S}_1^2}-C_{\text{S}_1^3}\right) \left(n\cdot S_1\right){}^2 q\cdot S_1 q\cdot
   S_2}{16 m_1 \left(\sigma ^2-1\right)} \nn
   \\ &\phantom{=asdfas} + \frac{m_2 \left(8 \sigma ^4-8 \sigma ^2+1\right) \left(C_{\text{S}_1^2}-C_{\text{S}_1^3}\right) n\cdot S_1
   n\cdot S_2 \left(q\cdot S_1\right){}^2}{16 m_1 \left(\sigma ^2-1\right)} \nn
   \\ &\phantom{=asdfasdf} + \frac{\left(8 \sigma ^4-8 \sigma ^2+1\right) \left(3 C_{\text{S}_1^2}+C_{\text{S}_1^3}\right) \left(n\cdot
   S_1\right){}^3 n\cdot S_2}{48 m_1^3 m_2 \left(\sigma ^2-1\right)^2}\,,
\\ c_{\square}^{S_1^2 S_2^2} &= \frac{ m_1^2 m_2^2 }{16} \left(-4 \left(\sigma ^2-1\right) \sigma ^2 C_{\text{S}_2^2}+C_{\text{S}_1^2}
   \left(\left(1-2 \sigma ^2\right)^2 C_{\text{S}_2^2}-4 \sigma ^4+4 \sigma ^2\right)+\left(1-2
   \sigma ^2\right)^2\right)\nn
   \\ &\phantom{=asdf} \times \left(q\cdot S_1\right)^2 \left(q\cdot S_2\right)^2 \nn
   \\ &\phantom{=} + \frac{ 4 \left(\sigma ^2-1\right) \sigma ^2 C_{\text{S}_2^2}+C_{\text{S}_1^2} \left(-\left(1-2 \sigma
   ^2\right)^2 C_{\text{S}_2^2}-4 \sigma ^4+4 \sigma ^2\right)+\left(1-2 \sigma ^2\right)^2 
   }{16 \left(\sigma ^2-1\right)}\nn
   \\ &\phantom{=asdf} \times \left(n\cdot S_2\right){}^2 \left(q\cdot S_1\right){}^2 \nn
   \\ &\phantom{=} + \frac{ 4 \left(\sigma ^2-1\right) \sigma ^2 C_{\text{S}_1^2}-C_{\text{S}_2^2} \left(\left(1-2 \sigma
   ^2\right)^2 C_{\text{S}_1^2}+4 \left(\sigma ^2-1\right) \sigma ^2\right)+\left(1-2 \sigma
   ^2\right)^2 }{16
   \left(\sigma ^2-1\right)}\nn
   \\ &\phantom{=asdf} \times  \left(n\cdot S_1\right){}^2 \left(q\cdot S_2\right){}^2\nn
   \\ &\phantom{=} + \frac{\left(\left(8 \sigma ^4-8 \sigma ^2+1\right) \left(C_{\text{S}_1^2}+1\right)
   \left(C_{\text{S}_2^2}+1\right)+\left(C_{\text{S}_1^2}-1\right) \left(C_{\text{S}_2^2}-1\right)\right)
   }{32 m_1^2 m_2^2 \left(\sigma
   ^2-1\right)^2}\nn
   \\ &\phantom{=asdf} \times \left(n\cdot S_1\right){}^2 \left(n\cdot S_2\right){}^2\,.
\el

\subsubsection{Triangle integral coefficients} \label{app:trigcoeff}
The integral coefficients are normalised by $i M = (16 \pi G)^2 c_{\bigtriangleup} \CI_{\bigtriangleup} + \cdots$ where elipsis denotes other contributions. The part of the Compton amplitude eqs.(\ref{eq:samehelBCFW}-\ref{eq:samehelBCFWconj2}) needed for the $S_1^n S_2^0$ and $S_1^n S_2^1$ sectors do not have spurious poles, and we compute full results for those sectors. The orthogonal vector $n^\m$ is defined as $n^\m = \e^{\a\b\g\m}p_{1\a} p_{2\b} q_{\g}$. Only the sectors containing nontrivial Wilson coefficients $C_{\text{S}^{n \ge 2}}$ are shown.
\bl
c_{\bigtriangleup}^{S_1^2 S_2^0} &= \frac{\left(C_{\text{S}_1^2}+1\right) \left(m_1^2 q^2 \left(35 \sigma ^4-30 \sigma ^2+3\right)
   \left(p_2\cdot S_1\right){}^2+\left(-95 \sigma ^4+102 \sigma ^2-15\right) \left(n\cdot
   S_1\right){}^2\right)}{64 \left(\sigma ^2-1\right)^2} \nn
   \\ &\phantom{=} + \frac{1}{16} \left(C_{\text{S}_1^2}-1\right) \left(3 m_1^2 m_2^2 \left(5 \sigma ^2-1\right)
   \left(q\cdot S_1\right){}^2-\frac{2 \left(\left(n\cdot S_1\right){}^2-m_1^2 q^2
   \left(p_2\cdot S_1\right){}^2\right)}{\left(\sigma ^2-1\right)^2}\right) \label{eq:tricoeff_2-0}\,,
\\ c_{\bigtriangleup}^{S_1^3 S_2^0} &= \frac{i \sigma  \left(3 C_{\text{S}_1^2}+C_{\text{S}_1^3}\right) n\cdot S_1 \left(m_1^2 q^2 \left(3-7
   \sigma ^2\right) \left(p_2\cdot S_1\right){}^2+\left(9 \sigma ^2-5\right) \left(n\cdot
   S_1\right){}^2\right)}{32 m_1^2 m_2 \left(\sigma ^2-1\right)^2} \nn
   \\ &\phantom{=asdfasdf} + \frac{i m_2 \sigma  \left(5 \sigma ^2-3\right) \left(C_{\text{S}_1^2}-C_{\text{S}_1^3}\right) n\cdot S_1
   \left(q\cdot S_1\right){}^2}{8 \left(\sigma ^2-1\right)}\,, \label{eq:tricoeff_3-0}
\\ c_{\bigtriangleup}^{S_1^4 S_2^0} &= \left(3 C_{\text{S}_1^2}^2+4 C_{\text{S}_1^3}+C_{\text{S}_1^4}\right) \left(-\frac{q^2 \left(49 \sigma ^4-46
   \sigma ^2+5\right) \left(n\cdot S_1\right){}^2 \left(p_2\cdot S_1\right){}^2}{1024 m_1^2
   m_2^2 \left(\sigma ^2-1\right)^3}\right. \nn
   \\ &\phantom{=asdf} \left.\phantom{}-\frac{\left(-239 \sigma ^4+250 \sigma ^2-35\right)
   \left(n\cdot S_1\right){}^4}{6144 m_1^4 m_2^2 \left(\sigma ^2-1\right)^3}-\frac{q^4
   \left(21 \sigma ^4-14 \sigma ^2+1\right) \left(p_2\cdot S_1\right){}^4}{6144 m_2^2
   \left(\sigma ^2-1\right)^3}\right) \nn
   \\ &\phantom{=} + \left(3 C_{\text{S}_1^2}^2-4 C_{\text{S}_1^3}+C_{\text{S}_1^4}\right) \left(-\frac{q^2 \left(n\cdot
   S_1\right){}^2 \left(p_2\cdot S_1\right){}^2}{128 m_1^2 m_2^2 \left(\sigma
   ^2-1\right)^3}+\frac{\left(n\cdot S_1\right){}^4}{256 m_1^4 m_2^2 \left(\sigma
   ^2-1\right)^3}\right. \nn
   \\ &\phantom{=asdf} \left.\phantom{asdf}-\frac{q^4 \left(p_2\cdot S_1\right){}^4}{768 m_2^2 \left(\sigma
   ^2-1\right)^3}+\frac{1}{256} m_2^2 \left(5 \sigma ^2-1\right) \left(q\cdot
   S_1\right){}^4\right) \nn
   \\ &\phantom{=} + \left(C_{\text{S}_1^2}^2-C_{\text{S}_1^4}\right) \left(\frac{\left(95 \sigma ^4-102 \sigma ^2+23\right)
   \left(n\cdot S_1\right){}^2 \left(q\cdot S_1\right){}^2}{512 m_1^2 \left(\sigma
   ^2-1\right)^2}\right. \nn
   \\ &\phantom{=asdf} \left.\phantom{asdf}+\frac{q^2 \left(-35 \sigma ^4+30 \sigma ^2-11\right) \left(p_2\cdot
   S_1\right){}^2 \left(q\cdot S_1\right){}^2}{512 \left(\sigma ^2-1\right)^2}\right) \,,
\\ c_{\bigtriangleup}^{S_1^2 S_2^1} &= \left(C_{\text{S}_1^2}+1\right) \left(-\frac{3 i q^2 \left(1-5 \sigma ^2\right) n\cdot S_1 p_1\cdot
   S_2 p_2\cdot S_1}{32 m_2 \left(\sigma ^2-1\right)^2}\right. \nn
   \\ &\phantom{=asdf} \left. \phantom{asdf}-\frac{5 i m_1 q^2 \sigma  \left(7
   \sigma ^2-3\right) n\cdot S_2 \left(p_2\cdot S_1\right){}^2}{64 m_2^2 \left(\sigma
   ^2-1\right)^2}-\frac{i \sigma  \left(51-95 \sigma ^2\right) \left(n\cdot S_1\right){}^2
   n\cdot S_2}{64 m_1 m_2^2 \left(\sigma ^2-1\right)^2}\right) \nn
   \\ &\phantom{=asdf} -\frac{3 i m_1 \sigma  \left(5 \sigma ^2-3\right) \left(C_{\text{S}_1^2}-1\right) n\cdot S_2
   \left(q\cdot S_1\right){}^2}{16 \left(\sigma ^2-1\right)}\,, \label{eq:tricoeff_2-1}
\\ c_{\bigtriangleup}^{S_1^3 S_2^1} &= \left(3 C_{\text{S}_1^2}+C_{\text{S}_1^3}\right) \left(\frac{q^2 \sigma  \left(13 \sigma ^2-9\right)
   \left(n\cdot S_1\right){}^2 p_1\cdot S_2 p_2\cdot S_1}{128 m_1^2 m_2^2 \left(\sigma
   ^2-1\right)^3} \right. \nn
   \\ &\phantom{=as} \left. \phantom{as}+\frac{q^4 \sigma  \left(7 \sigma ^2-3\right) p_1\cdot S_2
   \left(p_2\cdot S_1\right){}^3}{384 m_2^2 \left(\sigma ^2-1\right)^3}+\frac{\left(36
   \sigma ^4-37 \sigma ^2+5\right) \left(n\cdot S_1\right){}^3 n\cdot S_2}{128 m_1^3 m_2^3
   \left(\sigma ^2-1\right)^3} \right. \nn
   \\ &\phantom{=as} \left. \phantom{as}-\frac{q^2 \left(28 \sigma ^4-27 \sigma ^2+3\right) n\cdot S_1 n\cdot S_2
   \left(p_2\cdot S_1\right){}^2}{128 m_1 m_2^3 \left(\sigma ^2-1\right)^3}\right) \nn
   \\ &\phantom{=as} + \left(C_{\text{S}_1^2}-C_{\text{S}_1^3}\right) \left(\frac{\left(20 \sigma ^4-21 \sigma ^2+3\right) n\cdot
   S_1 n\cdot S_2 \left(q\cdot S_1\right){}^2}{32 m_1 m_2 \left(\sigma
   ^2-1\right)^2}\right. \nn
   \\ &\phantom{=asdf} \left. \phantom{asdf}+\frac{q^2 \sigma  \left(5 \sigma ^2-3\right) p_1\cdot S_2 p_2\cdot S_1
   \left(q\cdot S_1\right){}^2}{32 \left(\sigma ^2-1\right)^2} \right. \nn
   \\ &\phantom{=asdf} \left. \phantom{asdf} + \frac{q\cdot S_1 q\cdot S_2 \left(\left(n\cdot S_1\right){}^2-m_1^2 q^2 \left(p_2\cdot
   S_1\right){}^2\right)}{16 m_1 m_2 \left(\sigma ^2-1\right)^2}\right)\,, \label{eq:tricoeff_3-1}
\el
The triangle coefficient \eqc{eq:tricoeff_2-0} matches the corresponding factors in eq.(3.7) of ref.\cite{Kosmopoulos:2021zoq} up to Levi-Civita identities. Other sectors \emph{are} affected by the spurious poles of eqs.(\ref{eq:samehelBCFW}-\ref{eq:samehelBCFWconj2}), and only the spurious pole free contributions from opposite helicity graviton exchange channels $c_{\bigtriangleup, \pm\mp}^{S_1^m S_2^n} = c_{\bigtriangleup, +-}^{S_1^m S_2^n} + c_{\bigtriangleup, -+}^{S_1^m S_2^n}$ are reliably computable.
\bl
c_{\bigtriangleup, \pm\mp}^{S_1^0 S_2^2} &= -\frac{m_1^2 \left(15 \sigma ^4-15 \sigma ^2+2\right) \left(C_{\text{S}_2^2}+1\right) \left(n\cdot
   S_2\right){}^2}{16 m_2^2 \left(\sigma ^2-1\right)^2} + \frac{3}{16} m_1^4 \left(5 \sigma ^2-1\right) \left(C_{\text{S}_2^2}-1\right) \left(q\cdot
   S_2\right){}^2\,, \label{eq:tricoeff_0-2}
\\ c_{\bigtriangleup, \pm\mp}^{S_1^0 S_2^3} &= -\frac{i m_1 \sigma  \left(3 C_{\text{S}_2^2}+C_{\text{S}_2^3}\right) n\cdot S_2 \left(3 m_2^2 q^2
   \left(p_1\cdot S_2\right){}^2+\left(2-5 \sigma ^2\right) \left(n\cdot
   S_2\right){}^2\right)}{32 m_2^4 \left(\sigma ^2-1\right)^2} \nn
   \\ &\phantom{=asdf} + \frac{3 i m_1^3 \sigma  \left(5 \sigma ^2-3\right) \left(C_{\text{S}_2^2}-C_{\text{S}_2^3}\right) n\cdot
   S_2 \left(q\cdot S_2\right){}^2}{32 m_2^2 \left(\sigma ^2-1\right)}\,,
\\ c_{\bigtriangleup, \pm\mp}^{S_1^1 S_2^2} &= \left(C_{\text{S}_2^2}+1\right) \left(-\frac{i q^2 \sigma  n\cdot S_1 \left(p_1\cdot
   S_2\right){}^2}{8 m_2 \left(\sigma ^2-1\right)^2}-\frac{i m_1 q^2 \left(1-5 \sigma
   ^2\right) n\cdot S_2 p_1\cdot S_2 p_2\cdot S_1}{8 m_2^2 \left(\sigma
   ^2-1\right)^2} \right. \nn
   \\ &\phantom{=asdfasdf} \left. \phantom{asdf}-\frac{i \sigma  \left(5-10 \sigma ^2\right) n\cdot S_1 \left(n\cdot
   S_2\right){}^2}{8 m_2^3 \left(\sigma ^2-1\right)^2}\right) \nn
   \\ &\phantom{=asdf} -\frac{i m_1^2 \sigma  \left(5 \sigma ^2-3\right) \left(C_{\text{S}_2^2}-1\right) n\cdot S_1
   \left(q\cdot S_2\right){}^2}{4 m_2 \left(\sigma ^2-1\right)}\,,
\\ c_{\bigtriangleup, \pm\mp}^{S_1^0 S_2^4} &= \left(3 C_{\text{S}_2^2}^2+4 C_{\text{S}_2^3}+C_{\text{S}_2^4}\right) \left(-\frac{\sigma ^2 \left(4-5 \sigma ^2\right) \left(n\cdot
   S_2\right){}^4}{256 m_2^6 \left(\sigma ^2-1\right)^3}-\frac{q^4 \left(p_1\cdot
   S_2\right){}^4}{768 m_2^2 \left(\sigma ^2-1\right)^3} \right. \nn
   \\ &\phantom{=asdfasdf} \left. \phantom{asdf}-\frac{q^2 \left(3 \sigma ^2-2\right)
   \left(n\cdot S_2\right){}^2 \left(p_1\cdot S_2\right){}^2}{128 m_2^4 \left(\sigma
   ^2-1\right)^3}\right)\nn
   \\ &\phantom{=} + \left(C_{\text{S}_2^2}^2-C_{\text{S}_2^4}\right) \left(\frac{m_1^2 \left(15 \sigma ^4-15 \sigma
   ^2+2\right) \left(n\cdot S_2\right){}^2 \left(q\cdot S_2\right){}^2}{128 m_2^4
   \left(\sigma ^2-1\right)^2}\right. \nn
   \\ &\phantom{=asdfasdf} \left. \phantom{asdf}+\frac{m_1^2 q^2 \left(1-3 \sigma ^2\right) \left(p_1\cdot
   S_2\right){}^2 \left(q\cdot S_2\right){}^2}{128 m_2^2 \left(\sigma ^2-1\right)^2}\right) \nn
   \\ &\phantom{=} + \frac{m_1^4 \left(5 \sigma ^2-1\right) \left(3 C_{\text{S}_2^2}^2-4 C_{\text{S}_2^3}+C_{\text{S}_2^4}\right)
   \left(q\cdot S_2\right){}^4}{256 m_2^2}\,,
\\ c_{\bigtriangleup, \pm\mp}^{S_1^1 S_2^3} &= \left(3 C_{\text{S}_2^2}+C_{\text{S}_2^3}\right) \left(\frac{\left(20 \sigma
   ^4-19 \sigma ^2+2\right) n\cdot S_1 \left(n\cdot S_2\right){}^3}{96 m_1 m_2^5
   \left(\sigma ^2-1\right)^3} +\frac{q^4 \sigma  \left(p_1\cdot S_2\right){}^3 p_2\cdot
   S_1}{96 m_2^2 \left(\sigma ^2-1\right)^3} \right. \nn
   \\ &\phantom{=as} \left. \phantom{}+\frac{q^2 \sigma  \left(5 \sigma ^2-4\right) \left(n\cdot S_2\right){}^2
   p_1\cdot S_2 p_2\cdot S_1}{32 m_2^4 \left(\sigma ^2-1\right)^3}+\frac{q^2 \left(1-2 \sigma ^2\right) n\cdot S_1
   n\cdot S_2 \left(p_1\cdot S_2\right){}^2}{32 m_1 m_2^3 \left(\sigma
   ^2-1\right)^3} \right) \nn
   \\ &\phantom{=} + \left(C_{\text{S}_2^2}-C_{\text{S}_2^3}\right) \left(\frac{m_1 \left(20 \sigma ^4-21 \sigma ^2+3\right)
   n\cdot S_1 n\cdot S_2 \left(q\cdot S_2\right){}^2}{32 m_2^3 \left(\sigma
   ^2-1\right)^2}\right. \nn
   \\ &\phantom{=asdf} \left. \phantom{asdf}+\frac{m_1^2 q^2 \sigma  \left(5 \sigma ^2-3\right) p_1\cdot S_2 p_2\cdot
   S_1 \left(q\cdot S_2\right){}^2}{32 m_2^2 \left(\sigma ^2-1\right)^2}\right)\,,
\\ c_{\bigtriangleup, \pm\mp}^{S_1^2 S_2^2} &= \left(C_{\text{S}_1^2}+1\right) \left(C_{\text{S}_2^2}+1\right) \left(\frac{\left(95 \sigma ^4-95 \sigma
   ^2+12\right) \left(n\cdot S_1\right){}^2 \left(n\cdot S_2\right){}^2}{256 m_1^2 m_2^4
   \left(\sigma ^2-1\right)^3} \right. \nn
   \\ &\phantom{=a} + \frac{q^2 \left(3-7 \sigma ^2\right)
   \left(n\cdot S_1\right){}^2 \left(p_1\cdot S_2\right){}^2}{256 m_1^2 m_2^2 \left(\sigma
   ^2-1\right)^3}+\frac{q^2 \sigma  \left(15 \sigma ^2-11\right) n\cdot S_1 n\cdot S_2
   p_1\cdot S_2 p_2\cdot S_1}{64 m_1 m_2^3 \left(\sigma ^2-1\right)^3}  \nn
   \\ &\phantom{=} \left. \phantom{a} +\frac{q^2 \left(-35
   \sigma ^4+35 \sigma ^2-4\right) \left(n\cdot S_2\right){}^2 \left(p_2\cdot
   S_1\right){}^2}{256 m_2^4 \left(\sigma ^2-1\right)^3}+\frac{q^4 \left(1-5 \sigma
   ^2\right) \left(p_1\cdot S_2\right){}^2 \left(p_2\cdot S_1\right){}^2}{256 m_2^2
   \left(\sigma ^2-1\right)^3}\right) \nn
   \\ &\phantom{=as} +\left(C_{\text{S}_1^2}+1\right) \left(C_{\text{S}_2^2}-1\right) \left(\frac{\left(-95 \sigma ^4+102 \sigma
   ^2-15\right) \left(n\cdot S_1\right){}^2 \left(q\cdot S_2\right){}^2}{256 m_2^2
   \left(\sigma ^2-1\right)^2} \right. \nn
   \\ &\phantom{=asdf} \left. \phantom{asdf} +\frac{m_1^2 q^2 \left(35 \sigma ^4-30 \sigma ^2+3\right)
   \left(p_2\cdot S_1\right){}^2 \left(q\cdot S_2\right){}^2}{256 m_2^2 \left(\sigma
   ^2-1\right)^2}\right) \nn
   \\ &\phantom{=as} + \left(C_{\text{S}_1^2}-1\right) \left(C_{\text{S}_2^2}+1\right) \left(-\frac{\left(15 \sigma ^4-15 \sigma
   ^2+2\right) \left(n\cdot S_2\right){}^2 \left(q\cdot S_1\right){}^2}{64 m_2^2
   \left(\sigma ^2-1\right)^2}\right. \nn
   \\ &\phantom{=asdf} \left. \phantom{asdf}-\frac{q^2 \left(1-3 \sigma ^2\right) \left(p_1\cdot
   S_2\right){}^2 \left(q\cdot S_1\right){}^2}{64 \left(\sigma ^2-1\right)^2}\right) \nn
   \\ &\phantom{=as} + \frac{3}{64} m_1^2 \left(5 \sigma ^2-1\right) \left(C_{\text{S}_1^2}-1\right)
   \left(C_{\text{S}_2^2}-1\right) \left(q\cdot S_1\right){}^2 \left(q\cdot S_2\right){}^2\,. \label{eq:tricoeff_2-2}
\el
Although the results eqs.(\ref{eq:tricoeff_0-2}-\ref{eq:tricoeff_2-2}) are not full results, the coefficients provide full information for black holes $C_{\text{S}^{n}} = 1$ since the same helicity graviton exchange contributions $c_{\bigtriangleup, \pm\pm}$ do not contribute at this order in $\hbar$~\cite{Guevara:2017csg}. The vanishing of same helicity exchange contributions was confirmed explicitly for the sectors $S_1^n S_2^0$ and $S_1^n S_2^1$ provided in eqs.(\ref{eq:tricoeff_2-0}-\ref{eq:tricoeff_3-1}). The following remarks could be of interest.
\bn
\item The same helicity graviton exchange contributions for the sectors $S_1^3 S_2^0$ of \eqc{eq:tricoeff_3-0} and $S_1^2 S_2^1$ of \eqc{eq:tricoeff_2-1} were found to be nonzero, but $c_{\bigtriangleup, ++}$ contributions cancel $c_{\bigtriangleup, --}$ contributions for generic Wilson coefficients. It is unclear if the pattern persists to all $\CO(S^3)$ sectors. For other computed sectors of eqs.(\ref{eq:tricoeff_2-0}-\ref{eq:tricoeff_3-1}), the same helicity exchange contributions $c_{\bigtriangleup, \pm\pm} = c_{\bigtriangleup, ++} + c_{\bigtriangleup, --}$ are non-vanishing.
\item The Kerr limit $C_{\text{S}^{n}} = 1$ kills all terms with spin contractions of the form $(q \cdot S)$ in the triangle coefficients, where all spin contractions were expressed using the basis $\{ (q \cdot S)\,,\, (p \cdot S) \,,\,(n \cdot S)\}$. This partly explains the pattern observed in refs.\cite{Bern:2020buy,Kosmopoulos:2021zoq}, where terms of the form $q^2 S^2$ and $(q \cdot S)^2$ appeared in the Hamiltonian only through the combination
\bl
q^2 S_i \cdot S_j - (q \cdot S_i) (q \cdot S_j) \,, \nn
\el
in the Kerr limit. This is the combination obtained by applying Levi-Civita identities to $(n \cdot S_i)(n \cdot S_j)$. The observation extends this pattern to higher spin orders.
\en

\section{Solving MPD equations iteratively to linear order in test-body spin} \label{app:MPD}
\subsection{Spin vector description of MPD equations}
The Mathisson-Papapetrou-Dixon (MPD) equations describe motion of a spinning body on a curved background, and leading order equations are given as~\cite{Mathisson:1937zz,Papapetrou:1951pa,doi:10.1098/rspa.1970.0020}
\bl
\bld
\frac{Dp^\m}{D \t} &= - \half R^{\m}_{~\n\a\b} u^\n S^{\a\b}\,,
\\ \frac{DS^{\m\n}}{D \t} &= p^\m u^\n - p^\n u^\m\,,
\eld \label{eq:MPDorig}
\el
where $S^{\m\n}$ is the rank-2 spin tensor, $p^\m$ is the 4-momentum, and $u^\m = Dx^\m / D \t$ is the 4-velocity of the particle moving on the worldline $x^\m (\t)$ where $\t$ is the proper time. $\frac{D}{D\t}$ denotes parallel transport on the worldline which will be abbreviated by a dot above the tensor, i.e. $\dot{a}^\m = D a^\m /D \t = u^\n \nabla_\n a^\m$.

Using the covariant SSC (also known as the Tulczyjew-Dixon SSC~\cite{doi:10.1098/rspa.1970.0020,tulczyjew1959motion}) $S^{\m\n} p_\n = 0$ and only keeping leading and sub-leading orders in $S$, we find
\bl
0 = \frac{D \left( S^{\m\n} p_\n \right)}{D\t} = \dot{S}^{\m\n} p_\n + S^{\m\n} \dot{p}_\n = p^\m (u \cdot p) - u^\m p^2 + \CO(S^2) \,,
\el
which means $p^\m = m u^\m + \CO(S^2)$ where $m^2 = p^2$ is the rest mass. For the covariant SSC the spin tensor can be described using the spin vector instead, which is given by the relations
\bl
\bld
S^\m &= - \frac{1}{2m} \e^{\m\a\b\g} p_\a S_{\b\g}\,,
\\ S^{\m\n} &= - \frac{1}{m} \e^{\m\n\a\b} p_\a S_{\b}\,,
\eld
\el
and the equations of motion for the spin vector becomes\footnote{Inserting the condition $p^\m = m u^\m + \CO(S^2)$ into $p \cdot \dot{p}$ and using \eqc{eq:MPDorig} gives $p \cdot \dot{p} = \CO(S^3)$.
}
\bl
\dot{S}^\m &= - \frac{1}{m^2} \left( p^\m \dot{p}_\n S^\n - S^\m p_\n \dot{p}^\n \right) = - u^\m \dot{u}_\n S^\n + \CO(S^4) \,, \label{eq:FWderiv}
\el
which is just the Fermi-Walker transport\footnote{A generic Fermi-Walker transport is given by the equation $\dot{v}^\m = \dot{u}^\m u_\n v^\n - u^\m \dot{u}_\n v^\n$, but since $u \cdot S = 0$ the first term drops out.} along the worldline $x^\m (\t)$ when neglecting $\CO(S^4)$ terms. Using the rescaled spin $s^\m = S^\m / m$, the equations of motion are given as
\bl
\bld
\dot{u}^\m &= \half R^{\m}_{~\a\b\g} \e^{\b\g\delta\s} u^\a u_\delta s_\s + \CO(s^2) \,,
\\ \dot{s}^\m &= - u^\m \dot{u}_\n s^\n + \CO(s^4) \,.
\eld \label{eq:EOM}
\el

\subsection{The master perturbation equations} \label{app:MPDpert}
Solving the equations \eqc{eq:EOM} directly is not an easy task in general, so we attempt to solve it perturbatively using the following ansatz
\bl
\bld
x^\m (\t_0) &= b_0^\m + u_0^\m \t_0 + \delta x^\m (\t_0)\,,
\\ s^\m (\t_0) &= s_0^\m + \delta s^\m (\t_0)\,,
\eld
\el
where the worldline parameter $\t_0$ runs from $\t_0 = -\infty$ (incoming state) to $\t_0 = +\infty$ (outgoing state) and $u_0 \cdot s_0 = 0$ from the orthogonality condition $p \cdot S = 0$. The worldline parameter $\t_0$ is the proper time for the unperturbed worldline $x_0^\m = b_0^\m + u_0^\m \t_0$, and unperturbed 4-velocity $u_0^\m = (\frac{1}{\sqrt{1-v_0^2}}, \frac{\vec{v}_0}{\sqrt{1-v_0^2}})$ is normalised as $\eta_{\m\n} u_0^\m u_0^\n = 1$, where the metric $g_{\m\n} = \eta_{\m\n} + h_{\m\n}$ is decomposed into the flat metric $\eta_{\m\n}$ and weak gravitational field $h_{\m\n}$ which vanishes at infinity. We take the impact parameter $b_0^\m = (0, \vec{b}_0)$ to be a purely spatial vector which is orthogonal to $u_0^\m$; $u_0 \cdot b_0 = 0$.

We introduce a formal expansion parameter $\l$ that will later be identified with the Newton's constant $G$;
\bl
\bld
\delta x^\m (\t_0) &= \l \xi_{(1)}^\m (\t_0) + \l^2 \xi_{(2)}^\m (\t_0) + \cdots\,,
\\ \delta s^\m (\t_0) &= \l \s_{(1)}^\m (\t_0) + \l^2 \s_{(2)}^\m (\t_0) + \cdots\,.
\eld
\el
The equations of motion \eqc{eq:EOM} can be expanded as a power series in $\l$ and solved order by order in $\l$ iteratively. For this purpose it is better to reparametrise the worldline in terms of unperturbed worldline's proper time $\t_0$, which leads to
\bl
\bld
\frac{d}{d\t_0} \left( \frac{d x^\m}{d\t_0} \right) &= \frac{d x^\m}{d \t_0} \frac{d}{d\t_0} \log \left(\frac{d \t}{d \t_0} \right) - \G^\m_{\a\b} \frac{d x^\a}{d\t_0} \frac{d x^\b}{d\t_0} + \half \bar{R}^{\m}_{~\a\b\g} \frac{d x^\a}{d\t_0} \frac{d x^\b}{d\t_0} s^\g\,,
\\ \frac{d s^\m}{d\t_0} &= - \G^\m_{\a\b} \frac{d x^\a}{d\t_0} s^\b - \half \frac{d x^\m}{d\t_0} \left( \frac{d\t}{d\t_0} \right)^{-2} \left( \bar{R}_{\n\a\b\g} s^\n \frac{d x^\a}{d\t_0} \frac{d x^\b}{d\t_0} s^\g \right)\,,
\eld \label{eq:EOM2}
\el
where the dual Riemann tensor $\bar{R}^{\m}_{~\n\a\b}$ has been introduced to simplify the equations.
\bl
\bar{R}^{\m}_{~\n\a\b} &= R^{\m}_{~\n\l\s} \e^{\l\s}_{\phantom{\l\s}\a\b} \,.
\el
The raising and lowering of indices are done with respect to the full metric $g_{\m\n}$ and $g^{\m\n}$ up to this point. The exact differential proper time $d\t$ is given as
\bl
d\t^2 = g_{\m\n} dx^\m dx^\n && \Rightarrow && \frac{d\t}{d\t_0} = \left( (\eta_{\m\n} + h_{\m\n}) \frac{d x^\m}{d\t_0} \frac{d x^\n}{d\t_0} \right)^{1/2} \,.
\el

The boundary conditions for the spin deviations $\s^\m_{(i)}$ are simply
\bl
\s^\m_{(i)} (\t_0 \to -\infty) &= 0 \,.
\el
However, we cannot use the same boundary conditions for the position deviations $\xi^\m_{(i)}$: There is an IR divergence due to long-ranged nature of Coulomb-like potentials coming from integrating velocity deviations at large distances $(\vec{v} - \vec{v}_0) \propto \frac{1}{r}$. The resolution used in refs.\cite{Mogull:2020sak,Jakobsen:2021zvh} and related works was to fix the boundary conditions at $\t_0 = 0$ instead, which is defined as the time of closest encounter. This approach is rather inconvenient for our purposes, since we need to invert the relations between variables at $\t_0 = 0$ and variables at $\t_0 \to -\infty$ to specify the initial scattering conditions.

The approach we take is to allow arbitrary ``subtractions'' $\prescript{(0)}{} {\xi^\m_{(i)}}$ of logarithmic dependence on $\t_0$ which do not change the trajectory of the test particle at $\t_0 \to -\infty$;
\bl
\bgd
\xi^\m_{(i)} = \prescript{(0)}{} {\xi^\m_{(i)}} + \prescript{(1)}{} {\xi^\m_{(i)} } \,, \prescript{(1)}{} {\xi^\m_{(i)}} (\t_0 \to -\infty) = 0 \,,
\\ \prescript{(0)}{} {\xi^\m_{(i)}} (\t_0 \to -\infty) \propto ( c_1 \log \abs{\t_0} , c_2 \vec{v}_0 \log \abs{\t_0} ) \,.
\egd
\el
The conditions ensure that initial scattering conditions are unchanged; $\frac{d \xi^\m_{(i)}}{d\t_0} (\t_0 \to - \infty) = 0$. An analogous boundary condition was used in ref.\cite{Saketh:2021sri}.

For computing $\CO(G^2)$ order scattering variables, we only need to perform the subtraction to $\xi^\m_{(1)}$. The logarithmically divergent terms coming from indefinite integration of $\frac{d \xi^\m_{(1)}}{d\t_0}$ were observed to be one of the two combinations, both of which we separate out as $\prescript{(0)}{} {\xi^\m_{(i)}}$.
\bl
\log \left[ b_0^2 (1 - v_0^2) + v_0^2 \t_0^2 \right] \,,\, \log \left[ v_0 \t_0 + \sqrt{b_0^2 (1 - v_0^2) + v_0^2 \t_0^2} \right] \,.
\el

\subsection{Iterative solution on Kerr-Schild coordinates}
The Kerr metric in Kerr-Schild coordinates\footnote{For a review, consult e.g. ref.\cite{Stephani:2003tm}.} is
\bl
\bld
g_{\m\n} &= \eta_{\m\n} - f k_\m k_\n \,,
\\ f &= \frac{2 G M r^3}{r^4 + a^2 z^2}\,,
\\ k_\m &= \left( 1, \frac{rx + ay}{r^2 + a^2}, \frac{ry - ax}{r^2 + a^2}, \frac{z}{r} \right)\,,
\eld
\el
where $r$ is implicitly defined by the relation
\bl
1 = \frac{x^2 + y^2}{r^2 + a^2} + \frac{z^2}{r^2} 
\,.
\el
The explicit solution is given as
\bl
2 r^2 = x^2 + y^2 + z^2 - a^2 + \sqrt{(x^2 + y^2 + z^2 - a^2)^2 + 4 a^2 z^2} \,,
\el
which is determined by the asymptotic behaviour $r \approx \sqrt{x^2 + y^2 + z^2}$ at infinity.

We use $G$ as the perturbation parameter. The quantities involving $G$ are given as;
\bl
\bld
h_{\a\b} &= G \bar{h}_{\a\b}\,,
\\ \G^{\a}_{\b\g} &= G \prescript{(1)}{}\G^{\a}_{\b\g} + G^2 \prescript{(2)}{}\G^{\a}_{\b\g}\,,
\\ R^{\m}_{~\n\a\b} &= G \prescript{(1)}{}R^{\m}_{~\n\a\b} + G^2 \prescript{(2)}{}R^{\m}_{~\n\a\b} + \cdots\,,
\\ \bar{R}^{\m}_{~\n\a\b} &= G \prescript{(1)}{}{\bar{R}}^{\m}_{~\n\a\b} + G^2 \prescript{(2)}{}{\bar{R}}^{\m}_{~\n\a\b} + \cdots\,.
\eld
\el
Each term in \eqc{eq:EOM2} can be reorganised as a series in $G$;
\bl
\log \left(\frac{d \t}{d \t_0} \right) &= \half \log \left( 1 + G \left[ 2 u_0 \cdot \dot{\xi}_{(1)} + \bar{h} (u_0 , u_0) \right] \right. \nn
\\ &\phantom{=} \left. \phantom{asdfasdf} + G^2 \left[ \dot{\xi}_{(1)}^2 + 2 u_0 \cdot \dot{\xi}_{(2)} + 2 \bar{h}(u_0 , \dot{\xi}_{(1)}) \right] \right) \nn
\\ &= G \left[ u_0 \cdot \dot{\xi}_{(1)} + \half \bar{h} (u_0 , u_0) \right] + G^2 \left[ \frac{\delta_{(1)} \bar{h} (u_0 , u_0)}{2} + u_0 \cdot \dot{\xi}_{(2)} \right. \nn
\\ &\phantom{=} 
\left. \phantom{asdfasdfasdf} + \bar{h}(u_0 , \dot{\xi}_{(1)}) + \half \dot{\xi}_{(1)}^2 - \left( u_0 \cdot \dot{\xi}_{(1)} + \half \bar{h} (u_0 , u_0) \right)^2 \right] \nn\,,
\\ \G^\m_{\a\b} \frac{d x^\a}{d\t_0} \frac{d x^\b}{d\t_0} &= G \prescript{(1)}{} \G^\m_{\a\b} u_0^\a u_0^\b + G^2 \left[ \delta_{(1)} \prescript{(1)}{} \G^\m_{\a\b} u_0^\a u_0^\b + \prescript{(2)}{} \G^\m_{\a\b} u_0^\a u_0^\b + 2 \prescript{(1)}{} \G^\m_{\a\b} u_0^\a \dot{\xi}_{(1)}^\b \right] \nn\,,
\\ \bar{R}^{\m}_{~\a\b\g} \frac{d x^\a}{d\t_0} \frac{d x^\b}{d\t_0} s^\g &= G \prescript{(1)}{}{\bar{R}}^{\m}_{~\a\b\g} u_0^\a u_0^\b s_0^\g + G^2 \left[ \delta_{(1)} \prescript{(1)}{}{\bar{R}}^{\m}_{~\a\b\g} u_0^\a u_0^\b s_0^\g + \prescript{(1)}{}{\bar{R}}^{\m}_{~\a\b\g} \dot{\xi}_{(1)}^\a u_0^\b s_0^\g \right. \nn
\\ &\phantom{=asd} \left. \phantom{asdf} + \prescript{(1)}{}{\bar{R}}^{\m}_{~\a\b\g} u_0^\a \dot{\xi}_{(1)}^\b s_0^\g + \prescript{(1)}{}{\bar{R}}^{\m}_{~\a\b\g} u_0^\a u_0^\b \s_{(1)}^\g + \prescript{(2)}{}{\bar{R}}^{\m}_{~\a\b\g} u_0^\a u_0^\b s_0^\g \right] \nn \,,
\\ \G^\m_{\a\b} \frac{d x^\a}{d\t_0} s^\b &= G \prescript{(1)}{}\G^\m_{\a\b} u_0^\a s_0^\b + G^2 \left[ \delta_{(1)} \prescript{(1)}{}\G^\m_{\a\b} u_0^\a s_0^\b + \prescript{(1)}{}\G^\m_{\a\b} \dot{\xi}_{(1)}^\a s_0^\b \right. \nn
\\ &\phantom{=asdfasdf} \left. \phantom{asdfasdfasdf} + \prescript{(1)}{}\G^\m_{\a\b} u_0^\a \s_{(1)}^\b + \prescript{(2)}{}\G^\m_{\a\b} u_0^\a s_0^\b \right] \nn \,,
\\ \bar{R}_{\n\a\b\g} s^\n \frac{d x^\a}{d\t} \frac{d x^\b}{d\t} s^\g &= \left( \frac{d\t}{d\t_0} \right)^{-2} \left( \bar{R}_{\n\a\b\g} s^\n \frac{d x^\a}{d\t_0} \frac{d x^\b}{d\t_0} s^\g \right) \nn \,,
\\ &= G \prescript{(1)}{}{\bar{R}}_{\n\a\b\g} s_0^\n u_0^\a u_0^\b s_0^\g \nn
\\ &\phantom{=} + G^2 \left[ \delta_{(1)} \prescript{(1)}{}{\bar{R}}_{\n\a\b\g} s_0^\n u_0^\a u_0^\b s_0^\g + \prescript{(2)}{}{\bar{R}}_{\n\a\b\g} s_0^\n u_0^\a u_0^\b s_0^\g \right. \nn
\\ &\phantom{=asdf} + \prescript{(1)}{}{\bar{R}}_{\n\a\b\g} \s_{(1)}^\n u_0^\a u_0^\b s_0^\g + \prescript{(1)}{}{\bar{R}}_{\n\a\b\g} s_0^\n \dot{\xi}_{(1)}^\a u_0^\b s_0^\g \nn
\\ &\phantom{=asdf} + \prescript{(1)}{}{\bar{R}}_{\n\a\b\g} s_0^\n u_0^\a \dot{\xi}_{(1)}^\b s_0^\g + \prescript{(1)}{}{\bar{R}}_{\n\a\b\g} s_0^\n u_0^\a u_0^\b \s_{(1)}^\g \nn
\\ &\phantom{=asdf} \left. \phantom{asdf} - \left( 2 u_0 \cdot \dot{\xi}_{(1)} + \bar{h} (u_0 , u_0) \right) \prescript{(1)}{}{\bar{R}}_{\n\a\b\g} s_0^\n u_0^\a u_0^\b s_0^\g \right] \nn\,,
\el
where $v \cdot w := \eta_{\a\b} v^\a w^\b$, $\dot{v} := \frac{dv}{d\t_0}$, $\bar{h}(v,w) := \bar{h}_{\a\b} v^\a w^\b$, and $\delta_{(1)} f := \xi_{(1)}^\m \p_\m f$. Reorganising \eqc{eq:EOM2} in a power series of $G$, the leading perturbation term is given as
\bl
\bld
P^{\m}_{\phantom{\m}\n} \frac{d^2 \xi^\n_{(1)}}{d \t_0^2} &= \frac{u_0^\m}{2} \frac{d \bar{h}(u_0,u_0)}{d\t_0} - \prescript{(1)}{} \G^{\m}_{\a\b} u_0^\a u_0^\b + \half \prescript{(1)}{} {\bar{R}}^{\m}_{~\a\b\g} u_0^\a u_0^{\b} s_{0}^{\g}\,,
\\ \frac{d \s_{(1)}^\m}{d \t_0} &= - \prescript{(1)}{} \G^{\m}_{\a\b} u_0^\a s_0^\b - \frac{u_0^\m}{2} \prescript{(1)}{} {\bar{R}}_{\n\a\b\g} s_0^\n u_0^\a u_{0}^{\b} s_{0}^{\g}\,,
\eld \label{eq:EOMG1}
\el
where indices are raised and lowered using $\eta_{\a\b}$ and $\eta^{\a\b}$. The projection tensor is defined as $P^{\m}_{\phantom{\m}\n} := \delta^\m_\n - u_0^\m u_{0\n}$. The values of the background are determined by $f(\t_0) = f(x^\m(\t_0)) = f(b^\m + u_0^\m \t_0)$ up to this order. The position part of \eqc{eq:EOMG1} can be written in a more compact form by applying the projection tensor to both sides.
\bl
P^{\m}_{\phantom{\m}\n} \frac{d^2 \xi^\n_{(1)}}{d \t_0^2} &= - P^{\m}_{\phantom{\m}\n} \prescript{(1)}{} \G^{\n}_{\a\b} u_0^\a u_0^\b + \half P^{\m}_{\phantom{\m}\n} \prescript{(1)}{} {\bar{R}}^{\n}_{~\a\b\g} u_0^\a u_0^{\b} s_{0}^{\g}\,. \label{eq:EOMG1pos}
\el
The contents of the first line of \eqc{eq:EOMG1} and \eqc{eq:EOMG1pos} are actually equivalent, because contracting $u_0^\m$ to the RHS of first line of \eqc{eq:EOMG1} turns out to be identically zero at this perturbation order.

For the subleading order accelerations, we use the condition $u_0 \cdot \dot{\xi}_{(1)} = 0$ to simplify the equations. The position acceleration is given as;
\bl
\bld
P^{\m}_{\phantom{\m}\n} \frac{d^2 \xi^\n_{(2)}}{d \t_0^2} &= u_0^\m \frac{d}{d\t_0} \left[ \frac{\xi_{(1)}^\a \partial_\a \bar{h} (u_0 , u_0)}{2} + \bar{h}(u_0 , \dot{\xi}_{(1)}) + \half \dot{\xi}_{(1)}^2 - \frac{\bar{h} (u_0 , u_0)^2}{4} \right] + \frac{\dot{\xi}_{(1)}^\m}{2}  \frac{d \bar{h}(u_0,u_0)}{d\t_0}
\\ &\phantom{=asdfasd} - \left( \xi_{(1)}^\a \p_\a \prescript{(1)}{} \G^\m_{\a\b} \right) u_0^\a u_0^\b - \prescript{(2)}{} \G^\m_{\a\b} u_0^\a u_0^\b - 2 \prescript{(1)}{} \G^\m_{\a\b} u_0^\a \dot{\xi}_{(1)}^\b
\\ &\phantom{=as} + \half \left[ \left( \xi_{(1)}^\a \p_\a \prescript{(1)}{}{\bar{R}}^{\m}_{~\a\b\g} \right) u_0^\a u_0^\b s_0^\g + \prescript{(2)}{}{\bar{R}}^{\m}_{~\a\b\g} u_0^\a u_0^\b s_0^\g \right.
\\ &\phantom{=as} \left.\phantom{asdfas} + \prescript{(1)}{}{\bar{R}}^{\m}_{~\a\b\g} \dot{\xi}_{(1)}^\a u_0^\b s_0^\g + \prescript{(1)}{}{\bar{R}}^{\m}_{~\a\b\g} u_0^\a \dot{\xi}_{(1)}^\b s_0^\g + \prescript{(1)}{}{\bar{R}}^{\m}_{~\a\b\g} u_0^\a u_0^\b \s_{(1)}^\g \right] \,,
\eld \label{eq:EOMG2pos}
\el
while the spin evolution equation is given as;
\bl
\bld
\frac{d \s_{(2)}^\m}{d \t_0} &= - \left( \xi_{(1)}^\a \p_\a \prescript{(1)}{}\G^\m_{\a\b} \right) u_0^\a s_0^\b - \prescript{(1)}{}\G^\m_{\a\b} \dot{\xi}_{(1)}^\a s_0^\b - \prescript{(1)}{}\G^\m_{\a\b} u_0^\a \s_{(1)}^\b - \prescript{(2)}{}\G^\m_{\a\b} u_0^\a s_0^\b
\\ &\phantom{=as} - \frac{u_0^\m}{2} \left[ \left( \xi_{(1)}^\s \p_\s \prescript{(1)}{}{\bar{R}}_{\n\a\b\g} \right) s_0^\n u_0^\a u_0^\b s_0^\g + \prescript{(2)}{}{\bar{R}}_{\n\a\b\g} s_0^\n u_0^\a u_0^\b s_0^\g + \prescript{(1)}{}{\bar{R}}_{\n\a\b\g} \s_{(1)}^\n u_0^\a u_0^\b s_0^\g \right.
\\ &\phantom{=asdfasd} + \prescript{(1)}{}{\bar{R}}_{\n\a\b\g} s_0^\n \dot{\xi}_{(1)}^\a u_0^\b s_0^\g + \prescript{(1)}{}{\bar{R}}_{\n\a\b\g} s_0^\n u_0^\a \dot{\xi}_{(1)}^\b s_0^\g + \prescript{(1)}{}{\bar{R}}_{\n\a\b\g} s_0^\n u_0^\a u_0^\b \s_{(1)}^\g
\\ &\phantom{=asdf} \left. \phantom{asd} - \bar{h} (u_0 , u_0) \prescript{(1)}{}{\bar{R}}_{\n\a\b\g} s_0^\n u_0^\a u_0^\b s_0^\g \right] - \frac{\dot{\xi}_{(1)}^\m}{2} \left[ \prescript{(1)}{} {\bar{R}}_{\n\a\b\g} s_0^\n u_0^\a u_{0}^{\b} s_{0}^{\g} \right]\,.
\eld
\el
We only integrate \eqc{eq:EOMG2pos} once, since we are only interested in scattering variables at $\CO(G^2)$ order.

\subsection{Example: equatorial motion with orthogonal initial spin} \label{app:MPDexample1}
We start with the unperturbed worldline and spin orientation\footnote{For brevity, we drop the subscripts of initial velocity and impact parameter in this section.}
\bl
\bld
x^\mu (\t_0) &= \left( \frac{\t_0}{\sqrt{1-v^2}} \,, b \,, \frac{v \t_0}{\sqrt{1-v^2}} \,, 0 \right) \Rightarrow u_0^\m = \left( \frac{1}{\sqrt{1-v^2}} \,, 0 \,, \frac{v}{\sqrt{1-v^2}} \,, 0 \right)\,,
\\ s_0^\m (\t_0) &= \left( 0 \,, s_0 \,, 0 \,, 0 \right)\,.
\eld
\el
Since the results are rather lengthy, we present the results component by component. We denote the worldline deflection by $\D v^\m = \left[G \dot{\xi}_{(1)}^\mu + G^2 \dot{\xi}_{(2)}^\mu \right] (+\infty)$ and the spin kick by $\D s^\m = \left[G \s_{(1)}^\mu + G^2 \s_{(2)}^\mu \right] (+\infty)$. The results were computed under the assumption $b > 0$, to $\CO(G^2, a^4, s_0^1)$ order.
\begin{align*}
\D v^0 &= {(GM)}^2 \left(-\frac{2 \left(v^2+1\right)^2}{b^2 v^2 \left(1-v^2\right)^{3/2}}+\frac{8
   \left(v^2+1\right) a}{b^3 v \left(1-v^2\right)^{3/2}}-\frac{4 \left(v^4+4 v^2+1\right)
   a^2}{b^4 v^2 \left(1-v^2\right)^{3/2}} \right.
\\  &\phantom{=} \left.\phantom{asdfasdfasdf}+\frac{16 \left(v^2+1\right) a^3}{b^5 v
   \left(1-v^2\right)^{3/2}}-\frac{\left(6 v^4+28 v^2+6\right) a^4}{b^6 v^2
   \left(1-v^2\right)^{3/2}} \right)\,,
\\ \D v^1 &= {GM} \left(-\frac{2 \left(v^2+1\right)}{b v \sqrt{1-v^2}}+\frac{4 a}{b^2
   \sqrt{1-v^2}}-\frac{2 \left(v^2+1\right) a^2}{b^3 v \sqrt{1-v^2}}+\frac{4 a^3}{b^4
   \sqrt{1-v^2}}-\frac{2 \left(v^2+1\right) a^4}{b^5 v
   \sqrt{1-v^2}}\right)
   \\ &\phantom{=}+{(GM)}^2 \left(-\frac{3 \left(\pi 
   \left(v^2+4\right)\right)}{4 \left(b^2 v \sqrt{1-v^2}\right)}+\frac{2 \pi  \left(3
   v^2+2\right) a}{b^3 v^2 \sqrt{1-v^2}}-\frac{3 \left(\pi  \left(15 v^4+72
   v^2+8\right)\right) a^2}{16 \left(b^4 v^3 \sqrt{1-v^2}\right)} \right.
   \\ &\phantom{=}\left. \phantom{asdfasdfasdfasdf}+\frac{3 \pi  \left(5
   v^2+4\right) a^3}{b^5 v^2 \sqrt{1-v^2}}-\frac{5 \left(\pi  \left(35 v^4+180
   v^2+24\right)\right) a^4}{32 \left(b^6 v^3
   \sqrt{1-v^2}\right)}\right)\,,
\\ \D v^2 &= {(GM)}^2 \left(-\frac{2 \left(v^2+1\right)^2}{b^2 v^3 \left(1-v^2\right)^{3/2}}+\frac{8
   \left(v^2+1\right) a}{b^3 v^2 \left(1-v^2\right)^{3/2}}-\frac{4 \left(v^4+4 v^2+1\right)
   a^2}{b^4 v^3 \left(1-v^2\right)^{3/2}}\right.
   \\ &\phantom{=}\left. \phantom{asdfasdfasdf}+\frac{16 \left(v^2+1\right) a^3}{b^5 v^2
   \left(1-v^2\right)^{3/2}}-\frac{\left(6 v^4+28 v^2+6\right) a^4}{b^6 v^3
   \left(1-v^2\right)^{3/2}}\right)\,,
\\ \D v^3 &= {GM} s_0 \left(\frac{4}{b^2 \sqrt{1-v^2}}-\frac{4 \left(v^2+1\right) a}{b^3 v
   \sqrt{1-v^2}}+\frac{12 a^2}{b^4 \sqrt{1-v^2}}-\frac{8 \left(v^2+1\right) a^3}{b^5 v
   \sqrt{1-v^2}}+\frac{20 a^4}{b^6 \sqrt{1-v^2}})\right)
   \\ &\phantom{=}+{(GM)}^2 s_0 \left(\frac{3 \pi  \left(3
   v^2+2\right)}{4 b^3 v^2 \sqrt{1-v^2}}-\frac{3 \left(\pi  \left(3 v^4+15
   v^2+2\right)\right) a}{4 \left(b^4 v^3 \sqrt{1-v^2}\right)}+\frac{3 \pi  \left(51
   v^2+44\right) a^2}{8 b^5 v^2 \sqrt{1-v^2}} \right.
   \\ &\phantom{=} \left. \phantom{asdfasdfasdf}-\frac{15 \left(\pi  \left(5 v^4+27
   v^2+4\right)\right) a^3}{8 \left(b^6 v^3 \sqrt{1-v^2}\right)}+\frac{15 \pi  \left(125
   v^2+114\right) a^4}{32 b^7 v^2 \sqrt{1-v^2}}\right)\,,
\\ \D s^0 &= {GM} s_0 \left(-\frac{2}{b v}+\frac{2 a}{b^2}-\frac{2 a^2}{b^3 v}+\frac{2
   a^3}{b^4}-\frac{2 a^4}{b^5 v}\right)
   \\ &\phantom{=}+{(GM)}^2 s_0 \left(-\frac{3 \pi }{2 \left(b^2
   v\right)}+\frac{3 \pi  \left(v^2+2\right) a}{2 b^3 v^2}-\frac{3 \left(\pi  \left(9
   v^2+2\right)\right) a^2}{4 \left(b^4 v^3\right)} \right.
   \\ &\phantom{=}\left. \phantom{asdfasdfasdf}+\frac{3 \pi  \left(5 v^2+12\right)
   a^3}{4 b^5 v^2}-\frac{15 \left(\pi  \left(15 v^2+4\right)\right) a^4}{16 \left(b^6
   v^3\right)} \right)\,,
\\ \D s^1 &= {(GM)}^2 s_0 \left(\frac{2-2 v^2}{b^2 v^2}+\frac{\left(2-2 v^2\right) a^2}{b^4
   v^2}+\frac{\left(2-2 v^2\right) a^4}{b^6 v^2}\right)\,,
\\ \D s^2 &= {GM} s_0 \left(\frac{2}{b}-\frac{2 a}{b^2 v}+\frac{2 a^2}{b^3}-\frac{2 a^3}{b^4
   v}+\frac{2 a^4}{b^5}\right)
   \\ &\phantom{=}+{(GM)}^2 s_0 \left(\frac{3 \pi 
   \left(v^2+2\right)}{4 b^2 v^2}-\frac{\left(\pi  \left(9 v^2+2\right)\right) a}{2
   \left(b^3 v^3\right)}+\frac{9 \pi  \left(5 v^2+12\right) a^2}{16 b^4 v^2} \right.
   \\ &\phantom{=}\left. \phantom{asdfasdfasdf}-\frac{3
   \left(\pi  \left(15 v^2+4\right)\right) a^3}{4 \left(b^5 v^3\right)}+\frac{25 \pi 
   \left(7 v^2+18\right) a^4}{32 b^6 v^2}\right)\,,
\\ \D s^3 &= 0\,.
\end{align*}
The orthogonality condition for spin is maintained, i.e. $\eta_{\m\n} (u_0^\m + \D v^\m) (s_0^\n + \D s^\n) = \CO (G^3)$.

\newpage

\bibliography{mybib}{}
\bibliographystyle{JHEP}
\end{document}